\documentclass[11pt]{article}
\usepackage{verbatim,amsmath,amssymb}
\usepackage{epsfig,float,color}
\usepackage[font=small,labelfont=bf]{caption}
\usepackage{geometry}
\usepackage{setspace}
\usepackage[table]{xcolor}
\usepackage{enumerate}
\usepackage{enumitem}
\usepackage{natbib}
\usepackage{amsthm,mathrsfs,amsfonts,dsfont} 
\usepackage{bm}
\usepackage{hyperref}
\usepackage[labelfont=bf]{caption}
\usepackage{lipsum} 												
\usepackage{booktabs}												
\usepackage{multirow}												
\usepackage{siunitx}
\usepackage{csquotes}

\definecolor{darkblue}{rgb}{0,0,1}
\definecolor{col1}{rgb}{1,0,1} 			
\definecolor{col2}{rgb}{0,0.5,0}		
\definecolor{col3}{rgb}{0.5,0,1}		
\definecolor{col4}{rgb}{0.1,.75,0}		

\hypersetup{pdftex=true, colorlinks=true, breaklinks=true, linkcolor=darkblue, menucolor=darkblue, pagecolor=darkblue, citecolor=darkblue, urlcolor=darkblue}

\newtheoremstyle{rem}
{6pt}
{6pt}
{\small}
{}
{\bf}
{:}
{.5em}
{}

\theoremstyle{rem}

\newcommand{\bitm}{\begin{itemize}}
\newcommand{\eitm}{\end{itemize}}
\newcommand{\bnumr}{\begin{enumerate}}
\newcommand{\enumr}{\end{enumerate}}

\newcommand{\errLe}{\delta_{L_2}}

\newcommand{\errM}{\delta_\mathrm{max}}
\newcommand{\errME}{\delta_\mathrm{max}^E}
\newcommand{\errMT}{\delta_\mathrm{max}^T}

\newcommand {\aab}{a^{\alpha\beta}}
\newcommand {\auab}{a_{\alpha\beta}}

\newcommand {\augd}{a_{\gamma\delta}}

\newcommand {\Aab}{A^{\alpha\beta}}
\newcommand {\Auab}{A_{\alpha\beta}}
\newcommand {\Agd}{A^{\gamma\delta}}
\newcommand {\Augd}{A_{\gamma\delta}}
\newcommand {\Abd}{A^{\beta\delta}}
\newcommand {\Abg}{A^{\beta\gamma}}
\newcommand {\Aad}{A^{\alpha\delta}}
\newcommand {\Aag}{A^{\alpha\gamma}}

\newcommand {\Mab}{M^{\alpha\beta}}

\newcommand {\bab}{b^{\alpha\beta}}

\newcommand {\Buab}{B_{\alpha\beta}}

\newcommand {\T}{\mathrm{T}}

\newcommand {\buab}{b_{\alpha\beta}}

\newcommand {\bugd}{b_{\gamma\delta}}
\newcommand {\Bugd}{B_{\gamma\delta}}

\newcommand {\tauab}{\tau^{\alpha\beta}}

\newcommand{\mcalS}{\mathcal{S}}
\newcommand{\mcalN}{\mathcal{N}}

\newcommand{\mrel}{\mathrm{el}}
\newcommand{\mrint}{\mathrm{int}}
\newcommand{\mrext}{\mathrm{ext}}
\newcommand{\mrect}{\mathrm{exact}}
\newcommand{\mrll}{\mathrm{ll}}
\newcommand{\mrlc}{\mathrm{lc}}
\newcommand{\mrex}{\mathrm{exp}}
\newcommand{\mrfe}{\mathrm{FE}}

\newcommand{\mrno}{\mathrm{no}}
\newcommand{\mrvr}{\mathrm{var}}
\newcommand{\mrop}{\mathrm{opt}}
\newcommand{\mrexc}{\mathrm{exact}}
\newcommand{\mrrf}{\mathrm{ref}}
\newcommand{\mrrg}{\mathrm{reg}}
\newcommand{\mrrd}{\mathrm{red}}

\newcommand{\brn}{\bar{n}}
\newcommand{\bre}{\bar{e}}
\newcommand{\brd}{\bar{d}}
\newcommand{\brN}{\bar{N}}
\newcommand{\brU}{\bar{U}}

\newcommand {\eqb}[1]{\begin{equation}\begin{array}{#1}}
\newcommand {\eqe}{\end{array}\end{equation}}

\newcommand {\esb}[1]{\begin{equation*}\begin{array}{#1}}
\newcommand {\ese}{\end{array}\end{equation*}}
\newcommand {\ds}{\displaystyle}

\newcommand {\pa}[2]{\frac{\partial{#1}}{\partial{#2}}}
\newcommand {\pad}[2]{\frac{\mathrm{d}{#1}}{\mathrm{d}{#2}}}
\newcommand {\paq}[2]{\frac{\partial^2{#1}}{\partial{#2}^2}}

\newcommand {\back}{\! \! \!}
\newcommand {\is}{\back &=& \back}
\newcommand {\dis}{\back &:=& \back}

\newcommand {\norm}[1]{\|#1\|}
\newcommand {\tr}{\mathrm{tr}\,}

\newcommand {\grad}{\mathrm{grad}\,}

\newcommand {\diag}{\mathrm{diag}\,}

\newcommand {\dif}{\mathrm{d}}

\newcommand {\II}{{I\kern-.3em I}}
\newcommand {\III}{{I\kern-.3em I\kern-.3em I}}

\newcommand {\into}{\int_\Omega}
\newcommand {\intoe}{\int_{\Omega^e}}

\newcommand {\intooe}{\int_{\Omega_0^e}}

\newcommand {\mra}{\mathrm{a}}
\newcommand {\mrb}{\mathrm{b}}
\newcommand {\mrc}{\mathrm{c}}

\newcommand {\mre}{\mathrm{e}}

\newcommand {\mrm}{\mathrm{m}}
\newcommand {\mrn}{\mathrm{n}}

\newcommand {\mrp}{\mathrm{p}}
\newcommand {\mrq}{\mathrm{q}}
\newcommand {\mrr}{\mathrm{r}}
\newcommand {\mrs}{\mathrm{s}}

\newcommand {\mru}{\mathrm{u}}

\newcommand {\mrP}{\mathrm{P}}
\newcommand {\mrR}{\mathrm{R}}

\newcommand {\mrF}{\mathrm{F}}
\newcommand {\mrU}{\mathrm{U}}

\newcommand {\mcc}{\mathbf{c}}

\newcommand {\mf}{\mathbf{f}}
\newcommand {\mg}{\mathbf{g}}

\newcommand {\mq}{\mathbf{q}}

\newcommand {\muu}{\mathbf{u}}

\newcommand {\mx}{\mathbf{x}}

\newcommand {\mLam}{\mathbf{\Lambda}}
\newcommand {\mmu}{\bm{\mu}}

\newcommand {\mbq}{\mathbf{\bar q}}

\newcommand {\mbN}{\mathbf{\bar N}}
\newcommand {\mbU}{\mathbf{\bar U}}

\newcommand {\mhq}{\mathbf{\hat q}}

\newcommand {\mhU}{\mathbf{\hat U}}

\newcommand {\ba}{\boldsymbol{a}}
\newcommand {\bb}{\boldsymbol{b}}

\newcommand {\bff}{\boldsymbol{f}}

\newcommand {\bn}{\boldsymbol{n}}

\newcommand {\bq}{\boldsymbol{q}}

\newcommand {\bt}{\boldsymbol{t}}
\newcommand {\bu}{\boldsymbol{u}}

\newcommand {\bx}{\boldsymbol{x}}
\newcommand {\by}{\boldsymbol{y}}

\newcommand {\blam}{\mbox{\boldmath$\lambda$}}

\newcommand {\bvphi}{\mbox{\boldmath$\varphi$}}

\newcommand {\bxi}{\mbox{\boldmath$\xi$}}

\newcommand {\tg}{\tilde{g}}
\newcommand {\tW}{\tilde{W}}
\newcommand {\tI}{\tilde{I}}
\newcommand {\tJ}{\tilde{J}}
\newcommand {\tB}{\tilde{B}}
\newcommand {\tbx}{\tilde{\boldsymbol{x}}}
\newcommand {\tbS}{\tilde{\boldsymbol{S}}}
\newcommand {\tbC}{\tilde{\boldsymbol{C}}}
\newcommand {\tbX}{\tilde{\boldsymbol{X}}}
\newcommand {\tbF}{\tilde{\boldsymbol{F}}}
\newcommand {\tbB}{\tilde{\boldsymbol{B}}}
\newcommand {\tbsig}{\tilde{\boldsymbol{\sigma}}}

\newcommand {\mB}{\mathbf{B}}

\newcommand {\mD}{\mathbf{D}}
\newcommand {\mE}{\mathbf{E}}
\newcommand {\mF}{\mathbf{F}}

\newcommand {\mH}{\mathbf{H}}
\newcommand {\mI}{\mathbf{I}}
\newcommand {\mJ}{\mathbf{J}}
\newcommand {\mK}{\mathbf{K}}
\newcommand {\mL}{\mathbf{L}}

\newcommand {\mN}{\mathbf{N}}

\newcommand {\mP}{\mathbf{P}}

\newcommand {\mS}{\mathbf{S}}
\newcommand {\mT}{\mathbf{T}}
\newcommand {\mU}{\mathbf{U}}

\newcommand {\mX}{\mathbf{X}}

\newcommand {\bA}{\boldsymbol{A}}
\newcommand {\bB}{\boldsymbol{B}}
\newcommand {\bC}{\boldsymbol{C}}

\newcommand {\bF}{\boldsymbol{F}}

\newcommand {\bI}{\boldsymbol{I}}

\newcommand {\bN}{\boldsymbol{N}}

\newcommand {\bX}{\boldsymbol{X}}

\newcommand {\eps}{\varepsilon}
\newcommand {\sig}{\sigma}

\newcommand {\bone}{\mathbf{1}}

\newcommand {\btsig}{\mbox{\boldmath$\tilde\sig$}}

\newcommand {\IR}{{\rm\kern.24em
   \vrule width.02em height1.53ex depth-.05ex
   \kern-.3em R}}
\newcommand {\ic}{{\rm\kern.20em
   \vrule width.02em height1.0ex depth-.05ex
   \kern-.22em c}}
\newcommand {\ia}{{\rm\kern.20em
   \vrule width.02em height1.05ex depth-.0ex
   \kern-.25em a}}
\newcommand {\IC}{{\rm\kern.24em
   \vrule width.02em height1.4ex depth-.05ex
   \kern-.26em C}}
\newcommand {\ID}{{\rm\kern.34em
   \vrule width.02em height1.5ex depth-.05ex
   \kern-.36em D}}
\newcommand {\IS}{{\rm\kern.24em
   \vrule width.02em height1.6ex depth.05ex
   \kern-.26em S}}
\newcommand {\IT}{{\rm\kern.50em
   \vrule width.02em height1.55ex depth-.05ex
   \kern-.52em T}}

\newcommand {\IE}{{\rm\kern.24em
   \vrule width.02em height1.55ex depth-.05ex
   \kern-.33em E}}
\newcommand {\IEa}{{\rm\kern.24em
   \vrule width.02em height1.55ex depth-.05ex
   \kern-.33em E}^{1}_{ijkl}}
\newcommand {\IEb}{{\rm\kern.24em
   \vrule width.02em height1.55ex depth-.05ex
   \kern-.33em E}^{2}_{ijkl}}

\newcommand {\sA}{\mathcal{A}}
\newcommand {\sB}{\mathcal{B}}

\newcommand {\sP}{\mathcal{P}}

\newcommand {\sR}{\mathcal{R}}
\newcommand {\sS}{\mathcal{S}}

\newcommand {\sV}{\mathcal{V}}

\newcommand {\vaua}{\ba_\alpha}
\newcommand {\vaa}{\ba^\alpha}

\newcommand {\vaub}{\ba_\beta}
\newcommand {\vab}{\ba^\beta}

\newcommand {\vauacb}{\ba_{\alpha,\beta}}

\newcommand {\vAua}{\bA_\alpha}
\newcommand {\vAa}{\bA^\alpha}

\newcommand {\vAb}{\bA^\beta}

\newcommand {\Ass}[2]{\kern 0.9ex \vrule width0.45em height0.2ex depth0ex \kern -2.1ex \bigwedge_{#1}^{#2}}
\newcommand {\ASS}[2]{\kern 1.45ex \vrule width0.5em height0.2ex depth0ex \kern -2.65ex \bigwedge_{#1}^{#2}}

\newcommand {\Iabgd}{{\mathrm{I}}^{\alpha\beta\gamma\delta}}

\newcommand {\cabgd}{{c}^{\alpha\beta\gamma\delta}}

\makeatletter
\newcommand*{\rom}[1]{\expandafter\@slowromancap\romannumeral #1@}
\makeatother

\begin{document}

\begin{center}
\Large{\bf{
Contact-based inverse analysis for nonlinear material identification in spatially heterogeneous solids		
}}\\

\end{center}

\renewcommand{\thefootnote}{\fnsymbol{footnote}}

\begin{center}
\large{Bart\l{}omiej \L{}azorczyk$^{\mra,\mrb}$
and Roger A.~Sauer$^{\mrb,\mra,\mrc}$\footnote[1]{corresponding author, email: roger.sauer@pg.edu.pl,  
roger.sauer@rub.de}
}\\
\vspace{4mm}

\small{\textit{
$^\mra$Department of Structural Mechanics, Gda\'{n}sk University of Technology, Gda\'{n}sk, Poland \\[1.1mm]
$^\mrb$Institute for Structural Mechanics, Ruhr University Bochum, Bochum, Germany \\[1.1mm]
$^\mrc$Department of Mechanical Engineering, Indian Institute of Technology Guwahati, Assam, India 
}}

\end{center}

\vspace{-4mm}

\renewcommand{\thefootnote}{\arabic{footnote}}

\rule{\linewidth}{.15mm}
{\bf Abstract}

This study presents a contact-based isogeometric Finite Element Model Updating~(FEMU) framework for identifying spatially varying constitutive parameters of nonlinear solids. The formulation considers large quasi-static deformations of hyperelastic 3D solids and thin shells due to mechanical contact. The proposed inverse approach utilizes full-field displacement measurements available at least on the free surface and, in the case of pure Dirichlet boundary conditions, the resultant contact forces as well. The nonuniform material parameter fields are discretized using low-order Lagrange interpolation independent of the isogeometric analysis mesh, providing control over the inverse problem size and potential discontinuities in the material. The FEMU least-squares objective is minimized using a trust-region reflective algorithm~--~a local gradient-based optimization approach. Computational efficiency is enhanced through the analytical derivatives of the objective and a material continuation strategy. The proposed framework is evaluated through three numerical examples based on synthetically generated data: a Canham shell strip on a rigid foundation, indentation of a Koiter shell model of the human abdominal wall, and indentation of a Neo-Hookean block. The examples verify the ability of the proposed method to reconstruct inhomogeneous material via mechanical contact. Analytical derivatives improve the computational efficiency and facilitate conducting sensitivity and identifiability analyses of the material parameters. The presented approach is non-destructive and can be used for various inverse problems, such as in-vivo biomechanics of soft tissues and laboratory material characterization.


\vspace{-5mm}
\rule{\linewidth}{.15mm}

\section{Introduction}\label{s:intro}

Full-field measurement techniques, such as Digital Image Correlation~(DIC)~\citep{sutton2009} and Digital Volume Correlation~(DVC)~\citep{Buljac2018}, have become well-established measurement methods in the past few decades, leading to a substantial development in material testing procedures~\citep{Pierron2021, Hild2025}. These advances have accelerated the development of several classical inverse approaches for material identification, including the Finite Element Model Updating~(FEMU) method~\citep{Kavanagh1971}, Virtual Fields Method~(VFM)~\citep{pierron2012virtual, Avril2026}, Constitutive Equation Gap Method~\citep{Latourte2008, Florentin2010}, and Equilibrium Gap Method~\citep{HAUSTRATE2025}. Over the years, these classical methods have been systematically reviewed and revisited, gradually revealing connections among them~\citep{Avril2007, Avril2008, Martins2018, Roux2020, Rmer2025}. With the rapid progress in machine learning, many novel frameworks for material identification have emerged. While some are based on the classical approaches mentioned above, others introduce fundamentally new directions. Notable contributions include data-driven material modeling~\citep{Leygue2018}, constitutive model discovery (see e.g.~EUCLID~\citep{FLASCHEL2021}), and Physics-Informed Neural Networks~(PINNs)~\citep{Haghighat2021}.

Material identification is an inverse problem, a class of inherently ill-posed problems for which the existence, uniqueness, and stability of the solution cannot be guaranteed~\citep{nelles2020}. The lack of stability renders inverse problems particularly susceptible to inaccuracies in experimental data, such as measurement noise. Furthermore, parameters often exhibit poor identifiability due to the insensitivity of the mathematical model to parameter variations or to compensating interactions between them~\citep{Hartmann2018, Zhang2022_02}. When spatially heterogeneous material fields are identified, their discretization often yields a high-dimensional parameter space. For iterative methods such as FEMU, the high dimensionality of the parameter space not only amplifies issues related to uniqueness and stability but also poses a computational challenge, requiring efficient optimization algorithms and practically prohibiting the use of finite differences to evaluate objective function derivatives.     

The aforementioned challenges make the identification of spatially varying material fields particularly difficult. Although most studies focus on homogeneous materials, the reconstruction of inhomogeneous material fields has recently received increasing attention. Beyond the classical approaches discussed above, Integrated-DIC~\citep{kirchhoff2024} treats image registration and material identification as a single inverse problem. Bayesian inference~\citep{KOUTSOURELAKIS2009, THANANJAYAN2023} also provides a powerful framework for reconstructing spatially varying material properties; however, its high computational cost requires using specialized techniques or surrogate models. Recently, PINNs have become a promising tool for identifying material fields in solids; see, for example,~\citet{Kamali2023, Srikitrungruang2025, Bouclier2026, Xu2026}.

FEMU and VFM are the most widely used classical techniques for reconstructing heterogeneous material fields in solids. The main concept of FEMU is to minimize the discrepancy between measurements and the Finite Element~(FE) model response. This discrepancy is typically formulated in a global least-squares sense. FEMU is straightforward to implement and is a well-established technique known for its robustness and low sensitivity to noise~\citep{Chen2024}. Moreover, FEMU provides a calibrated FE model of the body and can operate on partial experimental data. However, the iterative nature of FEMU, together with the frequent need for objective function derivatives, makes it computationally costly. This limitation is often alleviated by deriving analytical derivatives~\citep{Avril2007,Gokhale2008,Goenezen2011}. Over the years, FEMU has been employed to identify heterogeneous parameters in constitutive models including isotropic~\citep{Khalil2006} and anisotropic~\citep{Liu2005} linear elasticity, isotropic~\citep{Goenezen2012,SIMONALLUE2017} and anisotropic~\citep{Kroon2009} hyperelasticity, elastoplasticity~\citep{Campos2020}, and damage~\citep{Liu2019}. Recently,~\citet{LAZORCZYK2026} proposed an elastodynamic FEMU framework for identifying unknown stiffness and density fields in Bernoulli--Euler beams. In~\citet{lazorczyk2025}, this approach was further extended to the simultaneous identification of elasticity parameters and prestressing force.~\citet{Mei2026} introduced a sequential inverse approach for identification of heterogeneous shear modulus and stress-free configuration in prestressed incompressible Neo-Hookean solids.~\citet{TOUMINET2025} proposed a Bayesian extension of FEMU based on random fields for representing material inhomogeneity. By combining the adjoint method with automatic differentiation, the authors' approach enables an efficient identification of spatially varying material fields from full-field data.

Selecting a representation of inhomogeneous material fields is a fundamental task for their successful reconstruction. Prior knowledge of material heterogeneity, e.g.~by visual inspection, enables partitioning of the solid into regions with approximately homogeneous characteristics. However, such a priori knowledge is rarely available; thus, the identification problem also involves model selection, characterized by the bias/variance trade-off~\citep{nelles2020}. Several representation strategies have been proposed. In FEMU, a common approach is to define the unknown material parameters element-wise; however, this rapidly leads to very large inverse problems. A straightforward remedy is to use a coarser mesh or subdomain-based representation for the material field, as proposed, e.g., in~\citet{Kroon2008} and~\citet{Borzeszkowski2022}. Alternatively,~\citet{HAUSTRATE2025} proposed a covariance-weighted Equilibrium Gap method, treating material inhomogeneity as a random field represented by a Karhunen--Lo\`{e}ve expansion. More recently,~\citet{Li2026} introduced a FEMU framework that represents material heterogeneity via a neural network, enabling a flexible, self-regularized inverse formulation. 

In state-of-the-art material testing, specimen geometry and loading are optimized to generate heterogeneous strain fields, enabling identification of complex constitutive laws from a single load experiment~\citep{Pierron2021, Chen2024}. However, the geometry of inhomogeneous structures, such as the human abdominal wall or breast, cannot be modified; thus, boundary conditions, such as loading or physiological movements, remain the primary design variable. Using multiple load cases has been shown to facilitate material field identification~\citep{Borzeszkowski2022, LAZORCZYK2026}. In experimental biomechanics, pressure is regularly applied to induce deformation~\citep{Song2006, Lubowiecka2022} and has shown some success in inverse analyses~\citep{SIMONALLUE2017, Borzeszkowski2022}. Nevertheless, as a scalar quantity, it provides limited information. On the other hand, contact forces are 3D vectors that act at multiple locations and in multiple directions. Contact loading, for instance, through macro-indentation with a rigid probe, can generate much richer data. For example, probing at 10 locations and in 10 different directions gives 100 load cases. Varying the size of the contact probe enables even more combinations.

Mechanical contact is inherent to many medical procedures and therapies. An example is ultrasonography, where an ultrasound-emitting probe is brought into close contact with the skin. The probe contact can easily induce large deformation of the tissue without causing damage. Other examples of this kind include mammography, angioplasty, and cell probing. In all of these examples, the contact information is used to a limited degree. Basic quantities such as the contact force might be recorded, but the rich 3D deformation field induced by contact is usually not measured. Medical diagnosis, surgery, and therapy require knowledge about the properties and state of biological materials under in vivo conditions. Since standard testing of soft tissues can be inconvenient due to fragility, physiological conditions, and ethical and formal constraints~\citep{Evans2017, Navindaran2023}, non-invasive probing becomes particularly appealing.

Indentation techniques are widely used for assessing material properties. For example, surface measurements obtained from indentation were used to reconstruct heterogeneous material properties in porcine tissue and the human abdominal wall by~\citet{Pierrat2018} and~\citet{Remus2024}, respectively. The identifiability of hyperelastic constitutive parameters in homogeneous samples subjected to macro-indentation was investigated in~\citet{Oddes2023} and~\citet{Ashkenazi2026}. However, these studies assumed local material homogeneity. Studies on global identification of material fields using contact remain limited.~\citet{Hartmann2018} demonstrated that force measurements and surface deformation from indentation are sufficient to identify material properties of a two-layer Neo-Hookean block.~\citet{Fougeron2024} further investigated the identifiability of hyperelastic parameters in bi-layer soft tissues under macro-indentation, showing that it strongly depends on the selected constitutive law and that combining different experimental data types may deteriorate identification quality. Furthermore,~\citet{Lavigne2023} identified hyperelastic properties and frictionless contact traction fields from two deformed configurations without knowledge of a stress-free configuration. Recently,~\citet{Chen2026a} and~\citet{Chen2026b} employed neural operators to reconstruct spatially varying elastic properties in small-deformation contact. To the best of our knowledge, these are the only studies that use contact for the identification of heterogeneous material fields.

Every FEMU approach calls a corresponding finite element~(FE) solver several times during the analysis, covering a broad range of design parameters; thus, it requires an efficient and robust FE formulation. Isogeometric analysis~(IGA), pioneered by~\citet{Hughes2005}, has become a well-established tool for various computational tasks in solid mechanics. Regarding mechanical contact, IGA provides arbitrary smoothness across elements boundaries, which improves accuracy and robustness~\citep{DeLorenzis2014}. IGA was successfully used in various inverse problems, including material identification~\citep{dufour2015,Borzeszkowski2022,LAZORCZYK2026} and load reconstruction~\citep{Vubac2018,Vubac2019}.

This study demonstrates that simple contact probing can generate data with very high information content, enabling accurate identification of heterogeneous materials. In particular, the proposed approach facilitates a detailed characterization of complex material behavior. To this end, a Contact-based Inverse Analysis (CBIA) approach is proposed for reconstructing spatially heterogeneous constitutive parameters in nonlinear solids using isogeometric FEMU. The present work extends the study in~\citet{Borzeszkowski2022} by incorporating mechanical contact and substantially improving the reconstruction performance. The formulation considers large quasi-static deformations, which are commonly encountered in the modeling of soft biological tissues. While this study is restricted to isotropic constitutive models, its extension to anisotropy is straightforward. Spatially varying material properties are represented with a low-order Lagrange interpolation independent of the finite element~(FE) mesh, following the approach of~\citet{Borzeszkowski2022}. To the best of our knowledge, this is the first time such a contact-based FEMU framework is proposed for identifying spatially heterogeneous constitutive models in nonlinear solids and thin shells.
In summary, the proposed method is characterized by the following features:
\begin{itemize}[noitemsep,topsep=0pt]
	\item A general contact-based FEMU framework for identifying spatially distributed material properties. 
	\item A FEMU objective function based on displacements and resultant contact forces, with optional Tikhonov regularization.
	\item Gradient-based minimization of the objective function using analytical derivatives, a Newton--Raphson scheme with material incrementation, and a precomputed initial guess.
	\item Discretization of the unknown material fields with a low-order Lagrange mesh.	
	\item An isogeometric finite element formulation for nonlinear solids and thin shells. 
	\item A rigid, frictionless, and adhesionless mechanical contact formulation. 
	\item A feasibility study of 2D inclusion reconstruction in a heterogeneous solid from surface data only.
\end{itemize}
Key to the proposed identification approach are simple, minimally invasive, and nondestructive contact tests that generate information-rich data, enabling much more accurate inverse analyses than existing techniques. The proposed approach can be particularly suitable for soft tissues, such as the abdominal wall and blood vessels.

The remainder of this paper is organized as follows: Sec.~\ref{s:ForwardProb} introduces the forward problem considered here, including the finite element formulation and the discretization of the unknown material fields. Sec.~\ref{s:Inverse} presents the proposed inverse framework, which is followed by a set of numerical experiments shown in Sec.~\ref{s:Nex}. The paper concludes with Sec.~\ref{s:concl}. 

\section{Forward problem}\label{s:ForwardProb}

This section presents the forward problem that is used in the subsequent inverse analysis. The employed constitutive laws, contact formulation, and weak forms are introduced, followed by the finite element discretization and parametrization of the unknown material fields. For brevity, the general theory of nonlinear solids is omitted, and only the necessary kinematics is introduced for thin shells. In the following, the 3D quantities are denoted by tilde, e.g. $\tJ$, while their 2D or surface-related counterparts are written without any accents. Similarly, quantities associated with the reference configuration are indicated by uppercase symbols or the subscript \enquote{0}, e.g. $\sS_0$, while lowercase symbols refer to the deformed configuration.

\subsection{Curvilinear surface description for shells}\label{s:3Dsurf}

A point on the shell mid-surface\footnote{Or equivalently on the boundary of a solid body $\partial\sB$} $\sS$ is described by the mapping
\eqb{l}
	\bx = \bx(\xi^\alpha) \,,\quad \alpha = 1,\,2\,,
\label{}\eqe
where $\xi^\alpha$ are curvilinear coordinates associated with a 2D parametric domain $\sP$. The surface parametrization defines the covariant tangent vectors $\vaua := \partial\ba/\partial\xi^\alpha$, the covariant metric components $\auab := \vaua \cdot \vaub$, the surface unit normal $\bn := (\ba_1\times\ba_2)/J_a$, and the area element $\dif a := J_a\,\dif\xi^1\dif\xi^2$, where $J_a := \norm{\ba_1\times\ba_2}$. The dual basis vectors are defined as $\vaa := \aab\vaub$, satisfying $\vaa \cdot \ba_\beta = \delta^\alpha_\beta$, where $[\aab] := [\auab]^{-1}$, and $\delta^\alpha_\beta$ is the Kronecker delta. Repeated Greek indices imply summation over $\{1,2\}$ throughout the paper. 

The curvature tensor $\bb = \buab\,\vaa\otimes\vab$ is characterized by the Gauss--Weingarten formula
\eqb{l}
	\buab := \bn\cdot\vauacb = -\bn_{,\beta}\cdot\vaua \,,
\label{}\eqe
where a comma denotes the parametric derivative $\vauacb = \partial\vaua/\partial\xi^\beta$. Its invariants
\eqb{l}
	H := \ds\frac{1}{2}\,b^\alpha_\alpha = \frac{1}{2}\,\aab\buab \,,\qquad \kappa := \frac{\det{[\buab]}}{\det{[\auab]}} \,,
\label{}\eqe
represent the mean and Gaussian curvature, respectively. Analogous quantities are defined for a point $\bX(\xi^\alpha)$ in the undeformed shell configuration $\sS_0$, including $\vAua$, $\vAa$, $\Auab$, $\Aab$, $J_A$, $\bN$, and $\bB$.

The mapping between $\sS_0$ and $\sS$, represented by $\bx = \bvphi(\bX)$, is described by the surface deformation gradient
\eqb{l}
	\bF := \vaua\otimes\vAb \,,
\label{}\eqe
and the area change $\dif a = J\,\dif A$, where $J$ is the surface stretch given by $J = J_a/J_A$. The right surface Cauchy--Green tensor is defined by $\bC := \bF^\T\bF = \auab\,\vAa\otimes\vAb$ with invariants $I_1 = \Aab\,\auab$ and $\det\bC = J^2$. A more detailed description of thin shell kinematics and its variation can be found, e.g. in~\citet{sauer2018computational}. 

\subsection{Constitution}\label{s:const}

\subsubsection{3D solids}\label{s:const3D}

For heterogeneous hyperelastic solids, the constitutive law follows from the strain-energy density function
\eqb{l}
	\tW = \tW(\tbC,\tbX) \,,
\label{e:W3D}\eqe
from which the second Piola--Kirchoff and Cauchy stress tensors are obtained as
\eqb{l}
	\tbS = 2\ds\pa{\tW(\tbC,\tbX)}{\tbC} \qquad \text{and} \qquad \tbsig = 2\tJ^{-1}\tbF\ds\pa{\tW(\tbC,\tbX)}{\tbC}\tbF^\T \,,
\label{e:sig3D}\eqe
respectively, where $\tbF$ denotes the 3D deformation gradient, $\tbC$ is the right Cauchy--Green tensor in 3D, $\tJ = \det\tbF$. For 3D solids, this work employs an isotropic Neo-Hookean material model of the form  
\eqb{l}
	\tW(\tbX,\tI_1,\tJ) = \ds\frac{\tilde\mu(\tbX)}{2}\left(\tI_1 - 3 - 2\ln\tJ\right) + \frac{\tilde\Lambda(\tbX)}{2}(\ln\tJ)^2 \,,
\label{e:WNH}\eqe
which, after using Eq.~\eqref{e:sig3D}, yields
\eqb{l}
	\btsig = \ds\frac{\tilde\Lambda(\tbX)}{\tJ}\ln\tJ\bI + \frac{\tilde\mu(\tbX)}{\tJ}\left(\tbF\tbF^\T-\bI\right) \,,
\label{e:sigNH}\eqe
where $\bI$ is the standard identity in 3D and $\tI_1$ is the first invariant of $\tbC$, i.e., $\tI_1 = \tr{\tbC}$. The parameters, $\tilde\Lambda(\tbX)$ and $\tilde\mu(\tbX)$ denote the spatially varying Lam{\'e} parameters in 3D elasticity.

{\small \textbf{Remark 1}: For 2D solid models, as considered in Sec.~\ref{s:iron}, the strain-energy density in~\eqref{e:WNH} reduces to
\eqb{l}
	W(\bX,I_1,J) = \ds\frac{\mu(\bX)}{2}\left(I_1 - 2 - 2\ln J\right) + \frac{\Lambda(\bX)}{2}(\ln J)^2 \,,
\label{}\eqe
which leads to an analogous expression for the Cauchy stress tensor as in Eq.~\eqref{e:sigNH}.}\par

\subsubsection{Shells directly derived in surface form}\label{s:constShell}

Shell models can be derived using two different approaches: (1) the \textit{projection approach}, in which the shell is treated as a 3D continuum and the stress resultants are usually evaluated via numerical integration through the thickness; (2) the \textit{direct surface approach} adopted here, where the shell is defined with respect to its mid-surface, and the stresses and moments follow from the surface strain energy density function, see e.g.~\citet{Roohbakhshan2017}. For shells, the strain energy density can be split into membrane and bending contributions  
\eqb{l}
	W(\bX,\auab,\buab) = W_\mrm(\bX,\auab) + W_\mrb(\bX,\auab,\buab) \,,
\label{e:Wsurf}\eqe
where $W_\mrb$ typically depends only marginally on $\auab$. The stress and bending moment components are then
\begin{align}
	\tauab &= 2\ds\pa{W}{\auab} \,, \label{e:Stau} \\[1mm]
	\Mab_0 &= \ds\pa{W}{\buab} \,.  \label{e:SM}
\end{align}

\paragraph{Initially planar shells}\label{s:Can}

For initially planar shells, here, $W_\mrb$ is based on the \citet{Canham1970} model 
\eqb{l}
	W_\mrb(\bX,J,H,\kappa) = c(\bX)\,J\,(2H^2 - \kappa) \,,
\label{e:WCan}\eqe
where $c(\bX)$ denotes the Canham bending stiffness. Substitution of Eq.~\eqref{e:WCan} into Eq.~\eqref{e:SM} gives
\eqb{l}
	\Mab_0 = c(\bX)\,J\,\bab \,.
\label{e:MCan}\eqe
The membrane part, here, $W_\mrm$ follows from the incompressible Neo-Hookean model 
\eqb{l}
	W_\mrm(\bX,\tI_1,\tJ) = T\,\tW_\mathrm{NH} = \ds\frac{\mu(\bX)}{2}(\tI_1 - 3) + p(1-\tJ) \,.
\label{e:WSNH}\eqe
where $T$ is the initial shell thickness, $\tW_\mathrm{NH}$ is the 3D Neo-Hookean strain energy density, $\mu = T\,\tilde\mu$, and $p$ is the Lagrange multiplier enforcing the incompressibility constraint $\tg = 1 - \tJ$, which equals $p = \mu/J^2$ for plain strain~\citep{Sauer2016}. Substituting Eq.~\eqref{e:WSNH} into Eq.~\eqref{e:Stau} then yields\footnote{The contribution to $\tauab$ resulting from \eqref{e:WCan} is negligible for thin shells.}
\eqb{l}
	\tauab = \mu(\bX)\left(\Aab - \ds\frac{\aab}{J^2}\right) \,.
\label{e:tauNH}\eqe
The corresponding material tangents can be found in~\citet{Sauer2015}.

\paragraph{Initially curved shells}\label{s:koiter}

Here, the Koiter shell model~\citep{Ciarlet2005,Steigmann2013} is considered
\begin{flalign}
	W(\bX,\auab,\buab) = \ds\frac{1}{8}\left(\auab - \Auab\right)\cabgd\left(\augd - \Augd\right) + \frac{T^2}{24}\left(\buab - \Buab\right)\cabgd\left(\bugd - \Bugd\right),
\label{e:WKoit}\end{flalign}
where
\eqb{lll}
	\cabgd \is \Lambda(\bX)\,\Iabgd_\mathrm{dil} + 2\,\mu(\bX)\,\Iabgd_\mathrm{dev} \,, \\[2mm]
	\Iabgd_\mathrm{dil} \is \Aab\,\Agd \, \\[2mm]
	\Iabgd_\mathrm{dev} \is \ds\frac{1}{2}\left(\Aag\Abd + \Aad\Abg\right) \,,
\label{}\eqe
which, upon substitution into Eq.~\eqref{e:Stau}~and~\eqref{e:SM} leads to
\eqb{l}
	\tauab = \ds\frac{1}{2}\cabgd\left(\auab - \Auab\right) \,,
\label{e:tauKoit}\eqe
and 
\eqb{l}
	\Mab_0 = \ds\frac{T^2}{12}\cabgd\left(\buab - \Buab\right) \,.
\label{e:MKoit}\eqe

{\small \textbf{Remark 2}: $\Lambda$ and $\mu$ can be obtained by analytical integration of the 3D Saint Venant--Kirchhoff material model, resulting in
\eqb{l}
	\mu = T\tilde\mu \,, \quad\quad\quad \Lambda = T\ds\frac{2\,\tilde\Lambda\tilde\mu}{\tilde\Lambda + 2\tilde\mu} \,,
\label{}\eqe
which are related to Young's modulus $E$ and Poisson ratio $\nu$ through
\eqb{l}
	\tilde\mu = \ds\frac{E}{2(\nu + 1)} \,, \quad\quad\quad \tilde\Lambda = \frac{2\,\tilde\mu\,\nu}{1 - 2\nu} \,.
\label{}\eqe
}\par

\subsection{Contact formulation}\label{s:contact}

This work considers frictionless and adhesionless quasi-static contact. This simplification is justified because lubricants effectively eliminate dry friction in many medical procedures, such as ultrasonography. Furthermore, one of the contacting bodies is assumed to be rigid, which provides a sufficient preliminary model for a soft, deformable body (e.g., biological tissue) interacting with a much stiffer counterpart, such as a medical instrument. Normal contact is enforced through the impenetrability constraint
\eqb{l}
g_\mrn \leq 0 \,,	
\label{e:impenetrability}\eqe
where
\eqb{l}
g_\mrn := \left(\bx_\mrs - \bx_\mrp\right)\cdot\bn_\mrp \,,
\label{}\eqe
denotes the normal gap between a generic point\footnote{Representing a surface quadrature point in the discrete setting.} $\bx_\mrs$ on the contact surface of the deformable body $\partial\sB_{\mrc1}$ (or $\sS_{\mrc1} \subset \sS$ for shells) and the rigid obstacle surface $\partial\sB_{\mrc2}$. Here, $\bn_\mrp$ is the surface normal of $\partial\sB_{\mrc2}$ at $\bx_\mrp$, and $\bx_\mrp$ is the closest-point projection of $\bx_\mrs \in \partial\sB_{\mrc1}$ onto $\partial\sB_{\mrc2}$, obtained from the minimum distance problem
\eqb{l}
\bx_\mrp(\bx_\mrs) = \ds\left\{\by\,:\,\min_{\forall\by\in\partial\sB_{\mrc2}}\norm{\bx_\mrs - \by} \,\,\, \forall \bx_\mrs\in\partial\sB_{\mrc1} \right\} \,.
\label{}\eqe
This minimization problem may become significantly challenging for complex contact surface geometries~\citep{Wriggers2006}. The impenetrability constraint~\eqref{e:impenetrability} is enforced in the weak form (see Sec.~\ref{s:weak}) using the penalty method, which defines the normal contact traction as
\eqb{l}
\bt_\mrc = \left\{ \begin{array}{@{}ll@{}}
	-\epsilon_\mrn g_\mrn\,\bn_\mrn\,, \quad &g_\mrn < 0 \,,\\
	\boldsymbol{0}\,, \quad &g_\mrn \geq 0 \,, 
\end{array}\right.
\label{e:normalTraction}\eqe
where $\epsilon_\mrn$ is the penalty parameter. 

\subsection{Weak form}\label{s:weak}

For quasi-static problems, the weak form (or principle of virtual work) reads
\eqb{l}
G_\mrint + G_\mrc - G_\mrext = 0 \qquad \forall\,\delta\tbx\in\sV \,,
\label{e:weakform}\eqe
where $\delta\tbx\in\sV$ is a kinematically admissible variation of $\tbx$ (or $\delta\bx\in\sV$ for shells). Since only rigid contact is considered, the weak form of a two-body contact problem is reduced to a single deformable body. For a 3D solid, the internal virtual work is given by
\eqb{l}
	G_\mrint = \ds\int_\sB \btsig:\grad(\delta\tbx)\,\dif v\,,
\label{e:weakVolInt}\eqe
whereas, for Kirchhoff--Love (KL) thin shells, it takes the form
\eqb{l}
	G_\mrint = \ds\int_{\sS_0} \frac{1}{2}\delta\auab\,\tauab\dif A  + \ds\int_{\sS_0}\delta\buab\,\Mab_0\dif A\,,
\label{e:weakSurfInt}\eqe
where $\delta\auab$ and $\delta\buab$ denote the variations of $\auab$ and $\buab$, respectively, while $\tbsig$, $\tauab$, and $\Mab_0$ follows from the constitutive relations~\eqref{e:sigNH},~\eqref{e:tauNH},~\eqref{e:tauKoit},~\eqref{e:MCan}, and~\eqref{e:MKoit}. The contact contribution associated with the normal traction is written as
\eqb{l}
	G_\mrc = -\ds\int_{\partial\sB_{\mrc1}}\delta\bx\cdot\bt_\mrc\,\dif a \,,
\label{e:weakCont}\eqe
where $\bt_\mrc$ is defined in Eq.~\eqref{e:normalTraction}. The corresponding linearizations of the weak form, and the expressions for the external virtual work $G_\mrext$, can be found in~\citet{Holzapfel2000} for solids and in~\citet{Sauer2015} for shells.

\subsection{Finite element discretization}\label{s:FE}

\subsubsection{FE approximation}\label{s:WFFE}

The body $\sB$ (or shell $\sS$) is discretized using the \textit{isogeometric analysis} (IGA) framework proposed by~\citet{Hughes2005}. To recover the standard FEM structure for IGA shape function, the B\'{e}zier extraction operator is used~\citep{Borden2011}. Consequently, the geometry within an undeformed element $\Omega_0^e$ and deformed element $\Omega^e$ is approximated as
\eqb{l}
	\bX\approx\bX^h = \ds\sum_{I=1}^{n_\mre} N_I\mX_I = \mN\,\mX_e\,, \qquad \bx\approx\bx^h = \ds\sum_{I=1}^{n_\mre} N_I\mx_I = \mN\,\mx_e \,,
\label{e:xXFE}\eqe
where $\mX_e$ and $\mx_e$ are the positions of the control points of $\Omega_0^e$ and $\Omega^e$, respectively. Further, $\mN(\bxi) :=  [N_1\bone,N_2\bone,\dots,N_{n_\mre}\bone]$ is the matrix of $n_\mre$ NURBS-based shape functions $\left\{N_I(\bxi)\right\}^{n_\mre}_{I=1}$, and $\bone$ is the $d$-dimensional identity. The displacement field within the element is approximated similarly, 
\eqb{l}
	\bu\approx\bu^h = \ds\sum_{I=1}^{n_\mre} N_I\,\muu_I = \mN\,\muu_e \,, 
\label{e:uDisc}\eqe
where $\muu_e = \mx_e - \mX_e$. Discretization of the weak form~\eqref{e:weakform} then gives
\eqb{l}
	\ds\sum_{e=1}^{n_\mrel}\left(G^e_{\mrint} + G^e_{\mrc} - G^e_{\mrext}\right) = 0 \,\,\, \forall\,\delta\mx\in\sV^h \,,
\label{e:weakformDisc}\eqe
where $n_\mrel$ is the number of finite elements. The elemental contribution to the internal virtual work is written as
\eqb{l}
	G^e_\mrint = \delta\bx^\T_e\,\bff^e_\mrint \,.
\label{}\eqe
For 3D solids, substituting Eq.~\eqref{e:xXFE} into Eq.~\eqref{e:weakVolInt} gives 
\eqb{l}
	\ds\mf_\mrint^e = \into\mB^\T_e\tbsig\,\dif v\,,
\label{e:fint3D}\eqe
where $\mB_e^\T$ is the discrete strain-displacement matrix of an element in the current configuration~\citep{Wriggers2008}. For KL shells, the internal force vector can be decomposed into membrane and bending parts
\eqb{l}
	\bff^e_\mrint =  \bff^e_{\mrint \tau} + \bff^e_{\mrint M}\,.
\label{}\eqe
Using Eqs.~\eqref{e:xXFE}~and~\eqref{e:weakSurfInt}, these parts become
\eqb{l}
	\mf^{\,e}_{\mrint\tau} = \ds\intooe\tauab\, \mN_{,\alpha}^\T~\ba_\beta\,\dif A \,,
\label{e:FEvectortau}\eqe
and
\eqb{l}
	\mf^{\,e}_{\mrint M}= \ds\intooe \Mab_0\,\mN^\T_{;\alpha\beta}\ \bn\,\dif A \,,
\label{e:FEvectorM}\eqe
where $\mN_{;\alpha\beta} := \mN_{,\alpha\beta} -\Gamma_{\alpha\beta}^\gamma\mN_{,\gamma}$. Here, $\mN_{,\gamma}$ and $\mN_{,\alpha\beta}$ denote the first and the second derivatives of $\mN$ w.r.t.~$\xi^\alpha$, while $\Gamma^\gamma_{\alpha\beta} := \ba_{\alpha,\beta}\cdot\ba^\gamma$ are the Christoffel symbols. Discretization of the contact virtual work in Eq.~\eqref{e:weakCont} gives
\eqb{l}
	\mf^{\,e}_\mrc = - \ds\int_{\partial\sB_{c1}^e} \!\!\!\!\mN^\T \bt_\mrc \,\dif a \,.
\label{e:FEvectorCont}\eqe

{\small \textbf{Remark 3}: The impenetrability constraint~\eqref{e:impenetrability} is incorporated into the FE weak form through the standard \textit{active set} strategy. The active set $\sA$ consists of all quadrature points on $\sB^h_{c1}$ for which contact is detected. Since changes in $\sA$ render the contact force vector $\bff_\mrc$ non-differentiable, the objective function introduced in Sec.~\ref{s:objfcn} becomes non-differentiable as well. 
}\par

Consequently, the weak form~\eqref{e:weakformDisc} reads
\eqb{l}
	\delta\mx^\T\left(\mf_\mrint + \mf_\mrc - \mf_\mrext\right) = \bf0  \,\,\, \forall\,\delta\mx\in\sV \,,
\label{e:weakformDisc2}\eqe
where $\mf_\mrint$, $\mf_\mrc$, and $\mf_\mrext$ are the global FE vectors assembled from their elemental contributions. The kinematically admissible set $\sV^h$ contains all nodal variations satisfying the Dirichlet boundary conditions on $\partial_u\sB^h$ (or $\partial_u\sS^h$ for shells). Thus, for the remaining degrees-of-freedom~(dofs), Eq.~\eqref{e:weakformDisc2} implies
\eqb{l}
	\mf(\muu) = \mf_\mrint + \mf_\mrc - \mf_\mrext = \bf0 \,,
\label{e:weakFE}\eqe
which represents the discretized equilibrium equation. Its solution
\eqb{l}
\muu = \begin{bmatrix}
	\bu_1 \\ \bu_2 \\ \vdots \\ \bu_{n_\mrno} 
\end{bmatrix} \,,
\label{e:u}\eqe
is the global nodal displacement vector of size $d\,n_\mrno \times 1$. The elemental external force vectors $\bff^e_\mrext$, together with the elemental stiffness matrices resulting from the linearization of $G^e_\mrint$, $G^e_\mrc$, and $G^e_\mrext$ can be found in~\citet{Wriggers2006} for solids and in~\citet{Duong2017} and~\citet{Sauer2014} for the considered shell models.

\subsubsection{Discretization of the material parameters}\label{s:matFE}

\begin{figure}[h]
	\begin{center}
		\includegraphics[width=1\textwidth]{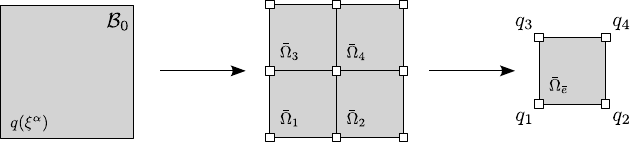}
		\caption{An example of a uniform material mesh consisting of four 4-node bilinear material elements $\bar{\Omega}_{\bar{e}}$ for a two-dimensional body $\sB_0$.} 
		\label{fig:matMeshExample}
	\end{center}
\end{figure}

Each unknown field of material parameters is defined over the body $\sB_0$ (or shell $\sS_0$) as a scalar field $q(\bxi)$. Following~\citet{Borzeszkowski2022}, $q(\bxi)$ is approximated within a \textit{material element} (ME), $\bar\Omega^{\bre}$, using $\brn_e$ nodal values and corresponding interpolation functions $\brN_I$ as 
\eqb{l}
q = q(\bxi) \approx q^h = \ds\sum_{I=1}^{\brn_e} \brN_I(\bxi)q_I = \mbN_{\bre}\,\mq_{\bre}  \,,
\label{e:qdiscr}\eqe
where $\mbN_{\bre} := [\brN_1,\brN_2,\dots,\brN_{\brn_e}]$ and $\mq_{\bre} := [q_1,q_2,\dots,q_{\brn_e}]^\T$ collect the material shape functions and the nodal values, respectively. The material elements collectively form the \textit{material mesh}, an example of which is shown in Fig.~\ref{fig:matMeshExample}. Herein, the material mesh consists of 1-node constant or 4-node bilinear Lagrange elements, which efficiently represent discontinuous and non-smooth fields. For higher-order material elements, a B-spline material mesh is recommended, since higher-order Lagrange interpolation may result in negative inter-nodal values. A comparison of Lagrange and B-spline material meshes in one dimension can be found in~\citet{LAZORCZYK2026}. For simplicity, all material fields are discretized using a common material mesh, leading to the global design vector
\eqb{l}
\mq = \begin{bmatrix}
	\bq_1 \\ \bq_2 \\ \vdots \\ \bq_{\brn_\mrno}
\end{bmatrix} \,,
\label{e:qfull}\eqe
with $n_\mrvr = \bar d\,\bar n_\mrno$ components, where $\bar n _\mrno$ is the number of material nodes and $\bar d$ is the number of unknown parameters per material node. Each nodal vector $\bq_I$ contains all parameters values (e.g., $[\Lambda_I,\mu_I]^\T$). An extension to multiple material meshes is straightforward.
To incorporate the material mesh into the FE analysis, a mapping between the material and FE meshes is required. Both meshes are defined in the parameter domain $\mathcal{P}(\xi^\alpha)$ using conforming rectangular elements, which satisfy the relation $\Omega_e^\square \subset \bar{\Omega}_{\bar{e}}^\square$. Here, $\Omega_e^\square$ and $\bar{\Omega}_{\bar{e}}^\square$ are the element domains in $\mathcal{P}$ for the FE mesh and material mesh, respectively. Further details on the mapping can be found in~\citet{Borzeszkowski2022}.

\section{Inverse analysis}\label{s:Inverse}

This section formulates the material identification as a constrained nonlinear least-squares (NLSQ) problem, followed by an overview of the proposed CBIA framework. To avoid the computational cost of finite differences, the identification procedure is accelerated using analytical derivatives of the objective function. For brevity, these are presented in Appendices~\ref{s:deriv} and \ref{s:sensi}.

\subsection{Objective function}\label{s:objfcn}

Following~\citet{Borzeszkowski2022} and~\citet{LAZORCZYK2026}, the considered FEMU inverse problem is formulated as a PDE-constrained minimization of the function
\eqb{l}
	\min\limits_{\mq} f(\muu(\mq),\mq) \,,
\label{e:minf}\eqe
such that the FE equilibrium~\eqref{e:weakFE} and box constraints on the design vector components $\mq$ are satisfied, i.e., 
\eqb{lll}
	\mf(\muu(\mq),\mq) = \bf0 \,,\\[2mm]
	0 < q_\mathrm{min} \leq q_I \leq q_\mathrm{max} \,.
\label{e:constraints}\eqe
The \textit{objective function} $f(\mq)$ quantifies the discrepancy between the experimental data and their numerical counterparts, with an optional regularization on $\mq$. It depends on the $n_\mrvr = \bar{d}\,\bar{n}_\mrno$ nodal values of $\bar{d}$ constitutive parameters $q(\xi^\alpha)$, discretized following Eq.~\eqref{e:qdiscr}. Here, $f(\mq)$ takes the least-squares form
\eqb{l}
	f(\mq) = \ds\sum_{i=1}^{n_\mrlc} \left[w_\mrU\frac{\left\| \mU_{\mrex\,i} - \mU_{\mrfe\,i}(\mq) \right\|^2}{\left\| \mU_{\mrex\,i} \right\|^2} + 
	w_\mrF\frac{\left\| \mF_{\mrex\,i} - \mF_{\mrfe\,i}(\mq) \right\|^2}{\left\| \mF_{\mrex\,i} \right\|^2}\right] + 
	\alpha^2\norm{\mL(\mq - \mcc)}^2 \,,
\label{e:objfnc}\eqe
where $n_\mrlc$ is the number of independent load cases and $w_\mrU$, $w_\mrF$ are weights associated with the displacements and the contact force contributions, respectively. For each load case, the vector
\eqb{l}
\mU_\mrex = \begin{bmatrix}
	\bu^\mrex_1 \\[1mm] \bu^\mrex_2 \\ \vdots \\ \bu^\mrex_{n_\mrex}
\end{bmatrix}
\label{e:Uexp}\eqe
contains $n_\mrex$ experimental measurements $\bu^\mrex_I$, $I=1,2,\dots,n_\mrex$, at location $\bx^\mrex_I\in\sB$ (or $\bx^\mrex_I\in\sS$ for shells) and 
\eqb{l}
\mU_\mrfe(\mq) = \begin{bmatrix}
	\bu^h(\bx_1^\mrex,\mq) \\[1mm] \bu^h(\bx_2^\mrex,\mq) \\ \vdots \\ \bu^h(\bx_{n_\mrex}^\mrex,\mq)
\end{bmatrix}
\label{e:UFE}\eqe
is a vector containing the corresponding FE displacements at $\bx^\mrex_I$, which follow from Eq.~\eqref{e:uDisc} as
\eqb{l}
	\bu^h(\bx^\mrex_I,\mq) = \mN(\bx^\mrex_I)\,\muu_e(\mq) \,.
\label{e:uFE}\eqe

To assign the same importance to all $n_\mrlc$ experiments all terms in the residual part of $f(\mq)$ are normalized w.r.t.~the Euclidean norm of the corresponding experimental vector. The weights are selected heuristically in the examples presented in Sec.~\ref{s:Nex}, and their optimization is beyond the scope of this paper. The last part of Eq.~\eqref{e:objfnc} corresponds to generalized Tikhonov regularization, where $\alpha$ is the regularization parameter, $\mcc$ is an approximation of the solution, and $\mL$ is a penalty matrix\footnote{For example a 2D finite difference Laplace operator, or any other approximate differential operator.}. The objective function includes the resultant of contact forces to ensure a determinable solution when pure Dirichlet boundary conditions are used, i.e., when the lack of force information allows one to identify $\mq$ only up to an additive constant. This happens in Sec.~\ref{s:abdo} and \ref{s:iron}, where the displacements of a rigid probe are prescribed\footnote{The presence of the contact forces in $f(\mq)$ may be alleviated by applying a known force to the probe instead of prescribing its displacements.}. Furthermore, all parts of Eq.~\eqref{e:objfnc} can be concatenated into a single vector, resulting in $f(\mq)$ reduced to a scalar product of two vectors, as shown in Appendix~\ref{s:deriv}.

\subsection{Overview of the inverse framework}\label{s:overview}

To minimize the objective function~\eqref{e:objfnc}, the proposed inverse framework employs the \textit{lsqnonlin} solver from the MATLAB Optimization Toolbox\textsuperscript{\tiny{TM}}, which uses the Trust-region reflective (TRR) approach~\citep{Coleman1996}. TRR is a gradient-based optimization method that iteratively constructs a local approximation of $f(\mq)$ within a trust-region around the current iterate $\mq_k$ and computes its minimum~\citep{Conn2000}. The algorithm terminates when the series $f(\mq_k)$ and $\mq_k$ converge, satisfying both
\begin{align}
	\left \| \mq_k - \mq_{k-1} \right\| &\le \epsilon \,, \label{e:qconv}\\[3pt]
	\left| f(\mq_k) - f(\mq_{k-1}) \right| &\le \epsilon\left(1 + \left|f(\mq_{k-1})\right| \right) \,, \label{e:fconv}
\end{align}
for a given small tolerance $\epsilon$. 

\begin{figure}[h]
	\begin{center}
		\includegraphics[width=1\textwidth]{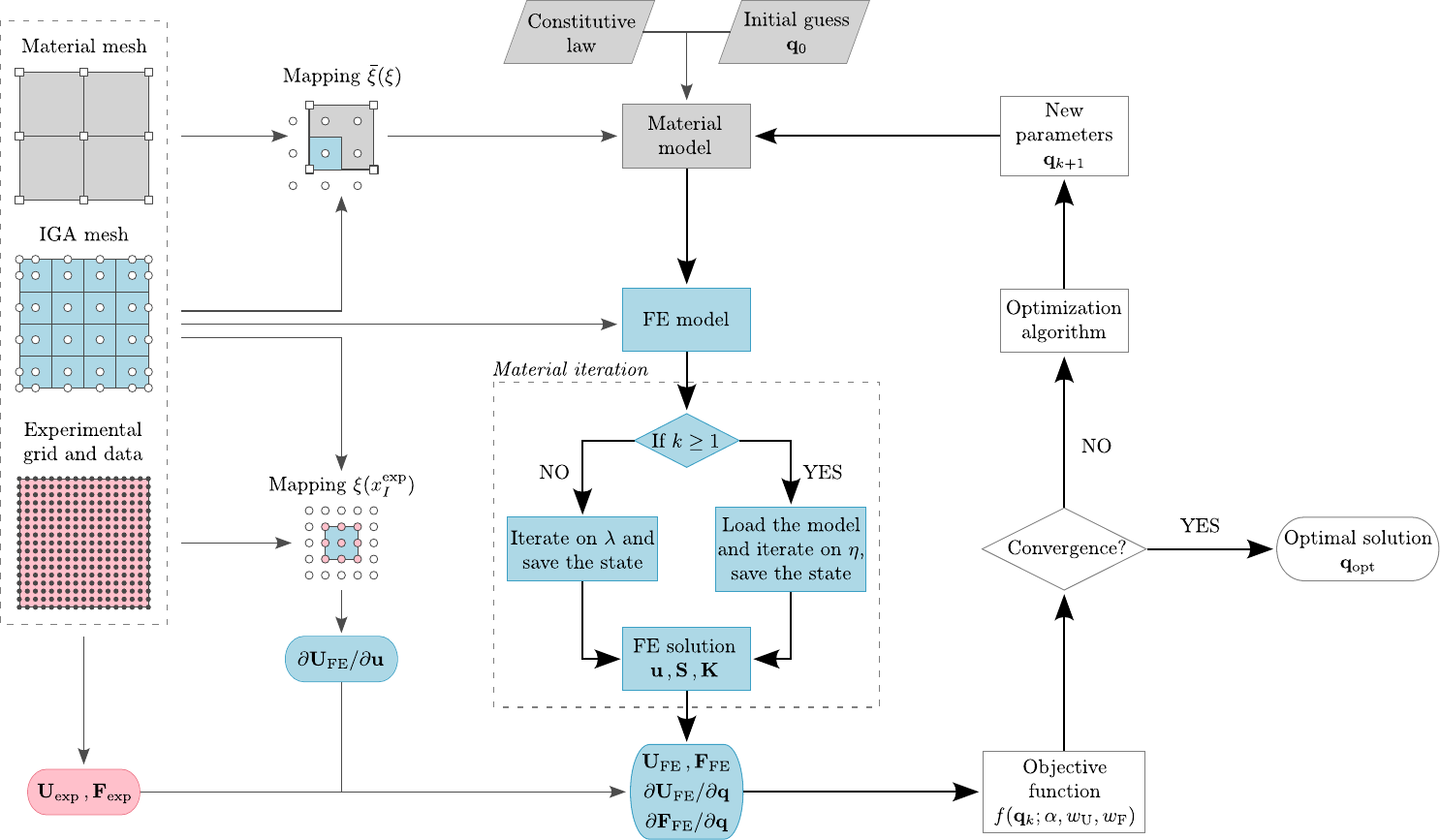}
		\caption{Flowchart of the proposed identification algorithm: provided the experimental data, the constitutive law, and the initial guess $\mq_0$, the algorithm determines the optimal solution $\mq_\mrop$ for a prescribed FE mesh, material mesh, and hyperparameters $\alpha,w_\mathrm{U},w_\mathrm{F}$}.
		\label{fig:theflowchart}
	\end{center}
\end{figure}
Fig.~\ref{fig:theflowchart} presents the flowchart of the proposed inverse framework. The inverse analysis starts with a given IGA mesh, material mesh, and experimental grid. Provided the experimental data and the constitutive law, the algorithm computes the optimal\footnote{The inverse solution $\mq_\mrop$ is optimal with respect to the mathematical objective in Eq.~\eqref{e:objfnc}.} material parameters $\mq_\mrop$ for a selected initial guess $\mq_0$ and hyperparameters $w_\mrU$, $w_\mrF$, and $\alpha$. Thanks to the material iteration strategy, described in Sec.~\ref{s:matIter}, the FE model iterates on load level $\lambda$ only at the $k = 0$ iteration. For $k \geq 1$, the model updates solely the material increment $\eta$, while successively saving the last equilibrated configuration. The initial guess $\mq_0$ may come from a preliminary inverse analysis with a homogeneous material distribution (see Sec.~\ref{s:iron}). Here, this approach is referred to as \textit{homogeneous preconditioning}. It provides a reasonable initial guess based on the averaged response of the FE model; thus, it may filter out a potential non-physical solution to the inverse problem. Further, this strategy may serve as a basis for a future automatic adaptive material mesh scheme.

{\small \textbf{Remark 4}:   
	A similar strategy, referred to as \textit{element partition}, was used in~\citet{Kroon2008} and succeeding papers. There, the inverse procedure refined the representation of the unknown material distributions a predefined number of times. For the numerical examples discussed in this work, precomputing the initial guess is sufficient, and no intermediate material mesh is needed. 
}\par

Analytically derived Jacobian of residuals, $\mJ$, is transferred from the FE model to the optimization algorithm in an explicit matrix form, or as a function computing the Jacobian-vector product (see Appendix.~\ref{s:deriv}). Along with $\mq_\mrop$, the algorithm returns $\mJ(\mq_\mrop)$, enabling uncertainty quantification of $\mq$~\citep{Roux2020}. 

Since the mechanical contact introduces physical and numerical discontinuities~\citep{Duong2019}, the objective function becomes non-differentiable at any point corresponding to a change in the active set. This is illustrated in Sec.~\ref{s:strip}. However, the risk of landing at a non-differentiable point is very low in practice. Furthermore, for forward problems where the active set is not sensitive to the material parameters, such as indentation with a probe, this issue is marginalized. A potential remedy is to use a smooth contact formulation.

The material mesh regularizes the inverse problem itself. Consequently, no additional regularization is required in most of the examples considered. Nevertheless, for dense material meshes or high noise levels, additional control over $\mq$ becomes necessary\footnote{Additional regularization may also be justified when prior information about the sought material distribution is available.}. The same holds when only partial experimental data are available. Furthermore, the initial estimate obtained from homogeneous preconditioning may serve as an approximation $\mcc$ of the solution $\mq$, in which case the regularization penalizes deviations from the homogeneous initial guess.

A risk of landing at a local minimum poses a significant threat to NLSQ problems, as they often have multiple solutions. To mitigate this, design variables are bounded, and the solution sensitivity to the initial guess is examined. In addition, the homogeneous preconditioning may reduce the risk of finding an undesired solution as discussed above. If multimodality of $f(\mq)$ is confirmed, one may use a multistart approach. However, the preliminary studies conducted on the numerical examples in Sec.~\ref{s:Nex} indicate that they converge to the same solution for different initial guesses. 

\subsection{Material iteration strategy}\label{s:matIter}

In a standard procedure for solving Eq.~\eqref{e:weakFE}, the Newton--Raphson (NR) method is applied at each increment of the load parameter $\lambda$. Neglecting higher-order terms, the expansion of~\eqref{e:weakFE} at the $i^{\text{th}}$ NR iteration gives
\eqb{l}
\mf(\muu_i + \Delta\muu,\lambda) = \mf(\muu_i,\lambda) + \mK(\muu_i,\lambda)\Delta\muu = \bf0 \,,
\label{e:NRstandard}\eqe
where $\mK$ denotes the tangent stiffness matrix, see, e.g.,~\citet{Wriggers2008}. From an alternative perspective~\citep{Gokhale2008}, the parameter $\lambda$ may be replaced by $\eta\in[0,1]$, which interpolates between the converged material state $\mq_k$ and the updated one $\mq_{k+1}$, i.e., $\mq(\eta) = \mq_k + \eta(\mq_{k+1} - \mq_k)$. Analogous to Eq.~\eqref{e:NRstandard}, one obtains
\eqb{l}
\mf(\muu_i + \Delta\muu,\eta) = \mf(\muu_i,\eta) + \mK(\muu_i,\eta)\Delta\muu = \bf0 \,,
\label{e:NRmaterial}\eqe
which preserves the same structure. The adopted computational strategy (see Fig.~\ref{fig:theflowchart}) is as follows: (1) at the initial optimization iteration the NR solver performs load incrementation on $\lambda$ and stores the converges state; (2) for the following iterations, the solver loads the previous state with fixed $\lambda$, solves \eqref{e:weakFE} for the material parameter $\eta$ and stores the converged state. If necessary, the interval $\eta \in [0,1]$ can be divided into multiple material steps. However, in most cases $\eta = 1$ is sufficient, in contrast to the typically more computationally demanding load incrementation. A limitation of the adopted approach is that it cannot be applied in the presence of path-dependent forces, such as friction. Moreover, as the number of load steps increases in the inverse analysis, the efficiency advantage gradually diminishes. Nevertheless, for all the considered examples in Sec.~\ref{s:Nex}, the adopted strategy achieves significant reduction in computational time.

Note that using homogeneous preconditioning as described in Sec.~\ref{s:overview} does not interrupt the material iteration strategy. The forward solver only needs to update the material mesh and recalculate the sensitivities (see Appendix~\ref{s:sensi}), while the last converged state at the end of homogeneous preconditioning can be passed to the target identification run with computed initial guess $\mq_0$. Thus, the forward solver still needs to iterate on the load level $\lambda$ only once.

\subsection{Experimental grid and data generation}\label{s:ExpDataGen}

\begin{figure}[h]
	\begin{center}
		\includegraphics[width=0.75\textwidth]{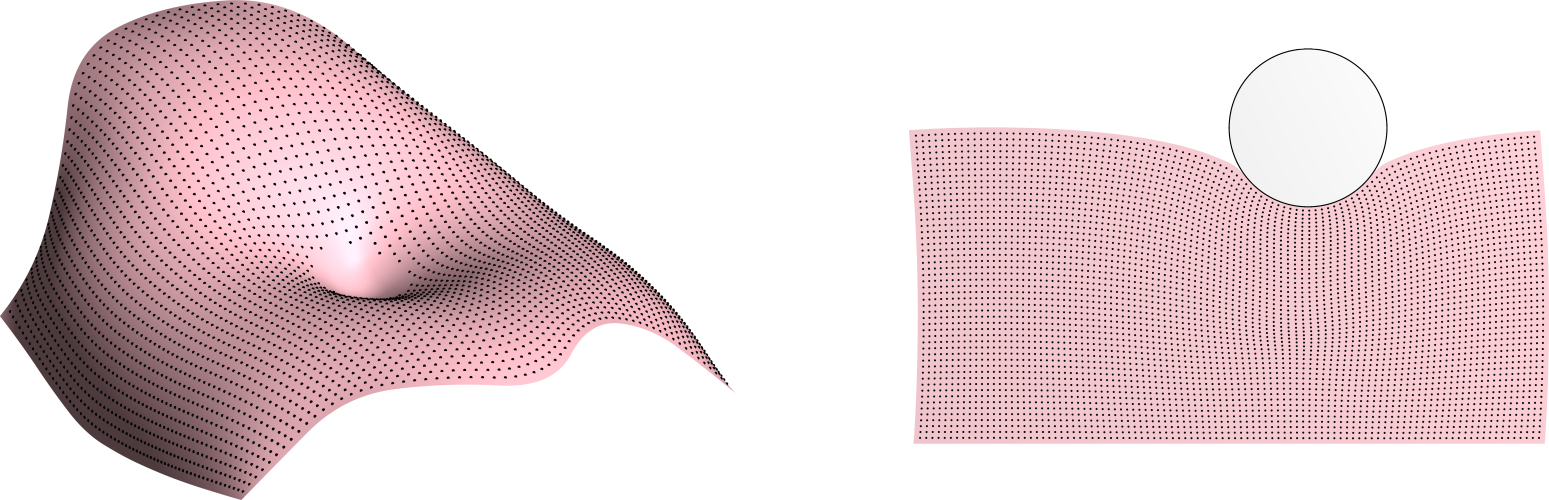}
		\caption{Illustration of experimental grids for the abdominal wall and block probing problems considered in Secs.~\ref{s:abdo} and~\ref{s:iron}, respectively.} 
		\label{fig:expgrid}
	\end{center}
\end{figure}

Since NURBS do not interpolate the control points, a direct comparison between FE nodal quantities and experimental measurements is not possible. Therefore, the control-point displacements are projected onto the experimental grid according to Eqs.~\eqref{e:uFE} and~\eqref{e:A1_10b}. Examples of the experimental grids employed in Sec.~\ref{s:Nex} are shown in Fig.~\ref{fig:expgrid}.

The quasi-experimental data used in Sec.~\ref{s:Nex} are generated synthetically using independent FE models. To avoid an obvious inverse crime~\citep{kaipio2005}, these models differ from those employed in the inversion. In this work, three main discrepancies between the forward and inverse solvers are introduced, representing sources of a systematic errors:
\begin{itemize}[noitemsep,topsep=0pt]
	\item \textit{Finite element discretization error}; The synthetic data are generated using a significantly finer IGA mesh than that used in the inversion. This constitutes the primary source of systematic error, particularly in contact problems exhibiting strong local deformations.
	\item \textit{Penalty parameter induced error}; The penalty method only approximates the exact contact conditions, allowing small penetrations. A higher value of penalty parameter $\epsilon_n$ is used for data generation to minimize penetration errors in the data.
	\item \textit{Material element discretization error}; The inverse procedure uses low-order Lagrange approximations for the material fields, while the synthetic data are generated using analytical material distributions.
\end{itemize}
In addition to these systematic errors, random noise is added to mimic measurement inaccuracies. Since systematic and random errors are of fundamentally different nature, they should be treated separately. A description of these error sources is provided in Sec.~\ref{s:Nex}, including FE mesh convergence studies and known material distributions. Other error sources occurring in practice are not considered here. Among them are:
\begin{itemize}[noitemsep,topsep=0pt]
	\item \textit{Constitutive error}; The same constitutive law is assumed in both data generation and inversion. 
	\item \textit{Kinematical simplification error}; For shells, synthetic data should originate from 3D continuum simulations; however, this is often impractical. As shown in~\citet{LAZORCZYK2026} for 1D structures, the additional error is small as long as the considered kinematical assumptions are valid. 
	\item \textit{Measurement error}; Inaccuracies associated with optical measurements are here reduced to Gaussian noise.
\end{itemize}
The investigation of these additional error sources is left for future work.

\section{Numerical examples}\label{s:Nex}

This section presents three numerical examples of increasing complexity, focusing on various aspects of the proposed CBIA framework. As the fundamental elements of the procedure have been presented in \citet{Borzeszkowski2022} and \citet{LAZORCZYK2026}, the discussion here is restricted to contact-related aspects. Firstly, a 1D cantilever shell strip deflecting onto a rigid foundation is examined in Sec.~\ref{s:strip}. In this simple inverse problem, the contact active set strongly depends on the material distribution; thus, it allows one to graphically study the effect of non-differentiability on the $f(\mq)$ landscape. Secondly, Sec.~\ref{s:abdo} extends the abdominal wall example from~\citet{Borzeszkowski2022} to contact-based synthetic experimental data. Finally, the reconstruction accuracy using only experimental surface data is investigated for a block made of rubber-like material in Sec.~\ref{s:iron}. While in the first example, only a single material field is identified, in the other two, two coupled material fields are reconstructed. All FE models utilize quadratic NURBS elements. For the FE convergence study, the discrete $L_2$-error of the displacements
\eqb{l}
e_u := \ds\frac{\norm{\muu_\mrexc - \muu_\mrfe}}{\norm{\muu_\mrexc}}
\label{e:FEerr}\eqe 
is examined, where $\muu_\mrect$ is a reference FE solution for a highly refined mesh. In Eq.~\eqref{e:FEerr}, both displacement vectors are evaluated on the example-specific experimental grid. In order to account for measurement inaccuracies, component-wise absolute noise is considered according to
\eqb{l}
u^\mrex_{Ii} = u_{\mrexc\,i}(\bx^\mrex_I) + u_{\mathrm{noise}\,i}(\bx^\mrex_I) \,,\quad u_{\mathrm{noise}\,i}\in\mcalN(0,\gamma_{Ii}^2) \,,
\label{e:noise}\eqe
where $i=1,2,3$ are the Cartesian components, $u_{\mrect\,i}$ is the reference solution, and $\mcalN(0,\gamma_{Ii}^2)$ is a normal distribution with zero mean and standard derivation $\gamma_{Ii}$, later simply referred to as noise. Resultant contact forces are always considered noise-free. All cases with random noise are repeated to obtain statistical estimates. The resulting errors in the following tables are presented as mean values with standard deviations in the form \textit{mean}~$\pm$~\textit{std}. The results in the subsequent examples report three indicators of the identification error: (1) the nodal relative error $\delta_I$, given by
\eqb{l}
\delta_I := \ds\left| \frac{q_{I,\mrrf} - q_{I,\mrop}}{q_{I,\mrrf}} \right| \,, \quad I=1,\dots,n_\mrvr \,, 
\label{e:error}\eqe
where $q_{I,\mrrf}$ are the reference values of the material parameters and $q_{I,\mrop}$ are the components of $\mq_\mrop$; (2) maximum nodal error $\errM := \max(\delta_I)$; and $L_2$-error, defined, e.g., for the Canham bending stiffness by
\eqb{l}
\errLe := \ds\frac{\norm{c_\mrop-c_\mrrf}_{L_2}}{\norm{c_\mrrf}_{L_2}} \,,\qquad \norm{c}_{L_2} := \ds\sqrt{\int_{\mcalS_0} |c(\bX)|^2 \,\dif A} \,,
\label{e:L2err}\eqe
where $c(\bX)$ follows from Eq.~\eqref{e:qdiscr}. The integration is computed element-wise and then summed. For the remaining material properties, $\errLe$ follows accordingly. 

All examples utilize the material iteration approach, as presented in Sec.~\ref{s:overview}. The size of the examined inverse problems varies from $n_\mrvr = 2$ material unknowns in Sec.~\ref{s:strip} to $n_\mrvr = 6498$ in Fig.~\ref{fig:abdoMEmeshConv} of Sec.~\ref{s:abdo}. For inverse problems with more than 400 unknowns (Fig.~\ref{fig:abdoMEmeshConv} and entire Sec.~\ref{s:iron}), it is more efficient to supply the TRR optimization algorithm with the Jacobian-vector product (see Appendix~\ref{s:jacvec}) instead of the full Jacobian. The stopping criteria tolerance in \eqref{e:qconv} and \eqref{e:fconv} is $\epsilon = 10^{-6}$. A smaller $\epsilon$ usually does not improve the identification. The number of load levels $n_\mrll$ always matches $n_\mrlc$ in \eqref{e:objfnc}. 

\subsection{Bending of a shell strip with contact constraint}\label{s:strip}

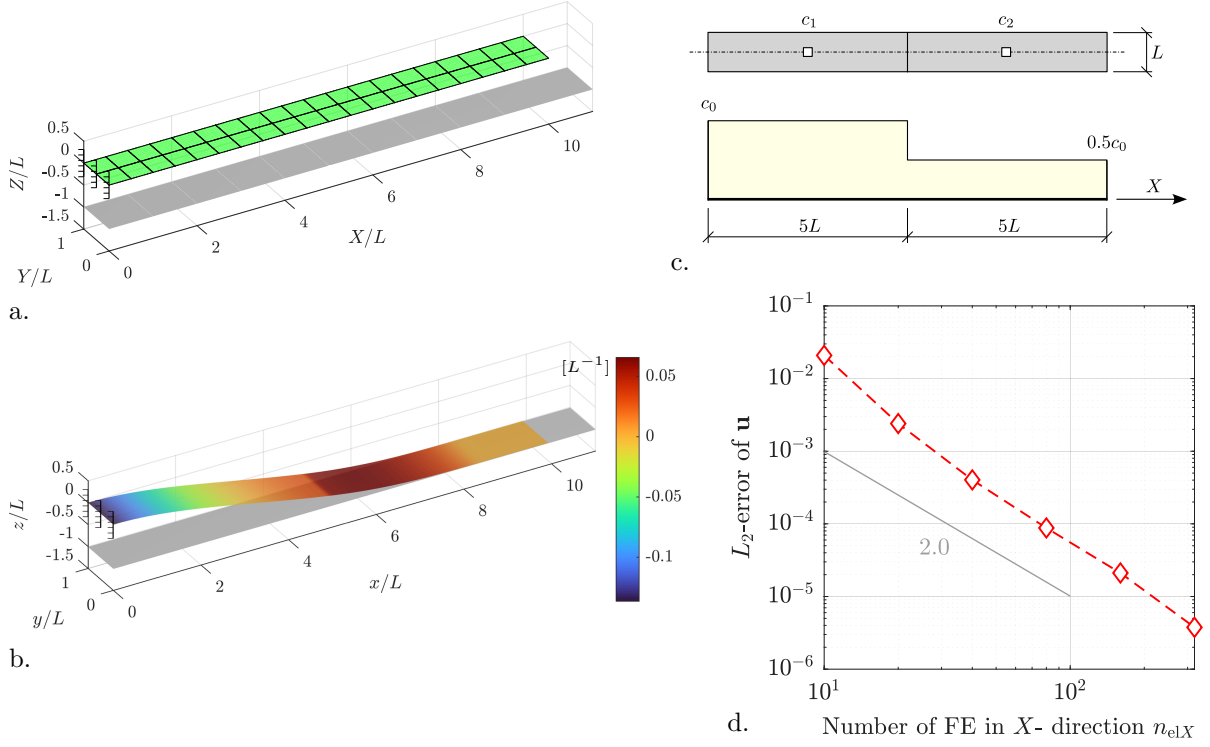
\begin{figure}[htb]
	\begin{center} \unitlength1cm
		\begin{picture}(0,9.8)
			\put(-8,6){\includegraphics[height=38mm]{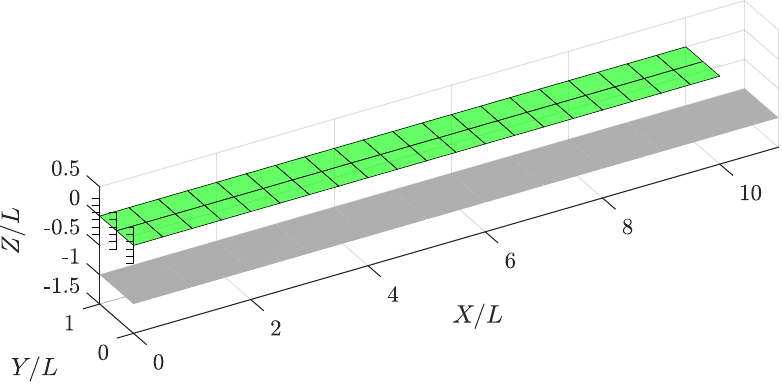}} 			
			\put(1,6.6){\includegraphics[height=30mm]{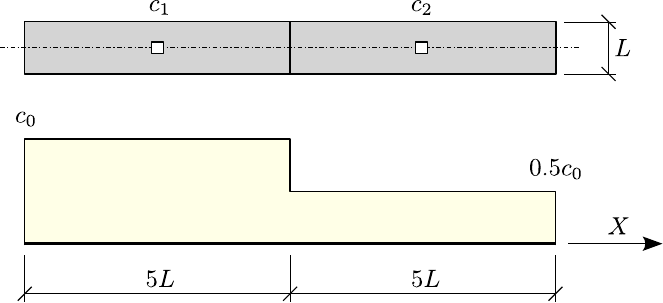}}
			\put(-8,1.5){\includegraphics[height=38mm]{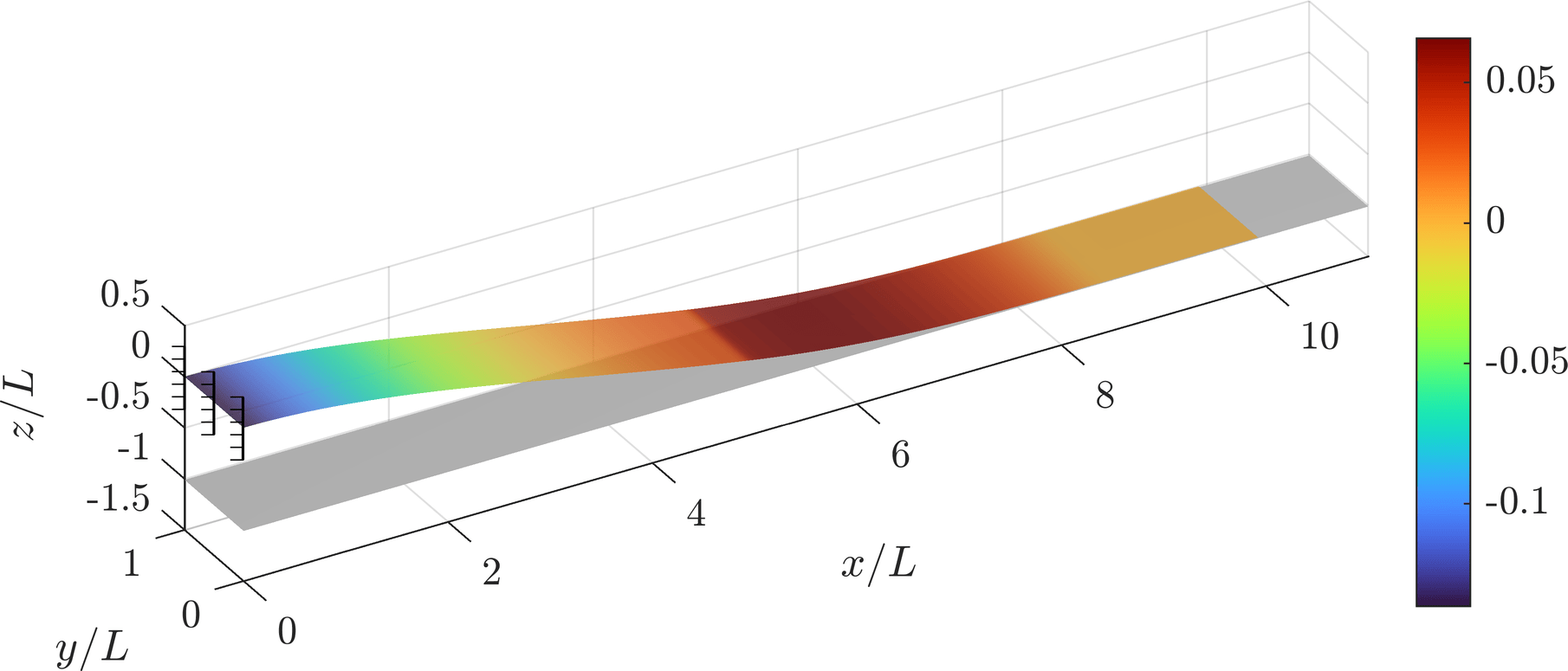}}  			 
			\put(1.5,0){\includegraphics[height=60mm]{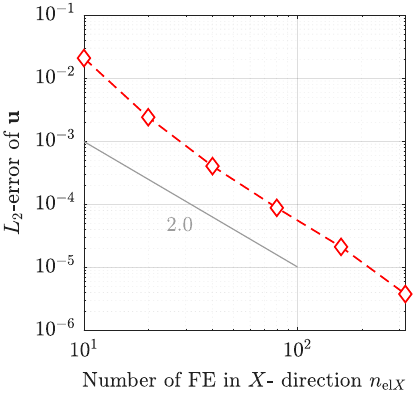}} 					
			\put(-8,5.6){\small{a.}}
			\put(0.75,6.25){\small{c.}}
			\put(-8,1){\small{b.}}
			\put(1.5,0.15){\small{d.}}
			\put(-0.7,4.9){\tiny{$[L^{-1}]$}}
		\end{picture}
		\caption{Bending of a shell strip: a.~undeformed configuration with boundary conditions; b.~the strip deformed by a uniform dead load, colored by curvature $\kappa_{11}$, ranging from $-0.137/L$ to $0.066/L$ (to avoid visible penetration, the strip is shifted by $0.01L$ upward); c.~material mesh with the reference distribution of the bending stiffness $c$; d.~FE convergence of the discrete $L_2$-error w.r.t.~the FE solution for 40960 elements.}
		\label{fig:strip1}	
	\end{center}
\end{figure}

The first example aims at presenting a low-dimensional inverse problem, which can be easily visualized, yet provides interesting insights into the identification problem. The example consists of an undeformed straight shell strip with length $L_x = 10L$, width $L_y = 1L$, and surface density $\rho = 10^{-5}\,m/L^2$, see Fig.~\ref{fig:strip1}a. The strip is clamped on the left side, while its deflection is constrained by a rigid, frictionless foundation at $Z = - 1L$. To accurately capture tip contact at the right end, separate line contact elements are used in addition to the surface contact elements. Fig.~\ref{fig:strip1}b presents the deformation of the strip under a dead load. The boundary conditions are chosen such that the contact active set strongly depends on the material distribution; thus allowing one to study the performance of the inverse framework for non-differentiable problems. 

The membrane response of the shell follows the incompressible Neo-Hookean model~\eqref{e:WSNH} with constant $\mu = 1 FL$. Here, $F$, $L$, and $m$ are normalization constants and need not be specified. The bending response is described by the Canham model~\eqref{e:WCan} with the bending stiffness distribution
\eqb{l}
c_\mrrf(X) = \left\{ \begin{array}{rcl}
	c_0 & \mbox{for} & X\in[0,5L] \,,\\
	0.5c_0 & \mbox{for} & X\in(5L,10L] \,,\\
\end{array}\,,\right.
\label{e:Ex1c}\eqe
where $c_0= 10^{-3}\,FL$, see Fig.~\ref{fig:strip1}c. For the basic material mesh, two constant ME are selected as Fig.~\ref{fig:strip1}c depicts. During the inverse analysis, only $c(X)$ is identified, while $\mu(X)$ is assumed to be known. Since the strip primarily bends, the influence of $\mu$ is marginal. Note that the material mesh matches the reference distribution perfectly, which is an example of an inverse crime. However, for the sake of this simple example, it is acceptable. The box constraints for $c_I$ are $c_\mathrm{min} = 0.01c_0$ and $c_\mathrm{max} = 10c_0$. For the cases without noise, the constant initial guess is $5.005c_0$. Otherwise, it is picked randomly from within the bounds.

A convergence study in Fig.~\ref{fig:strip1}d is performed to select a FE mesh for the quasi-experimental data generation. Following this study, the synthetic experimental data is generated using a mesh with $160\times16$ FE. For simplicity, the same $\epsilon_n = 10^{-1}\,F/L$ is used for both data generation and identification. All cases use the same uniform experimental grid with 101 and 11 points in $X$~and $Y$-~directions, respectively. Note that noise is applied along all directions in 3D.

\begin{figure}[htb]
	\begin{center} \unitlength1cm
		\begin{picture}(0,5.8)
			\put(-8,0){\includegraphics[height=58mm]{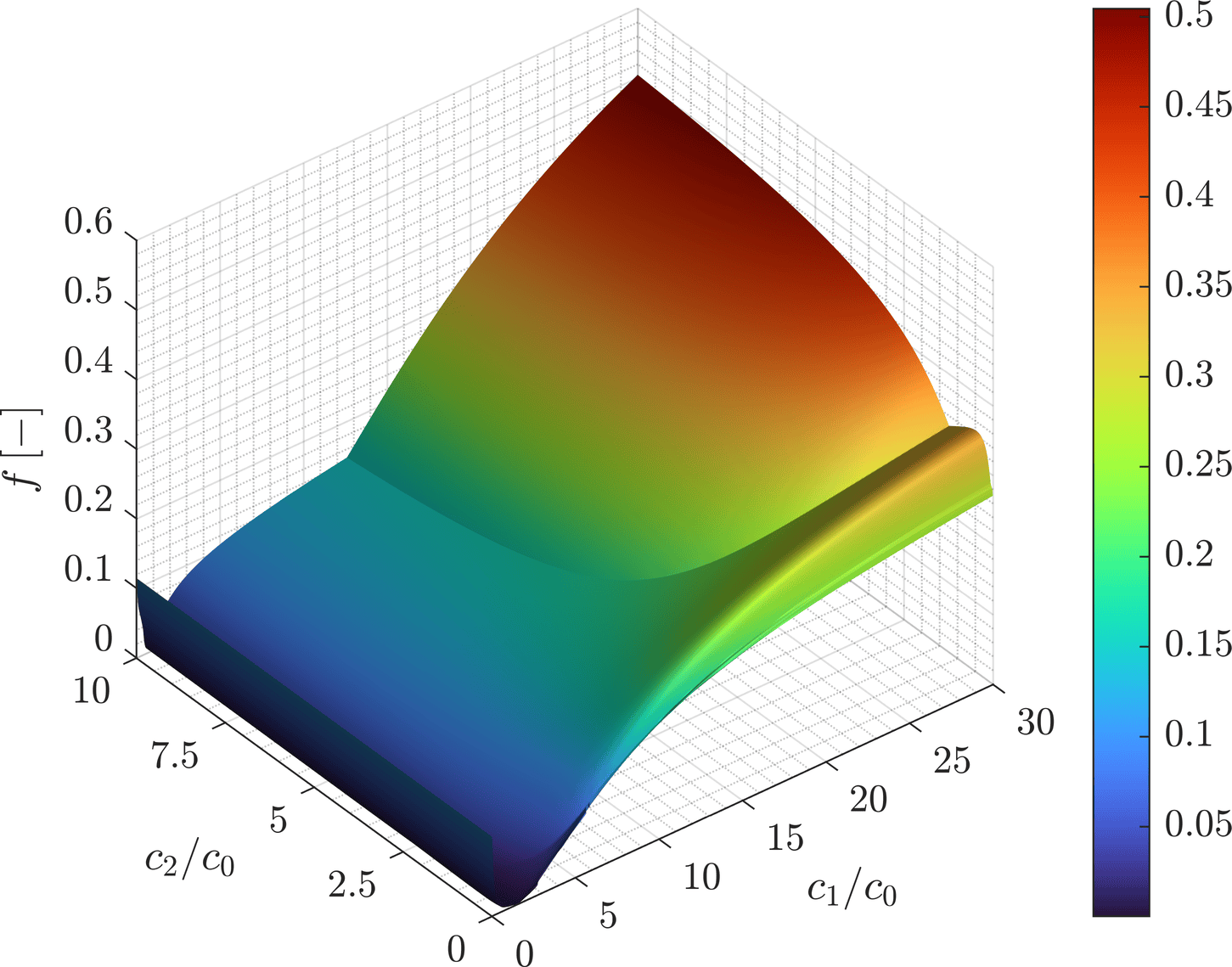}} 												
			\put(0.35,0){\includegraphics[height=58mm]{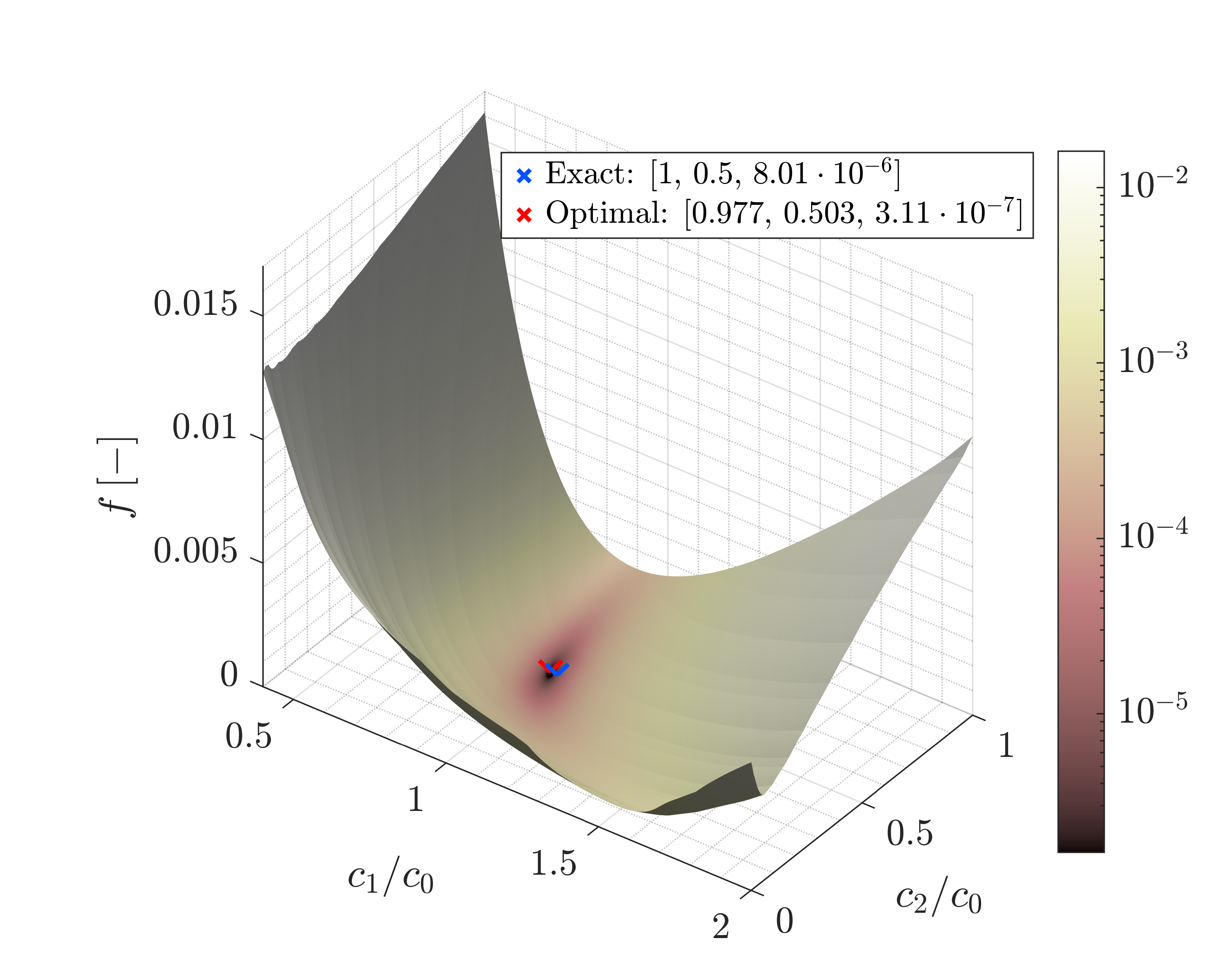}}					
			\put(-8,0){\small{a.}}
			\put(0.35,0){\small{b.}}
		\end{picture}
		\caption{Bending of a shell strip: a.~graph of $f(\mq)$ for Case 1.2 in Tab.~\ref{tab:Ex1_C} over the range $[0.001,30]\times[0.001,10]$; b.~zoom of the same graph over the range $[0.4,2]\times[0.001,1]$. The optimal solution $\mq_\mathrm{opt}$ and reference values are highlighted on the right surface.}
		\label{fig:strip_objfnc}	
	\end{center}
\end{figure}

\begin{figure}[htbp]
	\begin{center} \unitlength1cm
		\begin{picture}(0,5.4)
			\put(-8,0){\includegraphics[height=54mm]{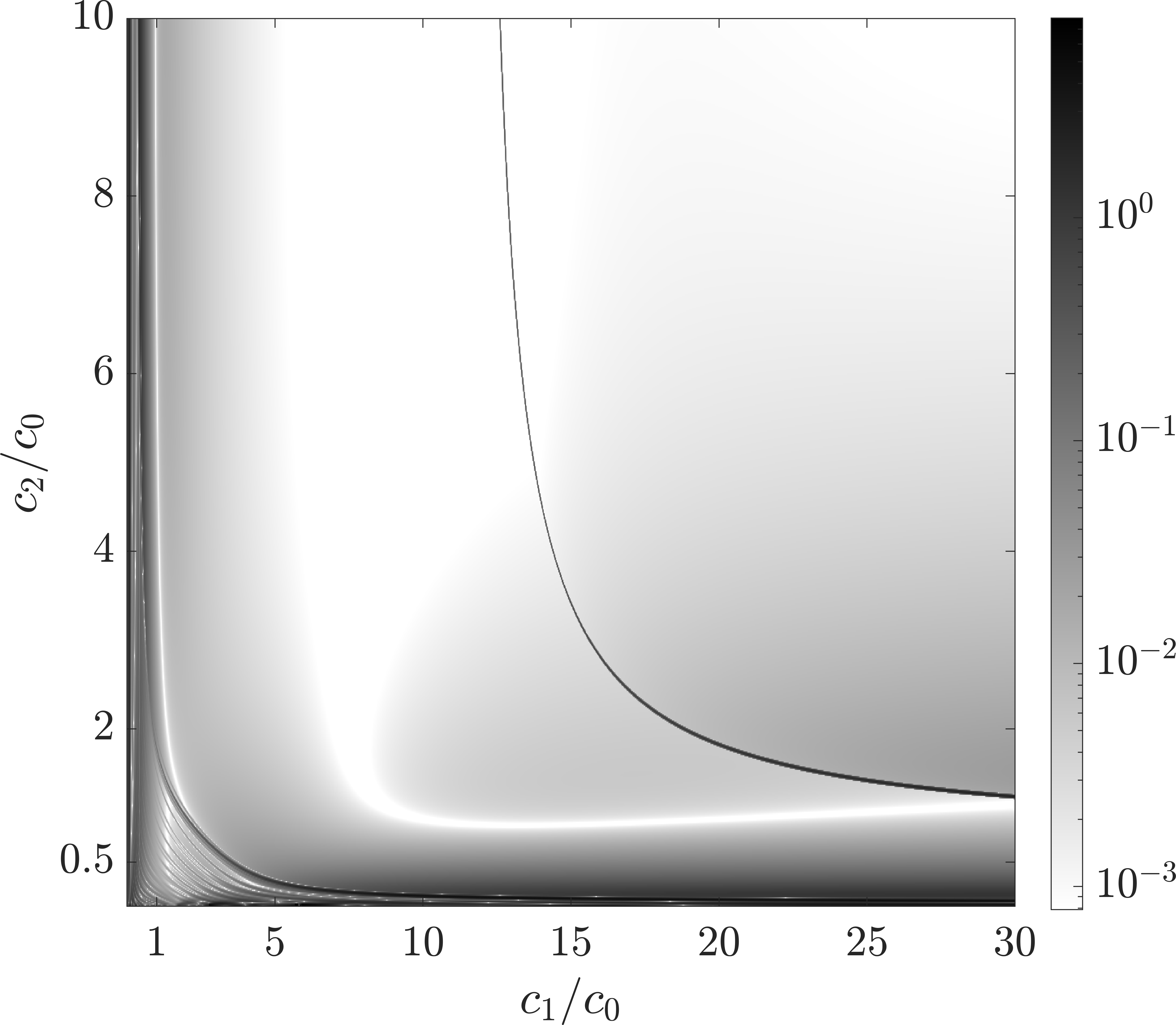}} 												
			\put(-1.2,0){\includegraphics[height=54mm]{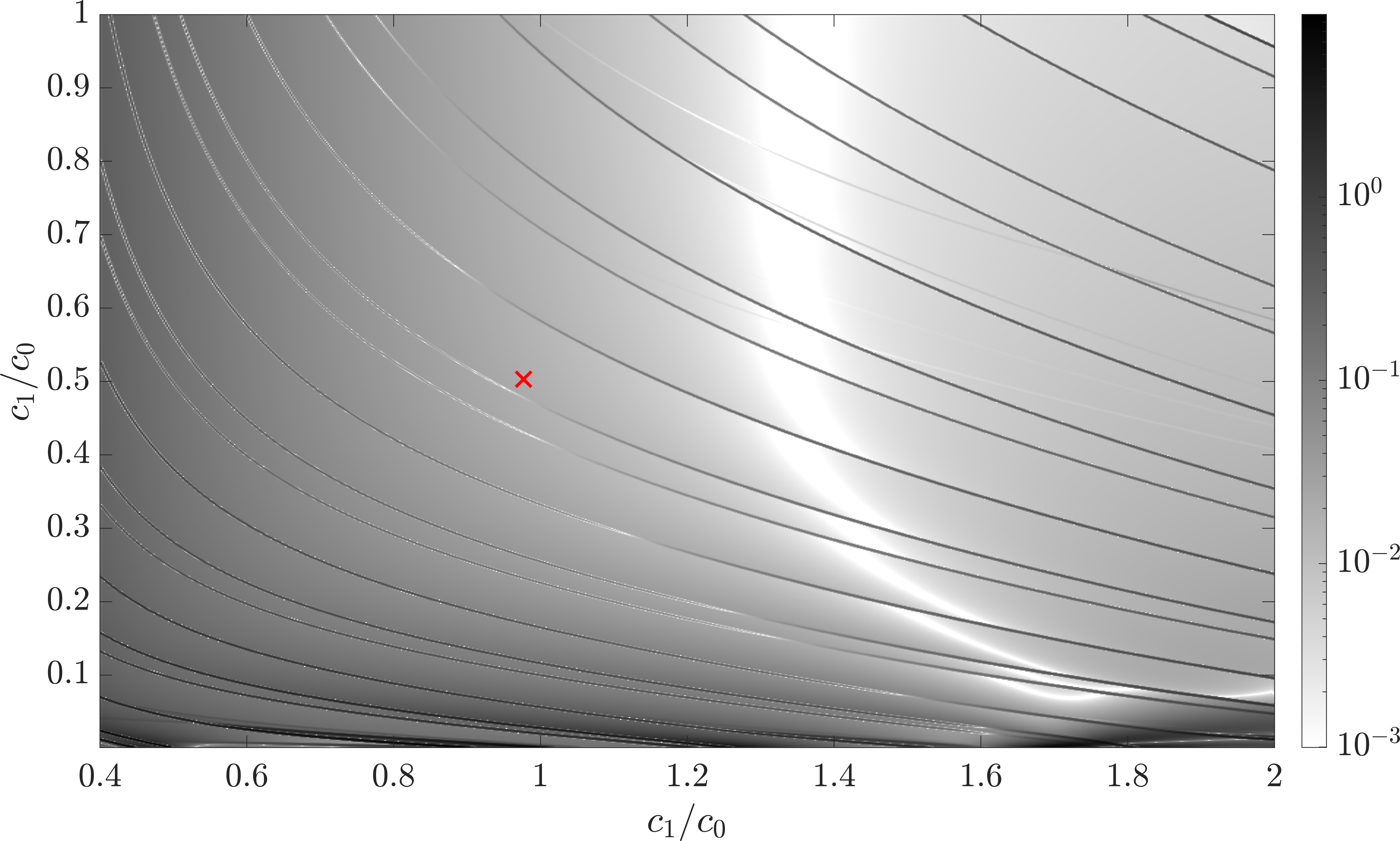}}					
			\put(-8,0){\small{a.}}
			\put(-1.35,0){\small{b.}}
		\end{picture}
		\caption{Bending of a shell strip: mean curvature maps for ranges $[0.001,30]\times[0.001,10]$~(a.)~and $[0.4,2]\times[0.001,1]$ (b.),~corresponding to the $f(\mq)$ landscapes from Fig.~\ref{fig:strip_objfnc}. The mean curvature is computed based on the finite difference derivatives of the $f(\mq)$. The red cross in b.~indicates the optimal solution.}
		\label{fig:strip_objfnc_lapl}	
	\end{center}
\end{figure}

\begin{table}[htb]
	\raggedright	
	\begin{tabular}{@{}llllllll@{}}
		\toprule
		Case & FE                & mat.                    & exp.                     & load              & noise      & $\errM$               & $\errLe$   			\\
			 & $n_\mrel$         & $\bar n_\mrel$          & $n_\mrex/n_\mrll$        & $n_\mrll$         & {[}$L${]}  & {[}\%{]}              & {[}\%{]}   			\\ \midrule
		1.1  & $10\times1$       & $2\times1$              & $101\times11$            & 1                 & 0          & 19.26                 & 17.23      		    \\
		1.2  & $20\times2$       & $2\times1$              & $101\times11$            & 1                 & 0          & 2.29   			   & 2.05     				\\ 
		1.3  & $40\times4$       & $2\times1$              & $101\times11$            & 1                 & 0          & 0.39                  & 0.35       			\\
		1.4  & $80\times8$       & $2\times1$              & $101\times11$            & 1                 & 0          & 0.069     			   & 0.063      			\\ 
		1.5  & $160\times16$     & $2\times1$              & $101\times11$            & 1                 & 0          & 0.012     			   & 0.0056     			\\ \midrule
		1.6  & $20\times2$       & $4\times1$              & $101\times11$            & 3                 & 0          & 2.96     			   & 1.88       			\\
		1.7  & $20\times2$       & $10\times1$             & $101\times11$            & 3                 & 0          & 15.56      		   & 3.86    				\\ 
		1.8  & $20\times2$       & $20\times1$             & $101\times11$            & 3                 & 0          & 42.86 			       & 6.78     				\\ \midrule
		1.9  & $20\times2$       & $2\times1$              & $101\times11$            & 3                 & 0.01       & $1.53\,\pm\,0.41$     & $1.23\,\pm\,0.18$      \\ 
		1.10 & $20\times2$       & $2\times1$              & $101\times11$            & 3                 & 0.02       & $1.89\,\pm\,0.86$     & $1.40\,\pm\,0.40$      \\ 
		1.11 & $20\times2$       & $2\times1$              & $101\times11$            & 3                 & 0.04       & $2.83\,\pm\,1.77$     & $1.80\,\pm\,0.80$      \\ \bottomrule
	\end{tabular}
	\caption{Bending of a shell strip: Cases of bending stiffness identification with their FE and material meshes, experimental grid density, no.~of load levels, \textit{std}. of noise, and identification errors $\errM$ and $\errLe$.} 
	\label{tab:Ex1_C}
\end{table}
The first five cases in Tab.~\ref{tab:Ex1_C} study the effect of FE mesh density on the identification results. In the absence of noise, the forward and inverse solvers differ only in the FE mesh. The identification errors decrease monotonically, reaching values around $0.01\%$ in Case 1.5, where both solvers are identical. These remaining values are likely due to numerical errors. Fig.~\ref{fig:strip_objfnc} shows the landscape of $f(\mq)$ for Case 1.2. The non-differentiability of $f(\mq)$ due to contact is clearly visible. This can also be seen from the mean curvature graphs in Fig.~\ref{fig:strip_objfnc_lapl}. Each curve in Fig.~\ref{fig:strip_objfnc_lapl} represents a fold in the $f(\mq)$ landscape; hence, a change in the active contact set. These folds can be divided into two categories: physical and artificial. The latter may occur due to discretization and contact formulation, while the former are independent of them~\citep{Duong2019}. An example of the former is the onset of the tip contact, corresponding to the most prominent quarter-round fold in Fig.~\ref{fig:strip_objfnc}a. Examples of the latter are the smaller folds corresponding to individual quadrature points entering the active set, shown in Fig.~\ref{fig:strip_objfnc}b and Fig.~\ref{fig:strip_objfnc_lapl}b. Alternative contact strategies, such as the popular mortar method, may partially alleviate the issue of artificial folds but do not eliminate it. Nevertheless, the objective function seems to have a unique minimum in the examined range of parameters. 
\begin{figure}[htbp]
	\begin{center} \unitlength1cm
		\begin{picture}(0,5.5)
			\put(-8,0){\includegraphics[height=55mm]{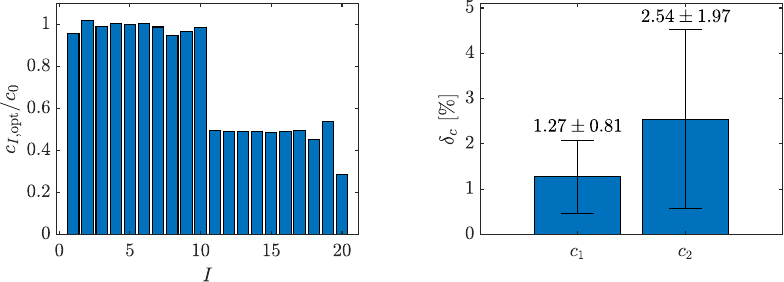}} 																
			\put(-7.6,0){\small{a.}}
			\put(0.7,0){\small{b.}}
		\end{picture}
		\caption{Bending of a shell strip: a.~identified nodal values of the bending stiffness $c$ for Case 1.8 in Tab.~\ref{tab:Ex1_C}; b.~average value and \textit{std}. of the nodal identification error of Case 1.11 in Tab.~\ref{tab:Ex1_C}.}
		\label{fig:strip2}	
	\end{center}
\end{figure}
\begin{figure}[htbp]
	\begin{center} \unitlength1cm
		\begin{picture}(0,5.5)
			\put(-7.2,0){\includegraphics[height=55mm]{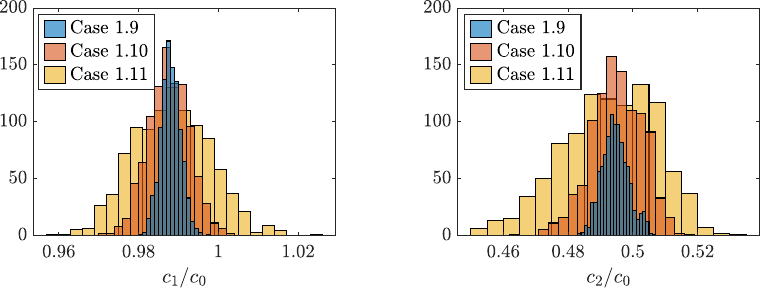}} 																
			\put(-7.6,0){\small{a.}}
			\put(0.7,0){\small{b.}}
		\end{picture}
		\caption{Bending of a shell strip: Histograms of the identified bending stiffness $c_1$ (a.) and $c_2$ (b.) for Cases 1.9--1.11 in Tab.~\ref{tab:Ex1_C}. Each histogram is based on $1000$ separate identification runs.}
		\label{fig:strip3}	
	\end{center}
\end{figure}

Cases 1.2 and 1.6--1.8 in Tab.~\ref{tab:Ex1_C} study the refinement of the material mesh in $X$-~direction using a fixed FE mesh. Using only a single load case with $100\%$ of the load (see Fig.~\ref{fig:strip1}b) would inevitably lead to an indeterminacy of the material unknowns at the right end of the strip. To avoid this, the subsequent cases in Tab.~\ref{tab:Ex1_C} use three ($n_\mrll = 3$) load cases with $10\%$, $50\%$, and $100\%$ of the original load level, where $10\%$ corresponds to contact initialization at the tip. As expected, the identification errors are consistently increasing with more ME. Refining the optimal $[2\times1]$ material mesh increases the flexibility of the spatial material representation, allowing it to compensate for discrepancies between the forward and inverse solvers. Fig.~\ref{fig:strip2}a shows that the identification error is particularly large at the right strip end. This is expected, since the material distribution at the tip affects the displacement only locally.

The last three examples in Tab.~\ref{tab:Ex1_C} investigate the effect of noise on the identification results, with $\gamma_{Ii}$ in Eq.~\eqref{e:noise} varying from $0.01L$ to $0.04L$. The identification errors and their standard deviations increase consistently with the noise level. Fig.~\ref{fig:strip2}b shows that $c_2$ dominates the identification error. As shown in Fig.~\ref{fig:strip_objfnc}, the deflection of the strip is less sensitive to $c_2$, which agrees with intuition. The histograms in Fig.~\ref{fig:strip3} and Tab.~\ref{tab:Ex1_C} reveal that increasing noise significantly affects the statistical dispersion of the identified values, while the mean values are affected less. The histograms exhibit an approximately normal distribution.

In all cases with $2\times1$ material mesh, the inverse solution is found in 7--8 iterations, while for denser material meshes it increases up to a maximum of 43 iterations.

\subsection{Abdominal wall}\label{s:abdo}

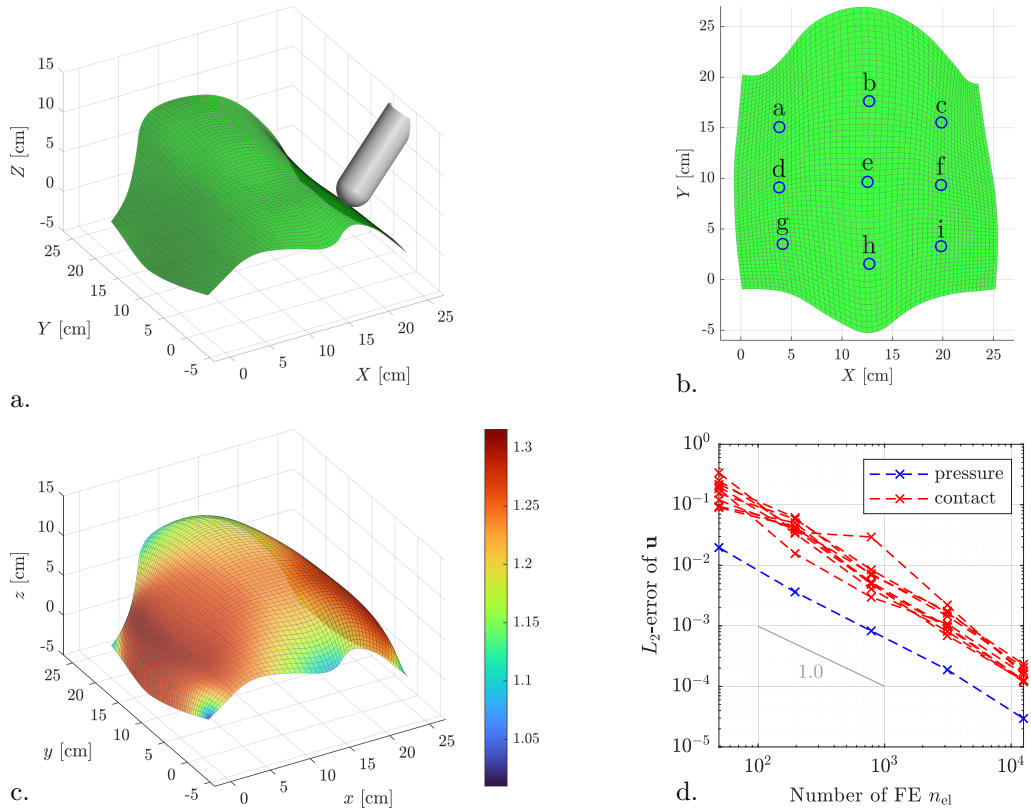
\begin{figure}[htb]
	\begin{center} \unitlength1cm
		\begin{picture}(0,10.6)
			\put(-7,5.6){\includegraphics[height=50mm]{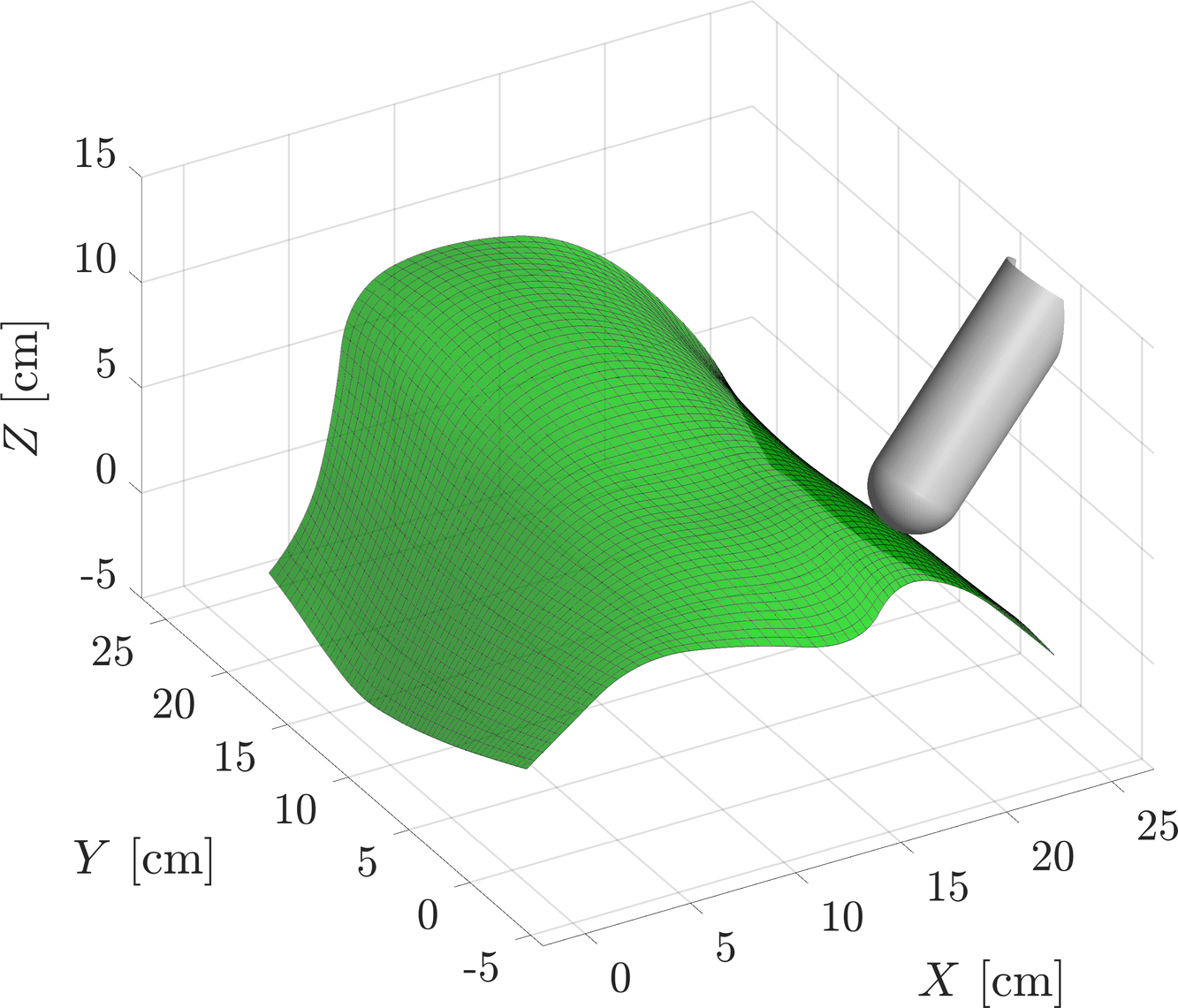}} 		
			\put(-7,0){\includegraphics[height=50mm]{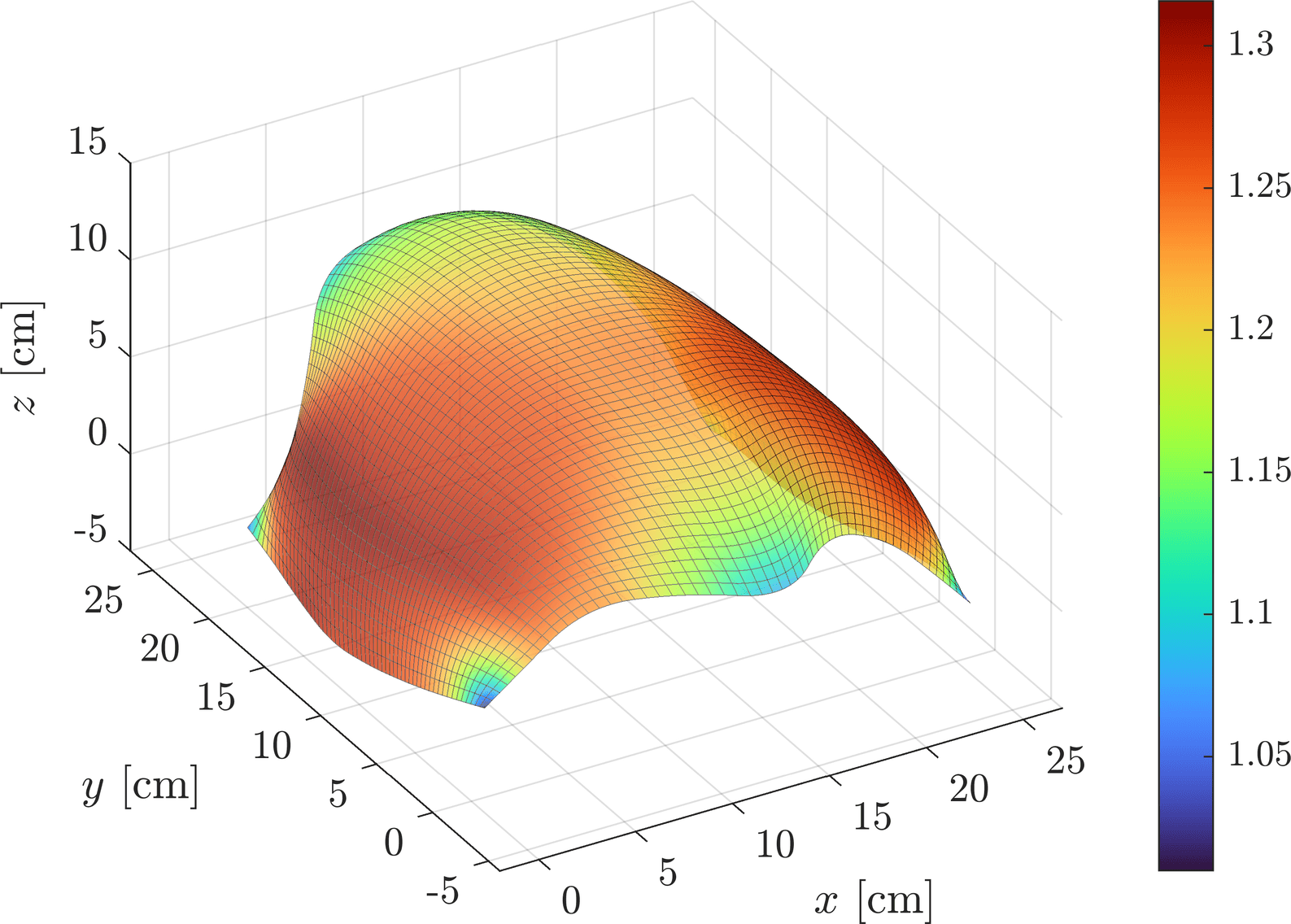}}  	
			\put(1.8,5.6){\includegraphics[height=50mm]{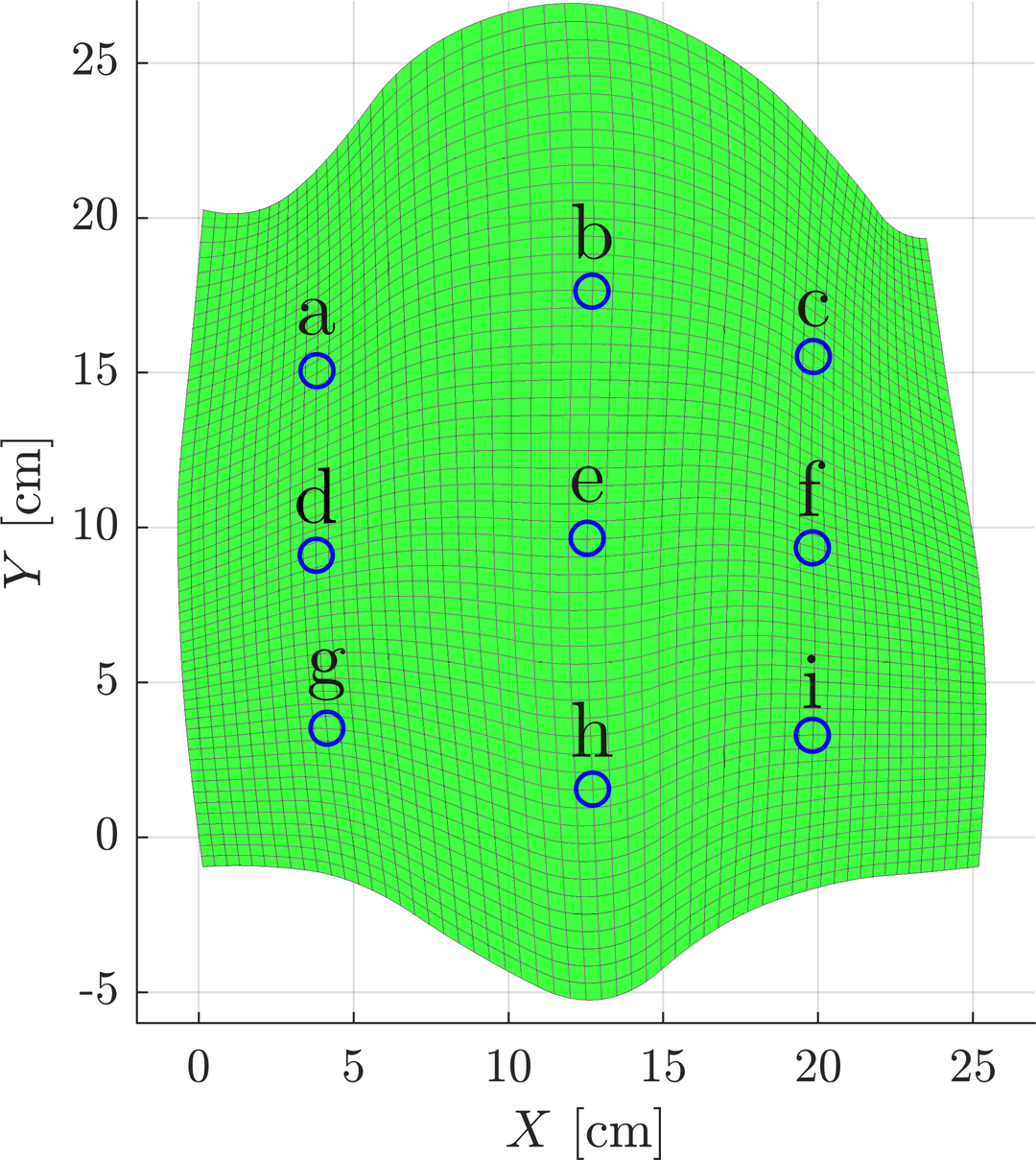}}  
			\put(1.3,0){\includegraphics[height=50mm]{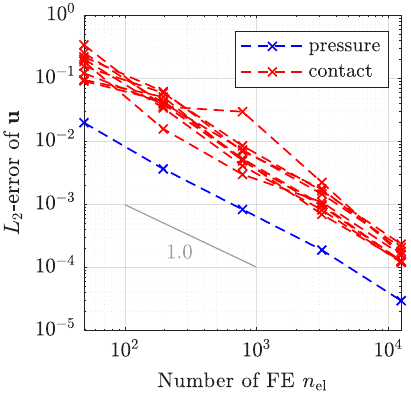}} 
			\put(-70mm,53mm){\small{a.}}
			\put(-70mm,1mm){\small{c.}}
			\put(18mm,55mm){\small{b.}}
			\put(18mm,1mm){\small{d.}}
		\end{picture}
		\caption{Abdominal wall: a.~Undeformed configuration with an exemplary position of the contact probe; b.~Top view of the indentation layout. The rigid probe indents the abdominal wall at nine positions, enumerated from~a to~i; c.~Deformed geometry of the abdominal wall under uniform pressure $p_\mathrm{intra}=\SI{1.6}{\kilo\pascal}$, without any contact. Colors represent the surface stretch $J$; d.~FE convergence of the discrete $L_2$-error w.r.t.~the FE solution for 28224 elements. The nine red dashed lines correspond to the nine indentation cases.}
		\label{fig:abdo1}	
	\end{center}
\end{figure}
The second example features the identification of Young's modulus and thickness distributions $E(\bX)$ and $T(\bX)$ in a shell model of the human abdominal wall, shown in Fig.~\ref{fig:abdo1}a. The model of~\citet{lubowiecka2018membrane} represents a real abdominal wall geometry, acquired using the measurement methodology of~\citet{SZYMCZAK2012}, and discretized with a single-patch NURBS following~\citet{Borzeszkowski2021}. The shell is pinned along its entire boundary and loaded through frictionless contact with a rigid probe of radius \SI{2}{cm}. Additionally, to approximate physiological conditions, the volume below the abdominal wall is constrained using the Lagrange multiplier method, with the uniform pressure $p_\mathrm{intra}$ acting beneath the shell chosen as the Lagrange multiplier. Consequently, the model behaves like a water-filled balloon\footnote{Neglecting the hydrostatic water pressure.}. In the initial contactless configuration, $p_\mathrm{intra} = 0$. Details of the volume constraint are given in~\citet{Sauer2014}. The active work of muscles is not considered here.

This study extends the numerical example presented in~\citet{Borzeszkowski2022}, referred here to as Pressure-based Inverse Analysis~(PBIA), where a nearly identical model was subjected to uniform pressure (see Fig.~\ref{fig:abdo1}c). That example was shown to have potential for material reconstruction. However, the pressurization of the abdominal wall requires invasive medical treatment, unlike the simple contact probing proposed here, which is noninvasive and does not require intra-abdominal pressure measurements. The model geometry used in~\citet{Borzeszkowski2022} is modified to avoid high identification errors in corners\footnote{The modifications include extending the model corners to obtain approximately right angles there.}. Hence, the PBIA is repeated here, now featuring absolute noise. The new results are summarized in Tab.~\ref{tab:abdo_pres}. 
\begin{figure}[htb] 
	\centering
	\includegraphics[height=36mm]{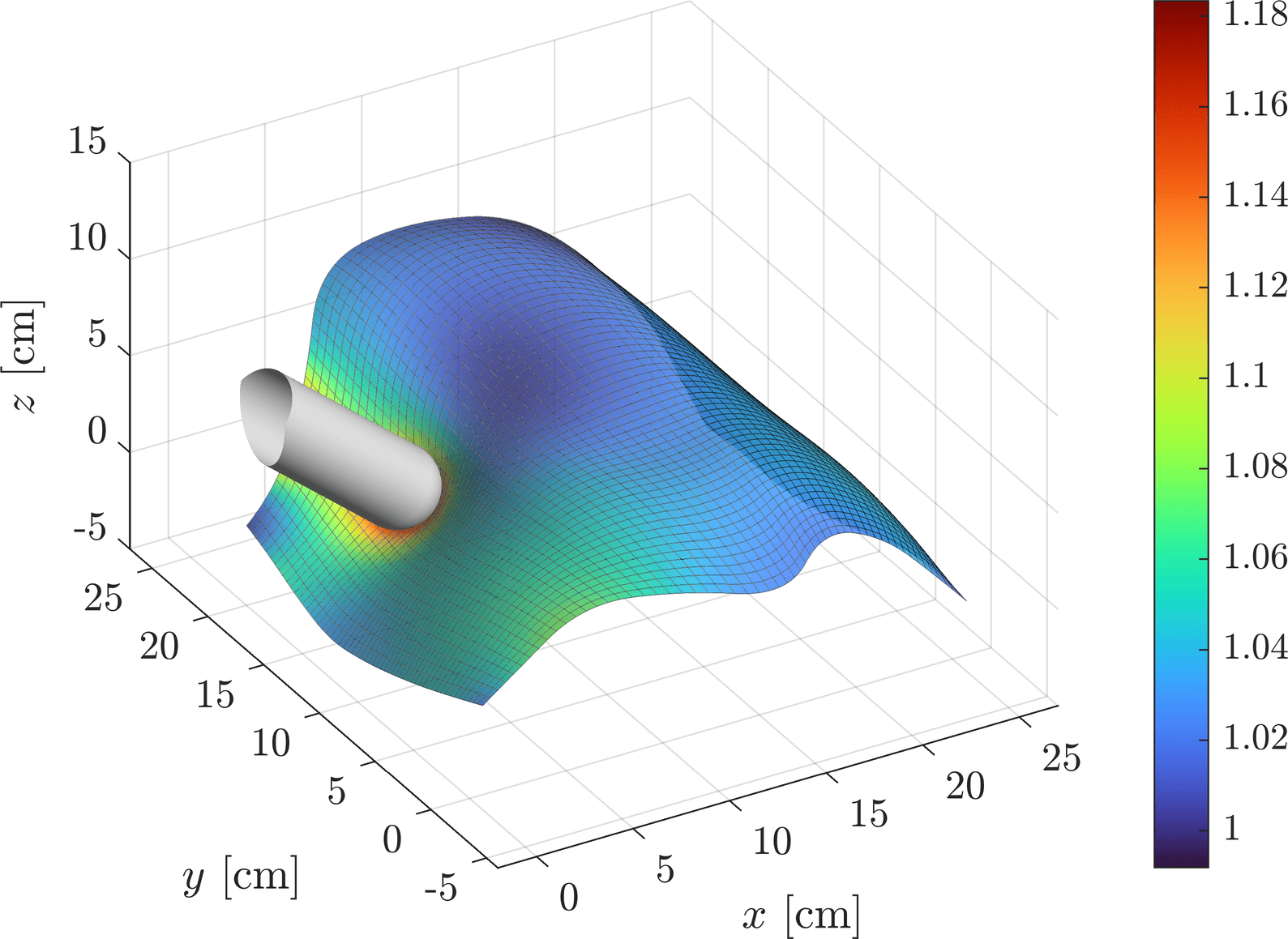}
	\put(-49mm,0mm){\small{a.}}
	\hfil 
	\includegraphics[height=36mm]{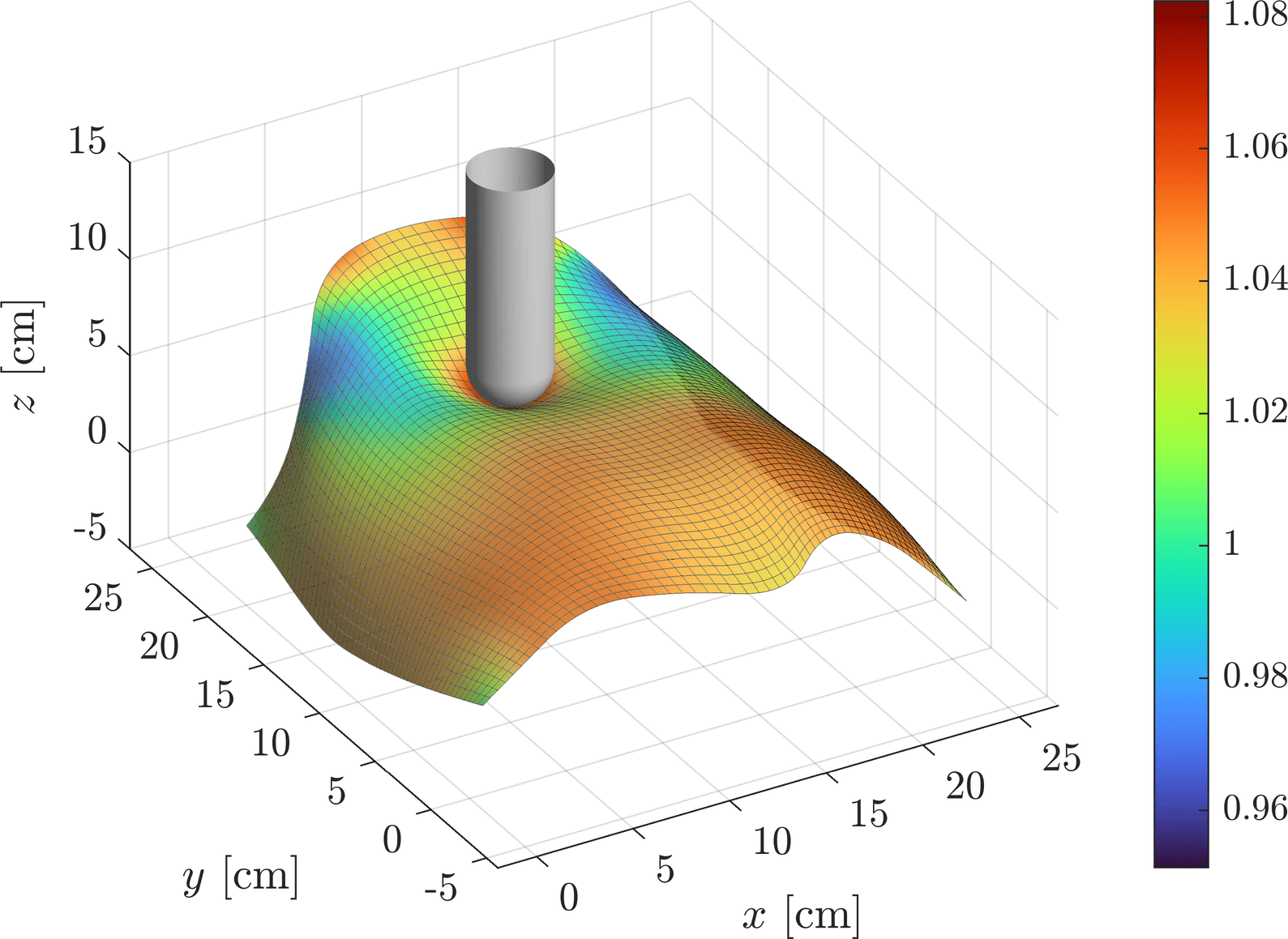}
	\put(-49mm,0mm){\small{b.}}
	\hfil 
	\includegraphics[height=36mm]{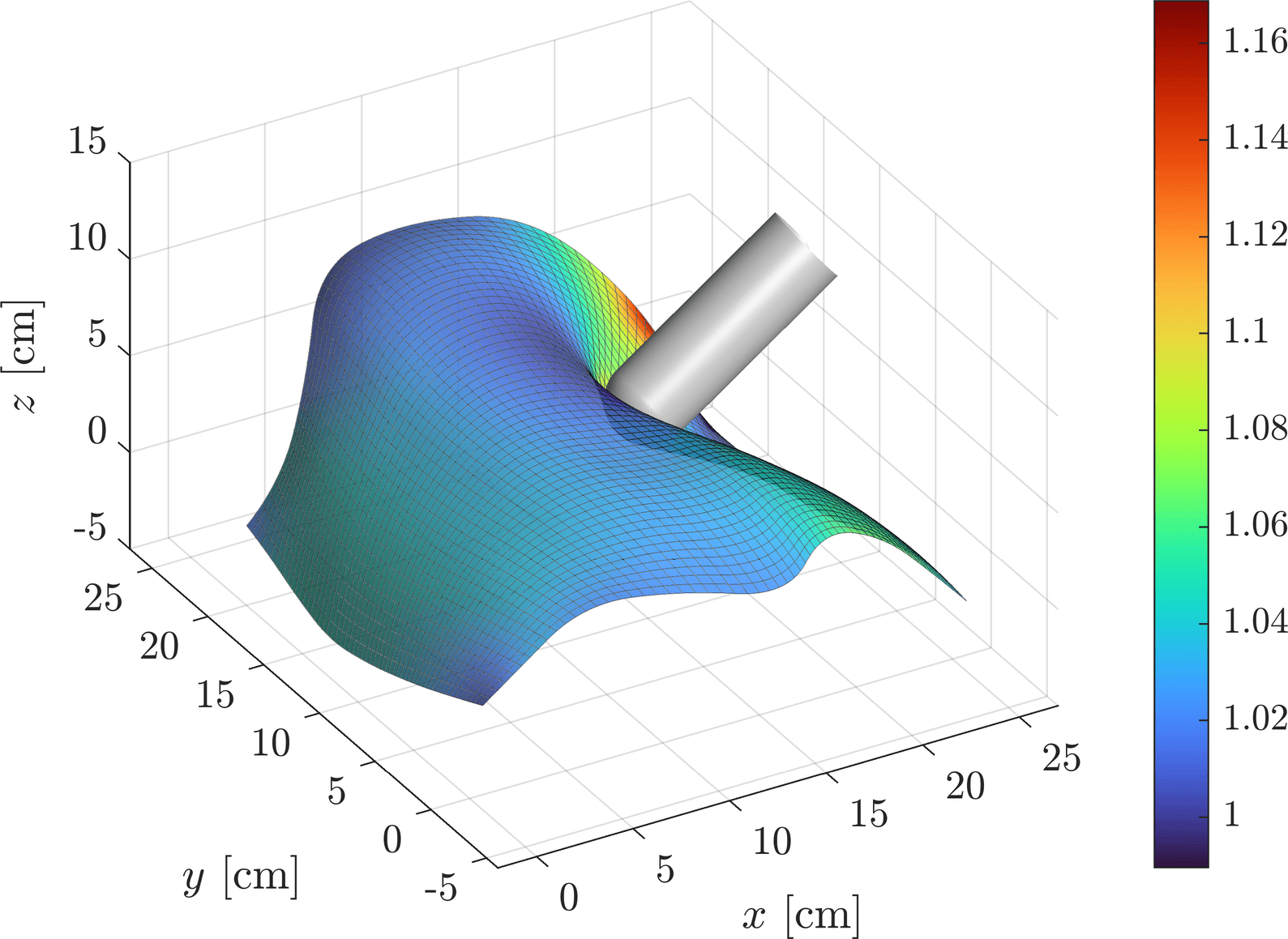} 
	\put(-49mm,0mm){\small{c.}}
	\vspace{3mm}
	\includegraphics[height=36mm]{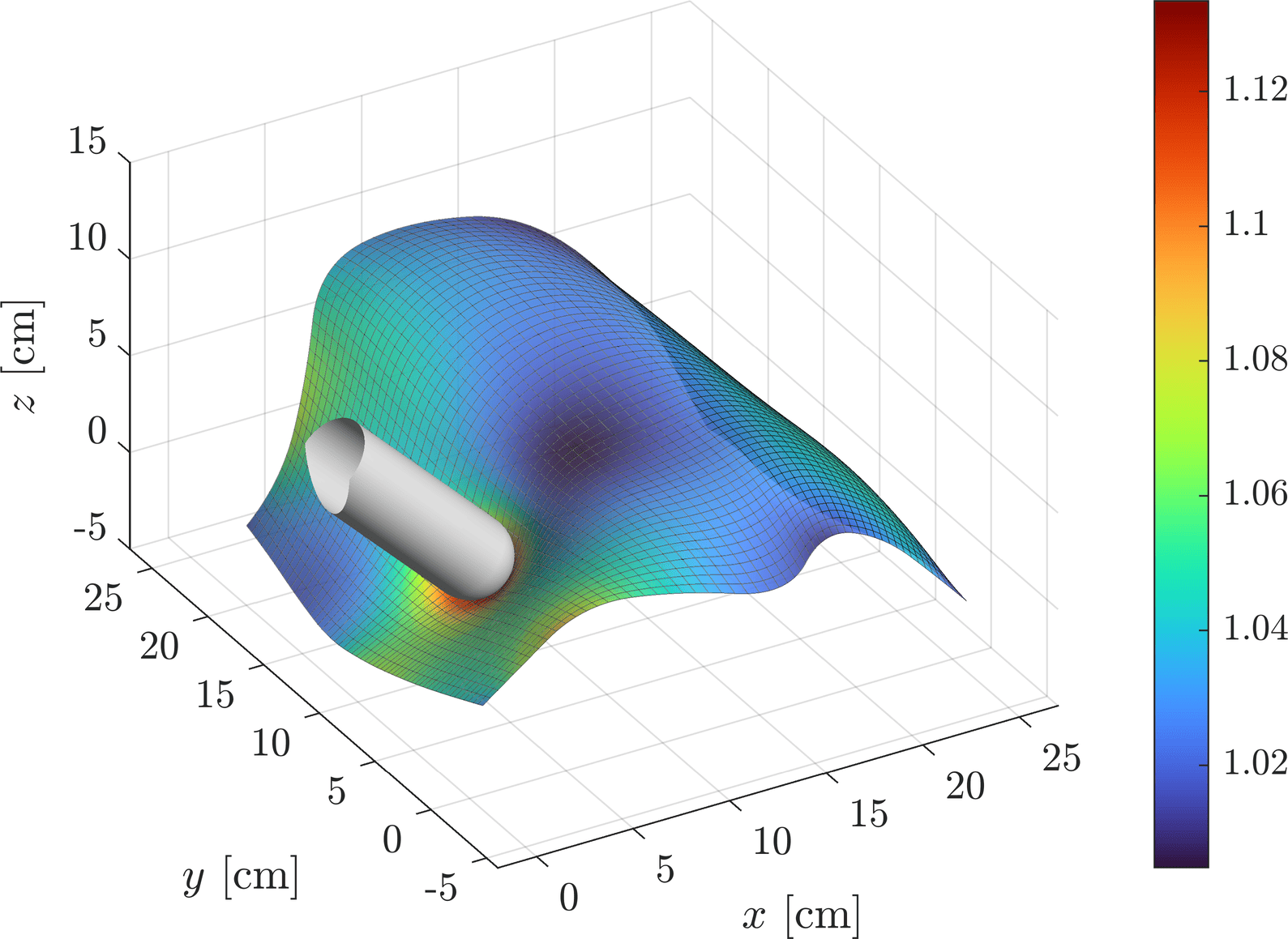}
	\put(-49mm,0mm){\small{d.}}
	\hfil 
	\includegraphics[height=36mm]{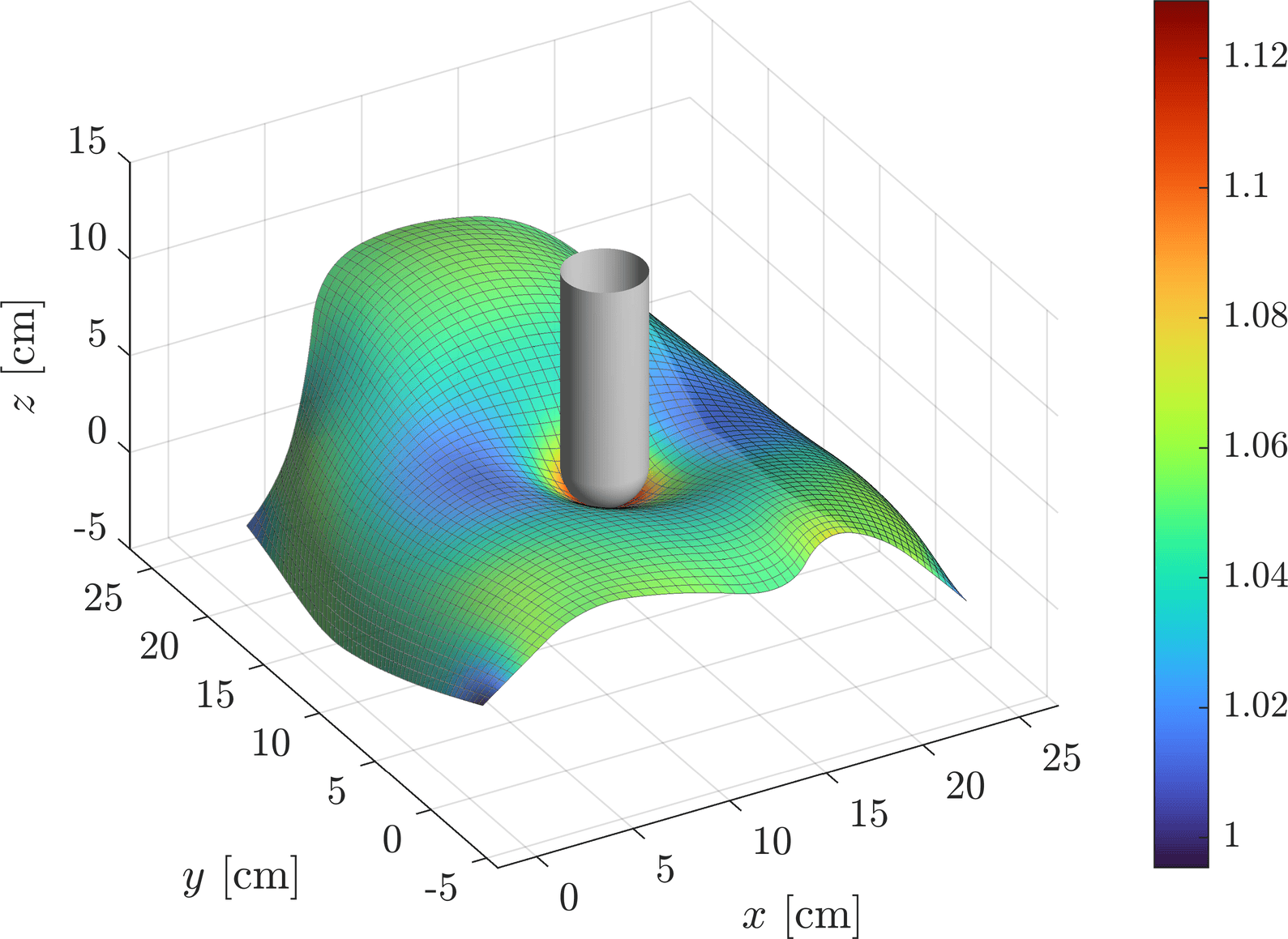}
	\put(-49mm,0mm){\small{e.}}
	\hfil 
	\includegraphics[height=36mm]{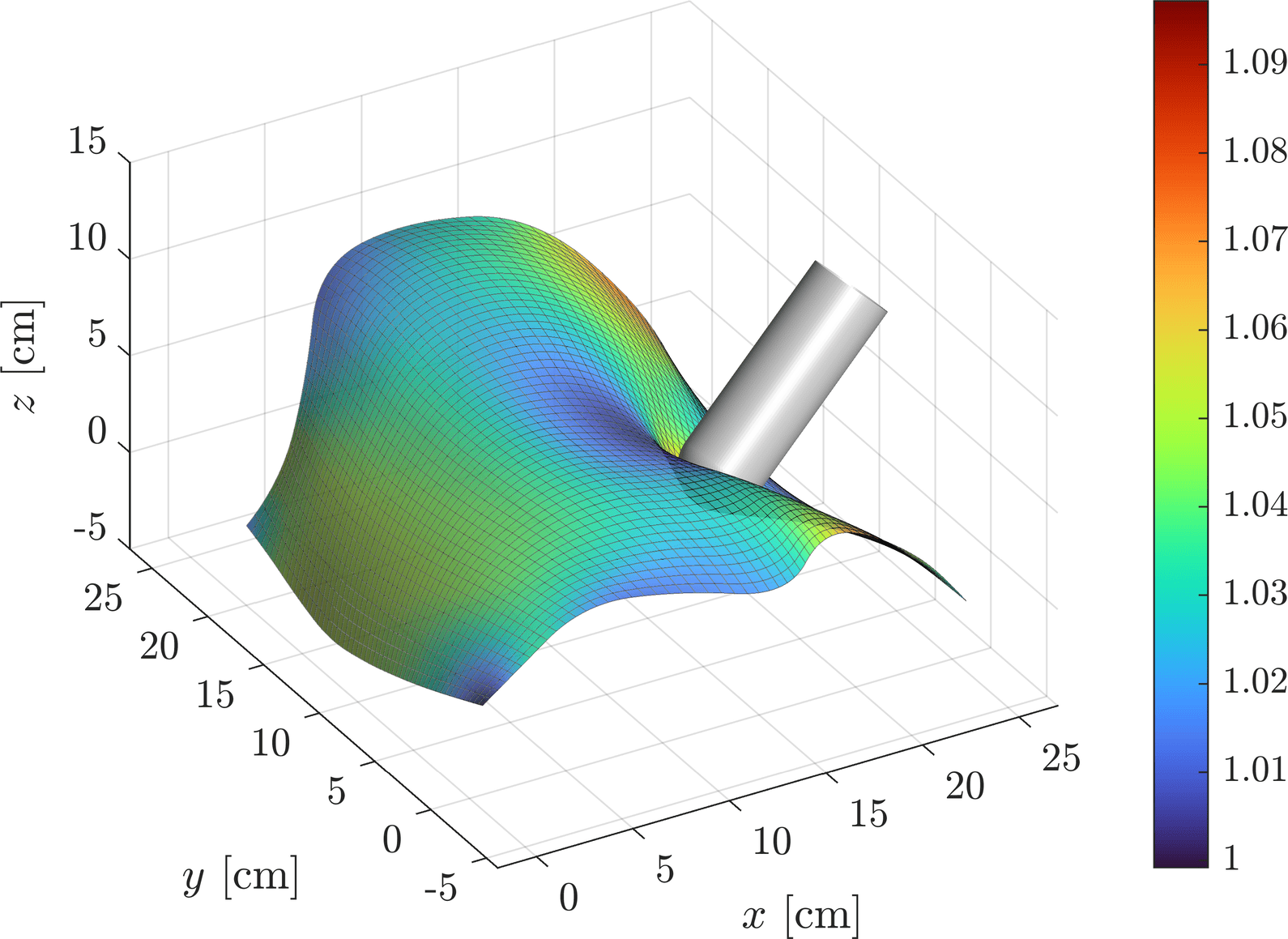} 
	\put(-49mm,0mm){\small{f.}}
	\vspace{3mm}
	\includegraphics[height=36mm]{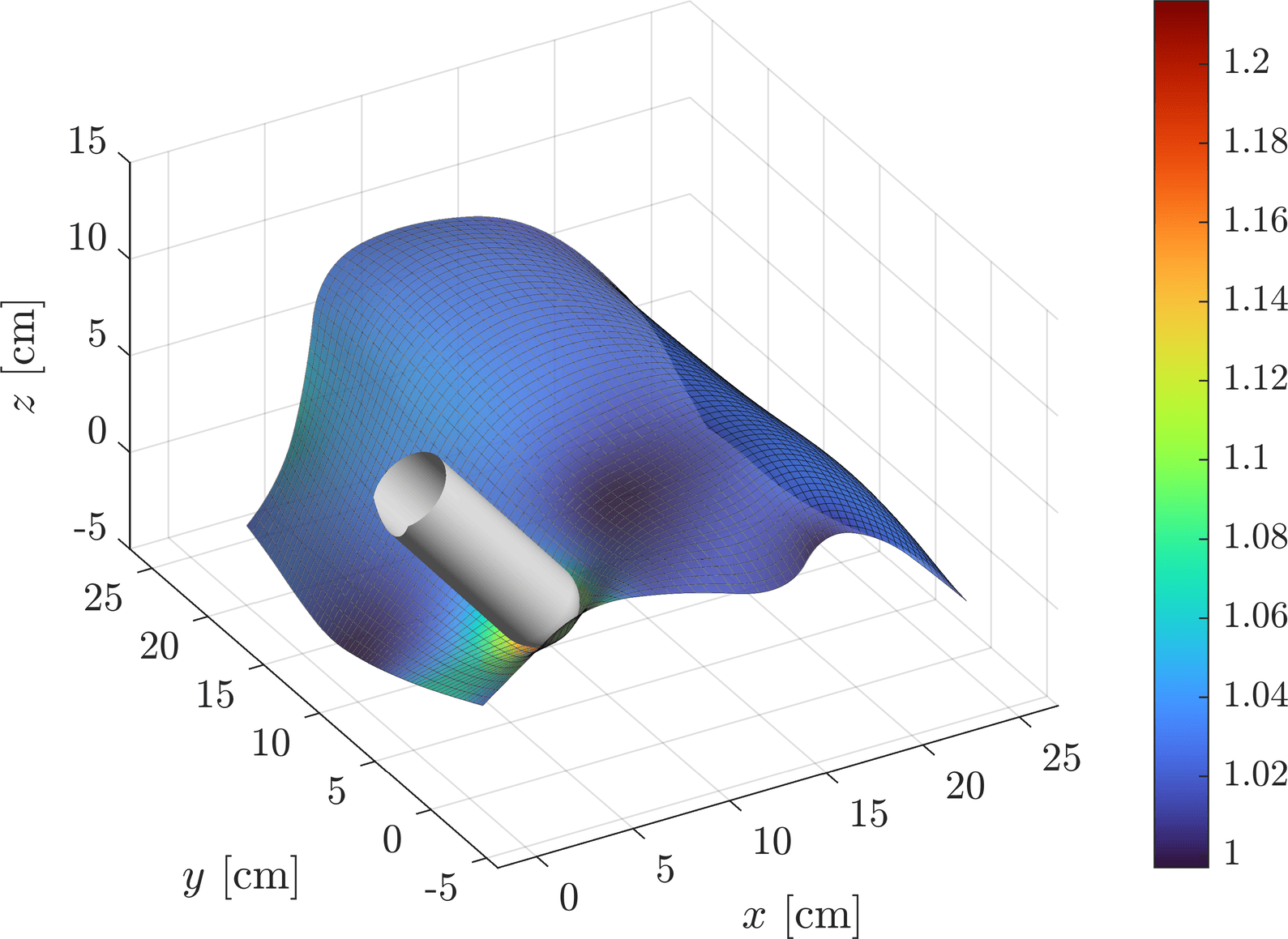}
	\put(-49mm,0mm){\small{g.}}
	\hfil 
	\includegraphics[height=36mm]{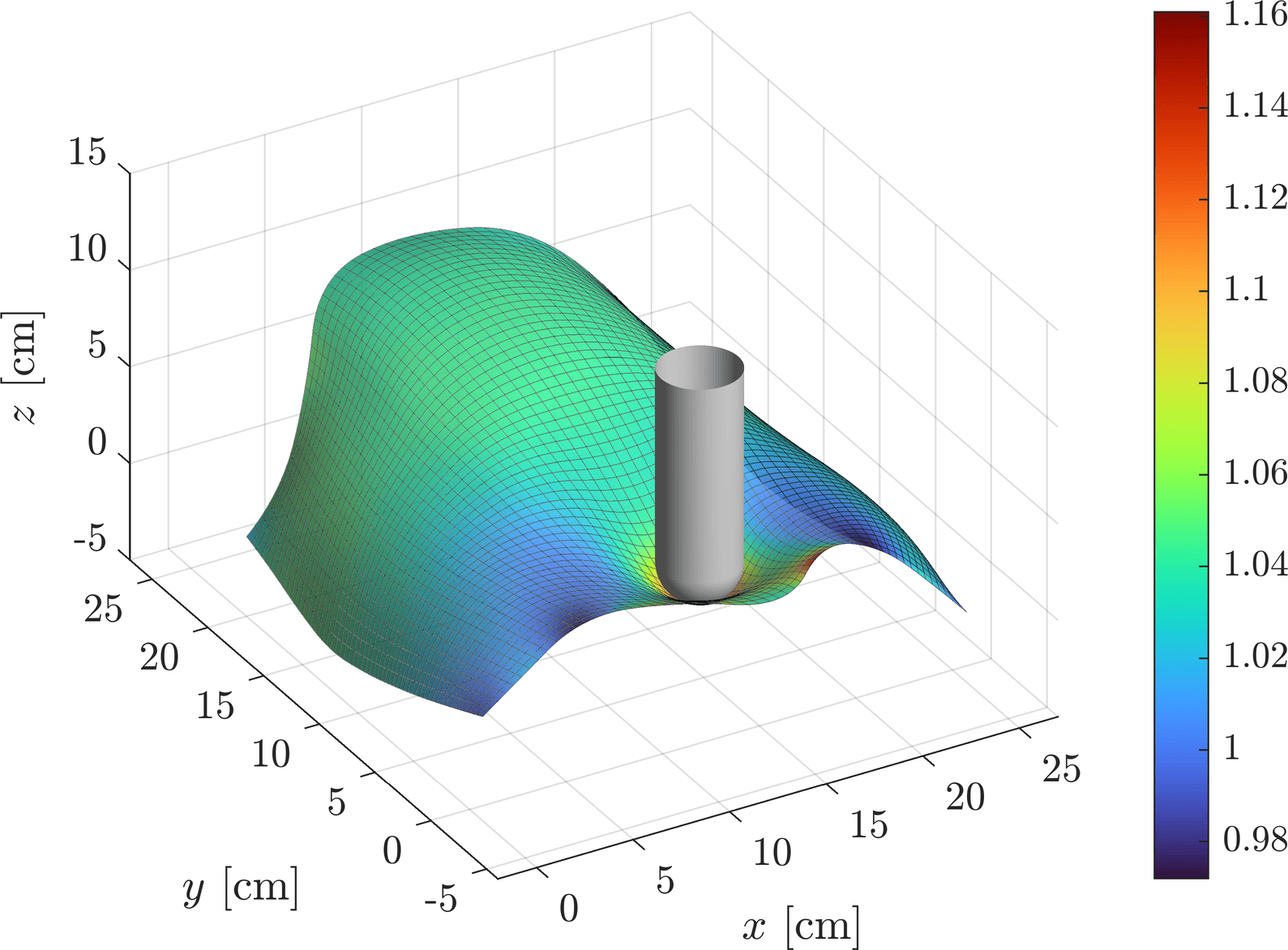}
	\put(-49mm,0mm){\small{h.}}
	\hfil 
	\includegraphics[height=36mm]{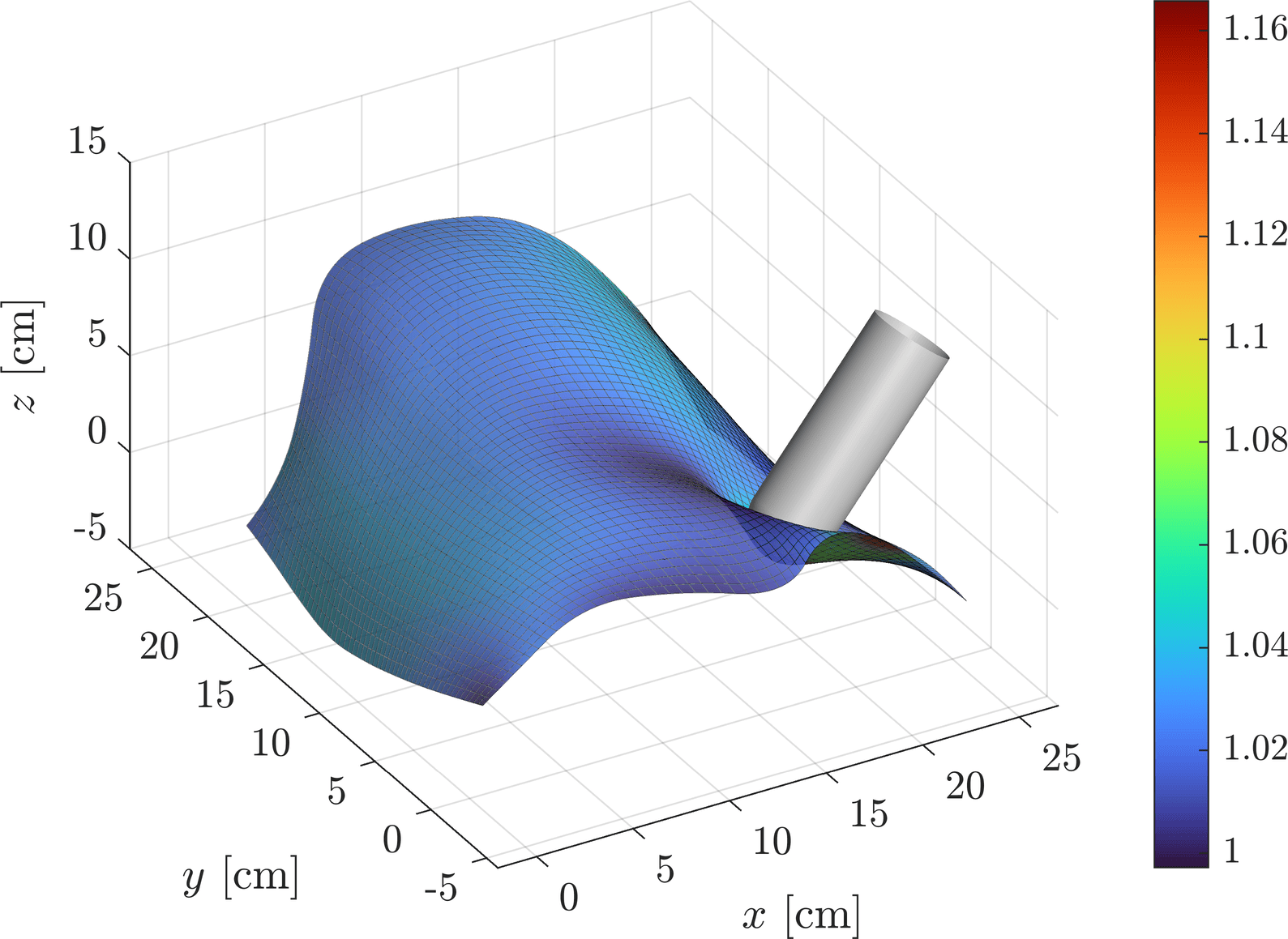} 
	\put(-49mm,0mm){\small{i.}}
	\caption{Abdominal wall: Deformed geometry under the nine contact indentation cases corresponding to Fig.~\ref{fig:abdo1}c. For cases b,~e,~and~h, the probe is pushed \qty{5}{cm} into the abdominal wall, whereas in the remaining cases it is pushed \qty{4}{cm}. Colors represent the surface stretch $J$.}
	\label{fig:abdo2}
\end{figure}

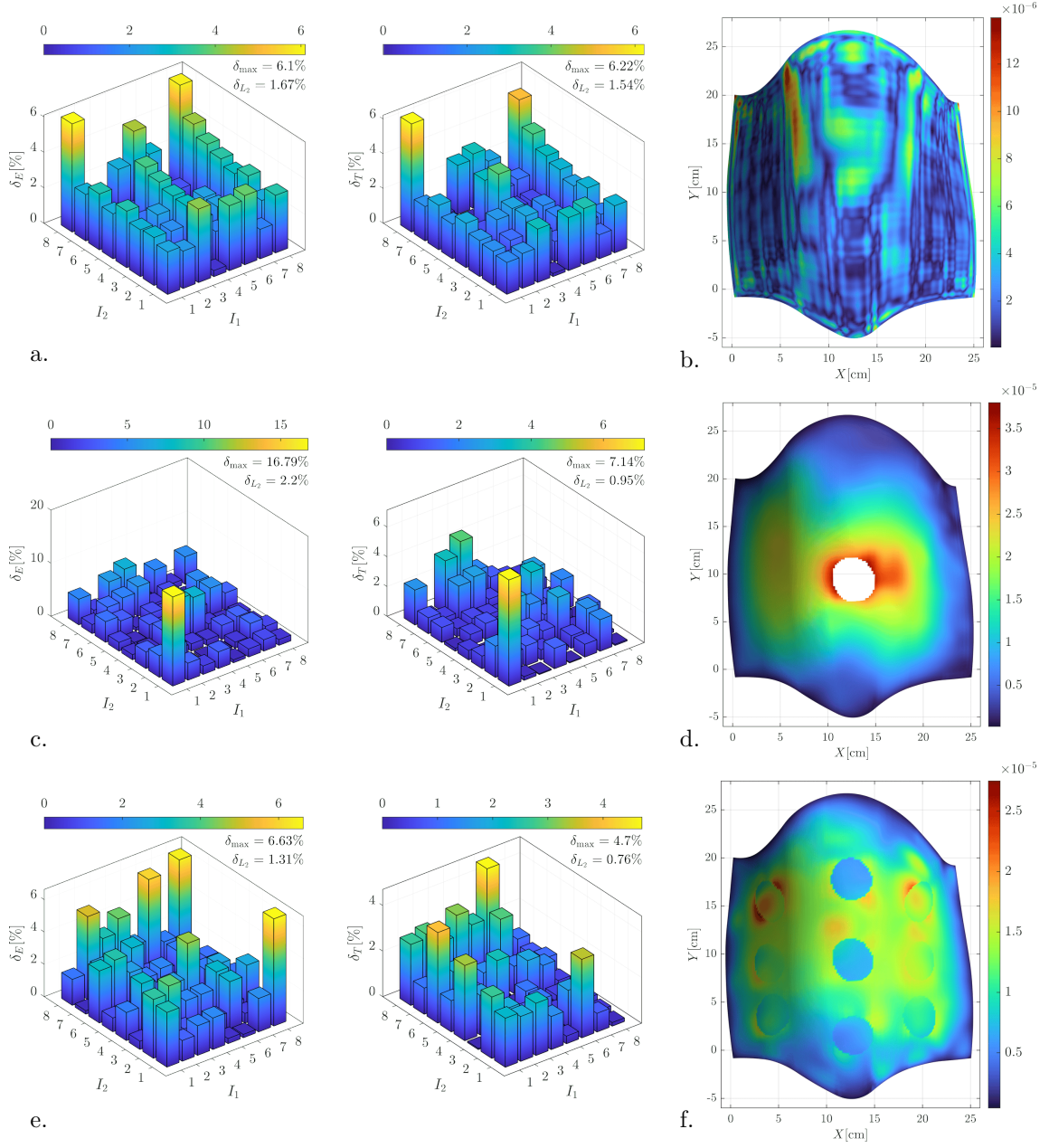
\begin{figure}[htbp] 
	\centering
	\includegraphics[width=150mm]{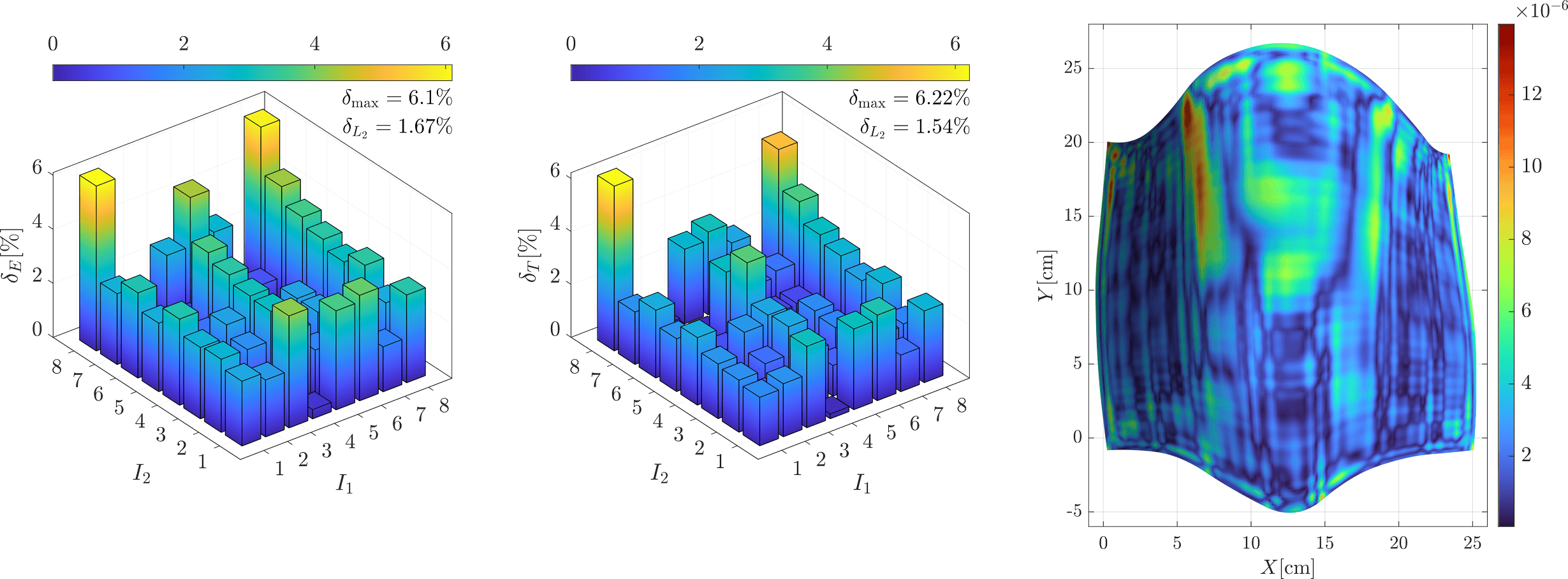}
	\put(-147mm,3mm){\small{a.}} 
	\put(-52mm,2mm){\small{b.}}
	\hfill
	\includegraphics[width=150mm]{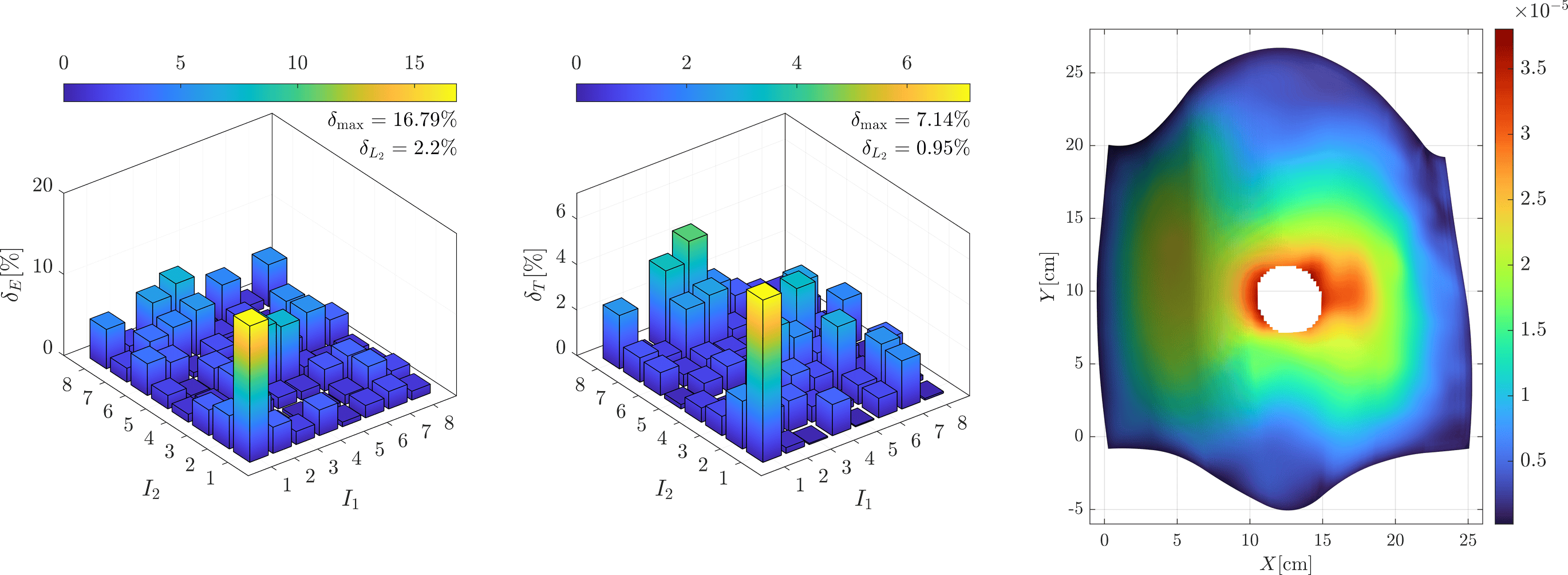}
	\put(-147mm,2mm){\small{c.}}
	\put(-52mm,2mm){\small{d.}}  
	\hfill
	\includegraphics[width=150mm]{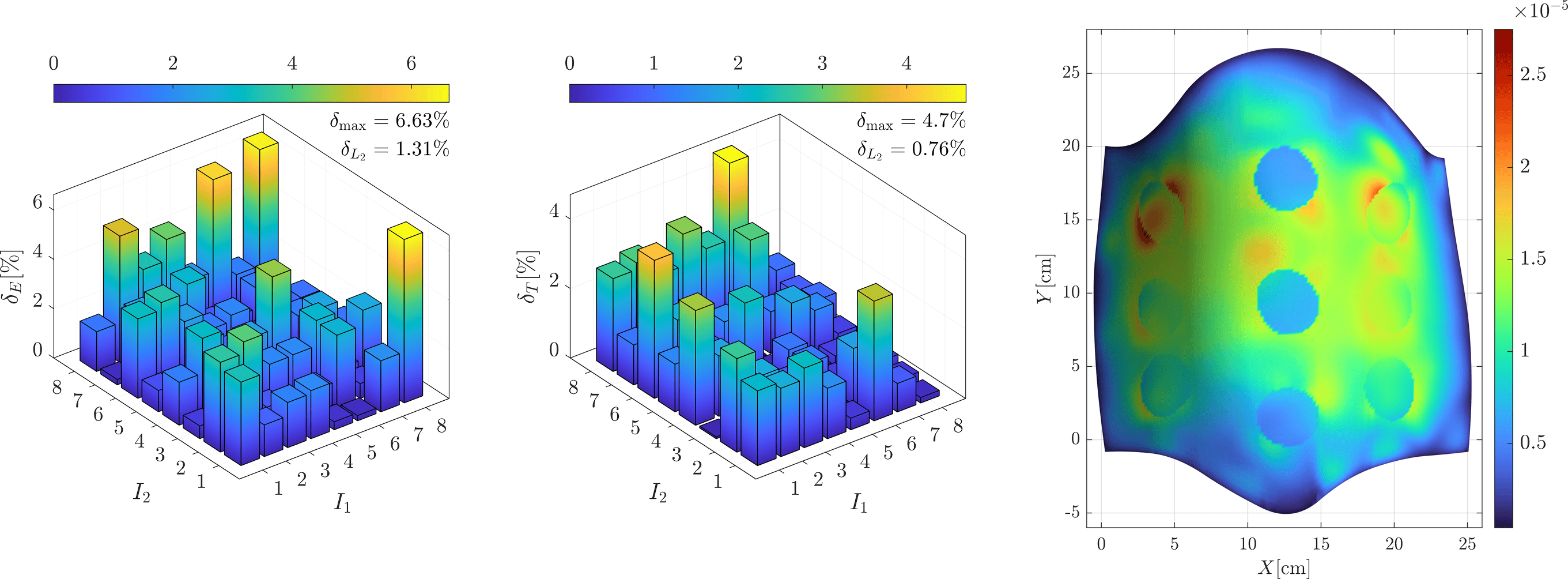}
	\put(-147mm,2mm){\small{e.}}
	\put(-52mm,2mm){\small{f.}} 
	\caption{Abdominal wall: A comparison of the identification results for Case 2.p1 (a.~and~b.), Case 2.1e (c.~and~d.), and Case 2.1all (e.~and~f.): a.,~c.,~and~e.,~ show the nodal identification errors for $E$ and $T$; b.,~d.,~and~f.,~show maps of the objective function residuals. The plotted residuals are root mean square (RMS) values of $\bar{\mU}_\mrR$ from all load cases mapped on the experimental grid. The characteristic holes visible in d.~and~f. correspond to the contact zones excluded from the experimental data. Note that the vertical axes have different scales across the bar plots.} 
	\label{fig:abdoPresContCompare}
\end{figure}

Fig.~\ref{fig:abdo1}b shows the nine chosen contact probe positions in the abdominal top view. The associated nine deformed configurations are presented in Fig.~\ref{fig:abdo2}. For Cases b,~e,~and~h, four load steps are generated, corresponding to \SI{2}{cm}, \SI{3}{cm}, \SI{4}{cm}, and \SI{5}{cm} of indentation. The remaining cases involve three load steps with \SI{2}{cm}, \SI{3}{cm}, and \SI{4}{cm} of indentation. The synthetic experimental data is generated using a model with $168\times168$ FE. A convergence study is conducted in Fig.~\ref{fig:abdo1}d to ensure comparable FE error levels for both PBIA and CBIA. Consequently, PBIA and CBIA employ $28\times28$ and $56\times56$ FE meshes, respectively. Note that the pressure and contact load cases exhibit similar convergence rates, but the contact load cases show higher initial FE error. The experimental data generation uses the contact penalty parameter $\epsilon_n = 10^{4}\,F/L$, while for the inverse analysis $\epsilon_n = 10^{3}\,F/L$. To mimic optical surface measurements, regions of the abdominal wall that are in contact with the probe are excluded from the data, as shown, for example, in Fig.~\ref{fig:abdoPresContCompare}d. 

\begin{figure}[htb] 
	\centering
	\includegraphics[height=55mm]{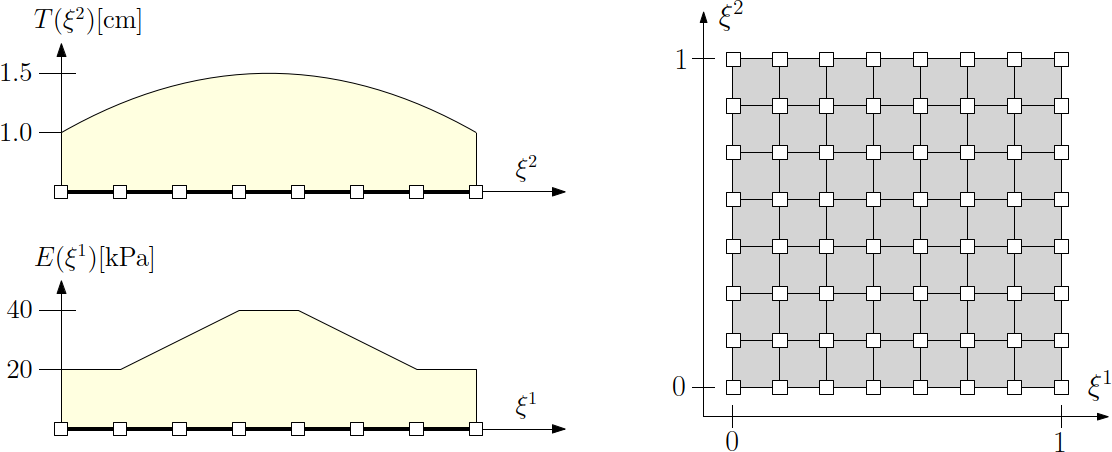}
	\put(-140mm,0mm){\small{a.}}
	\put(-60mm,0mm){\small{b.}} 
	\caption{Abdominal wall: a.~reference distributions for Young's modulus $E$ and thickness $T$; b.~material mesh. Adapted from~\citep{Borzeszkowski2022}.} 
	\label{fig:abdo_material} 
\end{figure}

The given material distributions, shown in Fig.~\ref{fig:abdo_material}, are defined on the parametric domain $\bxi$ and mapped onto the surface $\sS_0$. For the constitution, the Koiter shell model~\eqref{e:WKoit} is selected. The shell is assumed to be incompressible; thus, $\nu = 0.5$, and the Poisson ratio is not identified. For shell formulations directly derived on the mid-surface, $\nu = 0.5$ does not introduce any numerical difficulties. The reference distribution of Young's modulus is constant along $\xi^2 \mapsto Y$ and changes along $\xi^1 \mapsto X$ according to
\eqb{lll}
	E_\mrrf(\xi^1) = \left\{ \begin{array}{rcl}
		E_1 & \mbox{for} & \xi^1 \leq \frac{1}{7}\,  \vee \xi^1 \geq \frac{6}{7}\vspace{0.3em}\,,\\
		E_1+\frac{1}{2}(E_2-E_1) \cdot(7\xi^1-1) & \mbox{for} & \frac{1}{7} < \xi^1 < \frac{3}{7}\vspace{0.3em}\,,\\
		E_2 & \mbox{for} & \frac{3}{7} \leq \xi^1 \leq\frac{4}{7}\vspace{0.3em}\,,\\
		E_2-\frac{1}{2}(E_2-E_1)\cdot(7\xi^1-4) & \mbox{for} & \frac{4}{7}<\xi^1<\frac{6}{7}\,,\\
	\end{array}\right. \quad \xi^1 \in [0,1],
\label{e:abdoEref}\eqe
while the thickness distribution is constant along $\xi^1 \mapsto X$ and varies along $\xi^2 \mapsto Y$ as
\eqb{l}
	T_\mrrf(\xi^2) = T_2  -  (T_2 - T_1) (2\,\xi^2 -1)^2\,, \quad \xi^2 \in [0,1],
\label{e:abdoTref}\eqe
where $E_1=\SI{20}{\kilo\pascal}$, $E_2=\SI{40}{\kilo\pascal}$, $T_1 = \SI{1}{cm}$, and $T_2 = \SI{1.5}{cm}$, see Fig.~\ref{fig:abdo_material}a. The majority of the following cases use a $7\times7$ material mesh (see Fig.~\ref{fig:abdo_material}b) consisting of bilinear Lagrange elements, resulting in $n_\mrvr = 128$ discrete unknowns. Note that the material mesh is defined in $\boldsymbol{\xi}$ domain, meaning that the corresponding physical material elements are non-rectangular. 

The rigid probe, whose displacement is prescribed, induces contact pressure on the abdominal wall. As discussed in Sec.~\ref{s:Inverse}, the use of full Dirichlet boundary conditions requires the force data to be supplied in an alternative manner. Accordingly, the resultant contact force acting on the probe enters into the objective function. Its components are chosen to be unaffected by noise.

The selected weights in Eq.~\eqref{e:objfnc} are $w_\mru = 1$ and $w_\mrc = 0.001$. Additional regularization is omitted. All cases utilize the same $167\times167$ uniform experimental grid. The box constraints on the material parameters are $E_I \in [\SI{10}{\kilo\pascal},\SI{50}{\kilo\pascal}]$ and $T_I \in [\SI{0.5}{cm},\SI{2.0}{cm}]$, and the constant initial guess $E_0 = \SI{30}{\kilo\pascal}$, $T_0 = \SI{1.25}{cm}$ is used for the cases without noise. Otherwise, it is picked randomly within the box constraints.
\begin{table}[htb]
	\raggedright
	\begin{tabular}{@{}lllllllll@{}}
		\toprule
		Case    & FE                & mat.                    & exp.                               & load               & noise 		& $q(\bX)$        & $\errM / \Delta\errM$ 	  & $\errLe / \Delta\errLe$              \\
		& $n_\mrel$         & $\bar n_\mrel$          & $n_\mrex/n_\mrll$                  & $n_\mrll$                  & {[}$L${]}  	&                 & {[}\%{]}             & {[}\%{]}              \\ \midrule
		2.p1    & $28\times28$      & $7\times7$              & $167^2$                            & 4                  & 0     		& $E$             & $6.10$               & $1.67$                \\
		&                   &                         &                                    &                            &       		& $T$             & $6.22$               & $1.54$                \\ \midrule
		2.p2    & $28\times28$      & $7\times7$              & $167^2$                            & 4                  & 0.02     		& $E$             & $+\,5.46\pm3.14$    & $+\,0.48\pm0.26$     \\
		&                   &                         &                                    &                            &       		& $T$             & $+\,3.62\pm2.29$     & $+\,0.37\pm0.19$     \\ \midrule
		2.p3    & $28\times28$      & $7\times7$              & $167^2$                            & 4                  & 0.04     		& $E$             & $+\,15.15\pm6.21$    & $+\,1.57\pm0.41$     \\
		&                   &                         &                                    &                            &       		& $T$             & $+\,11.48\pm4.80$    & $+\,1.24\pm0.33$     \\ \midrule
		2.p4    & $28\times28$      & $7\times7$              & $167^2$                            & 4                  & 0.08     		& $E$             & $+\,32.91\pm11.02$   & $+\,4.15\pm0.83$     \\
		&                   &                         &                                    &                            &       		& $T$             & $+\,25.85\pm7.90$    & $+\,3.26\pm0.56$     \\ \bottomrule
	\end{tabular}
	\caption{Abdominal wall: Results of identifying $E(\xi^1,\xi^2)$ and $T(\xi^1,\xi^2)$ from PBIA, including: FE and material meshes, experimental grid resolution, no. of load levels, \textit{std}.~of noise, and identification errors. For Case 2.p1, the total identification errors $\errM$ and $\errLe$ are reported, while the cases with noise show only the corresponding identification error increases $\Delta\errM$ and $\Delta\errLe$ w.r.t.~Case 2.p1. All cases utilize the material mesh from Fig.~\ref{fig:abdo_material}b and four load cases corresponding to $25\%$, $50\%$, $75\%$, and $100\%$ of $p_\mathrm{intra}$ shown in Fig.~\ref{fig:abdo1}c. Cases 2.p2--2.p4 are repeated 99 times for statistical analysis.} 
	\label{tab:abdo_pres}
\end{table}

Tab.~\ref{tab:abdo_cont_no_noise1} presents the identification results of all nine CBIA cases performed separately, denoted as Cases 2.1a--i, without noise. A comparison of Tabs.~\ref{tab:abdo_pres} and~\ref{tab:abdo_cont_no_noise1} shows that PBIA and CBIA yield comparable identification $L_2$-errors $\errLe$. This is achieved by selecting a finer FE mesh for CBIA, as motivated by the FE convergence study in Fig.~\ref{fig:abdo1}d and discussed above. Nevertheless, the maximum identification errors, $\errME$, obtained with CBIA are up to three times larger than those with PBIA, whereas for $\errMT$, both increases and decreases are observed. The last column of Tab.~\ref{tab:abdo_cont_no_noise1} contains the results for all indentation cases simultaneously supplied to $f(\mq)$, referred to as Case 2.1all. This increases the number of load steps available from 4 (or 3) to 30. While $\errM$ is almost always reduced for both $E$ and $T$, $\errLe$ is a coarse average of the values for Cases 2.1a--i, particularly for $E$. For further analysis, Cases 2.1e and 2.1all are studied in more detail. Nevertheless, similar conclusions can be drawn from analyzing the remaining cases separately.  
\begin{table}[htbp]
	\raggedright
	\begin{tabular}{@{}llllllllllll@{}}
		\cmidrule(l){3-12}
		&       	& \multicolumn{10}{l}{Case from Fig.~\ref{fig:abdo1}c}                   		\\ \midrule
		Identification  			& $q(\bX)$ 	& a   	& b   	& c   	& d   	& e   	& f   	& g		& h		& i     & all	\\ 
		error {[}\%{]}				&			&		&		&		&		&		&		&		&		&	    		\\ \midrule
		\multirow{2}{*}{$\errM$} 	& $E$     	& 13.10	& 9.56	& 19.76	& 9.51	& 16.79	& 10.26	& 7.54	& 7.36	& 17.72 & 6.63	\\
		& $T$     	                            & 12.82	& 9.02	& 7.88	& 8.50	& 7.14	& 5.46	& 8.39	& 3.55	& 6.37  & 4.70  \\ \midrule
		\multirow{2}{*}{$\errLe$}  	& $E$     	& 1.36	& 1.88	& 1.87	& 1.01	& 2.20	& 1.23	& 0.81	& 0.66	& 1.29  & 1.31  \\
		& $T$     	                            & 1.27	& 1.08	& 1.15	& 1.11	& 0.95	& 0.92	& 0.74	& 0.42	& 0.83  & 0.76	\\ \bottomrule
	\end{tabular}
	\caption{Abdominal wall: Results of identifying $E(\xi^1,\xi^2)$ and $T(\xi^1,\xi^2)$ from CBIA without noise. All cases utilize $56\times56$ FE mesh, $7\times7$ material mesh, and $167\times167$ experimental grid. For all cases, the optimal solution is found in 12--15 iterations.}
	\label{tab:abdo_cont_no_noise1}
\end{table}

Figs.~\ref{fig:abdoPresContCompare}a,~c,~and~e compare the nodal identification errors $\delta_I$ for PBIA (Case 2.p1), single contact load case (Case 2.1e), and a set of contact load cases (Case 2.1all). As seen, $\errM$ dominates in the case of CBIA. Fig.~\ref{fig:abdoPresContCompare}c shows that the high $\errM$ values reported in Tab.~\ref{tab:abdo_cont_no_noise1} for Case 2.1e are caused by error peaks at the corners and edges of the material mesh. For Case 2.p1, the distribution of $\delta_I$ is more regular. Corresponding residual maps are presented in Figs.~\ref{fig:abdoPresContCompare}b,~d,~and~f. The residual maps for Case 2.p1 reveal patterns corresponding to FE and ME meshes, and their values are generally smaller than those of CBIA. For both CBIA cases, the residuals are higher near the contact areas, illustrating the concentration of FE errors there.

\begin{figure}[H] 
	\centering
	\includegraphics[width=157mm]{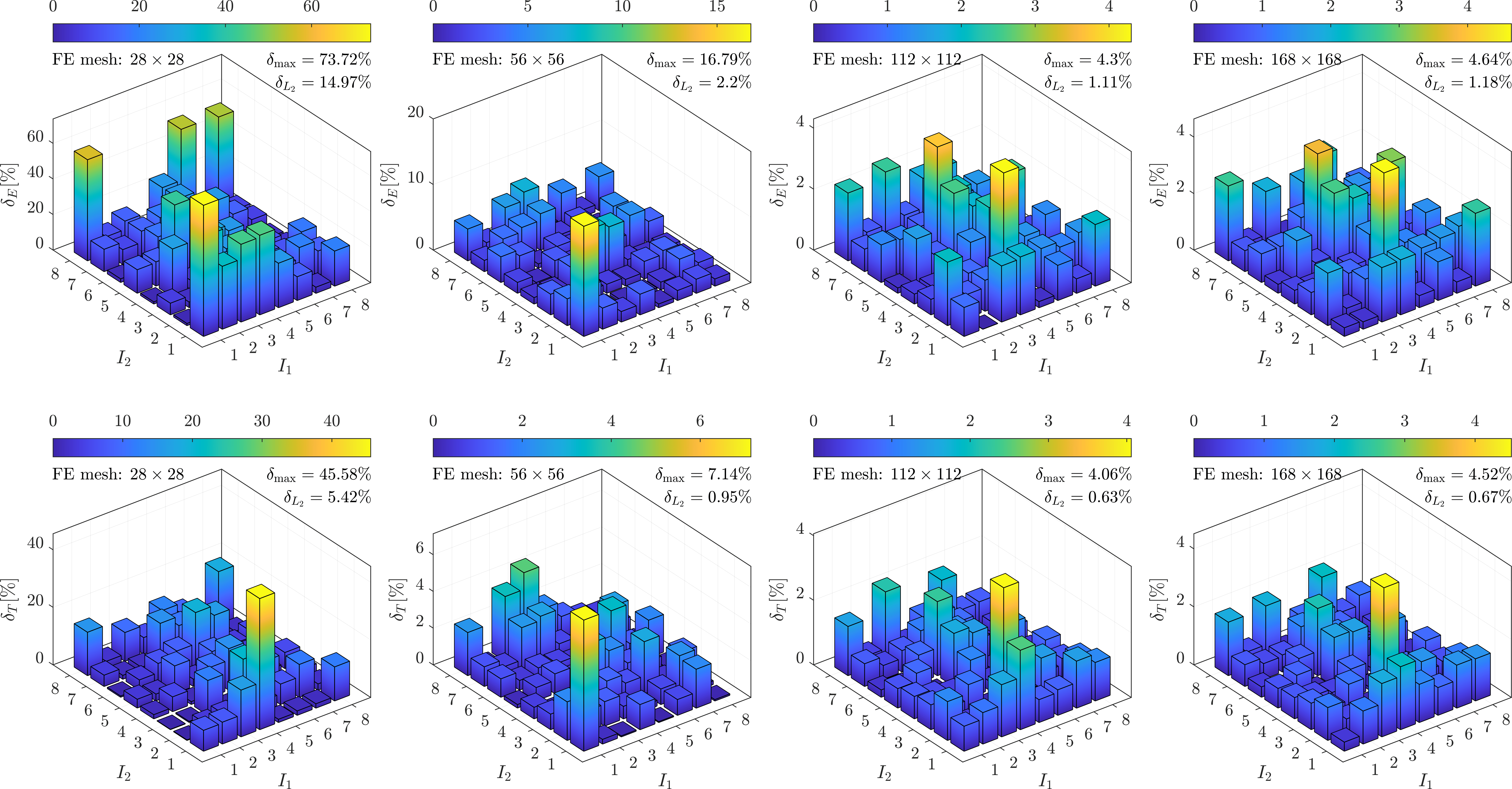}
	\put(-157mm,0mm){\small{a.}}
	\put(-157mm,44mm){\small{b.}} 
	\caption{Abdominal wall: Results of CBIA for the probe in position e from Fig.~\ref{fig:abdo1}b. The convergence of the nodal identification error $\delta_I$ for Young's modulus (a.) and thickness (b.) with a uniform FE mesh refinement. The cases shown here extend Case 2.1e from Tab.~\ref{tab:abdo_cont_no_noise1}, but do not appear explicitly there. Note that the vertical axes have different scales across the bar plots.} 
	\label{fig:abdoFEmeshConv}
\end{figure}
\begin{figure}[H] 
	\centering
	\includegraphics[width=157mm]{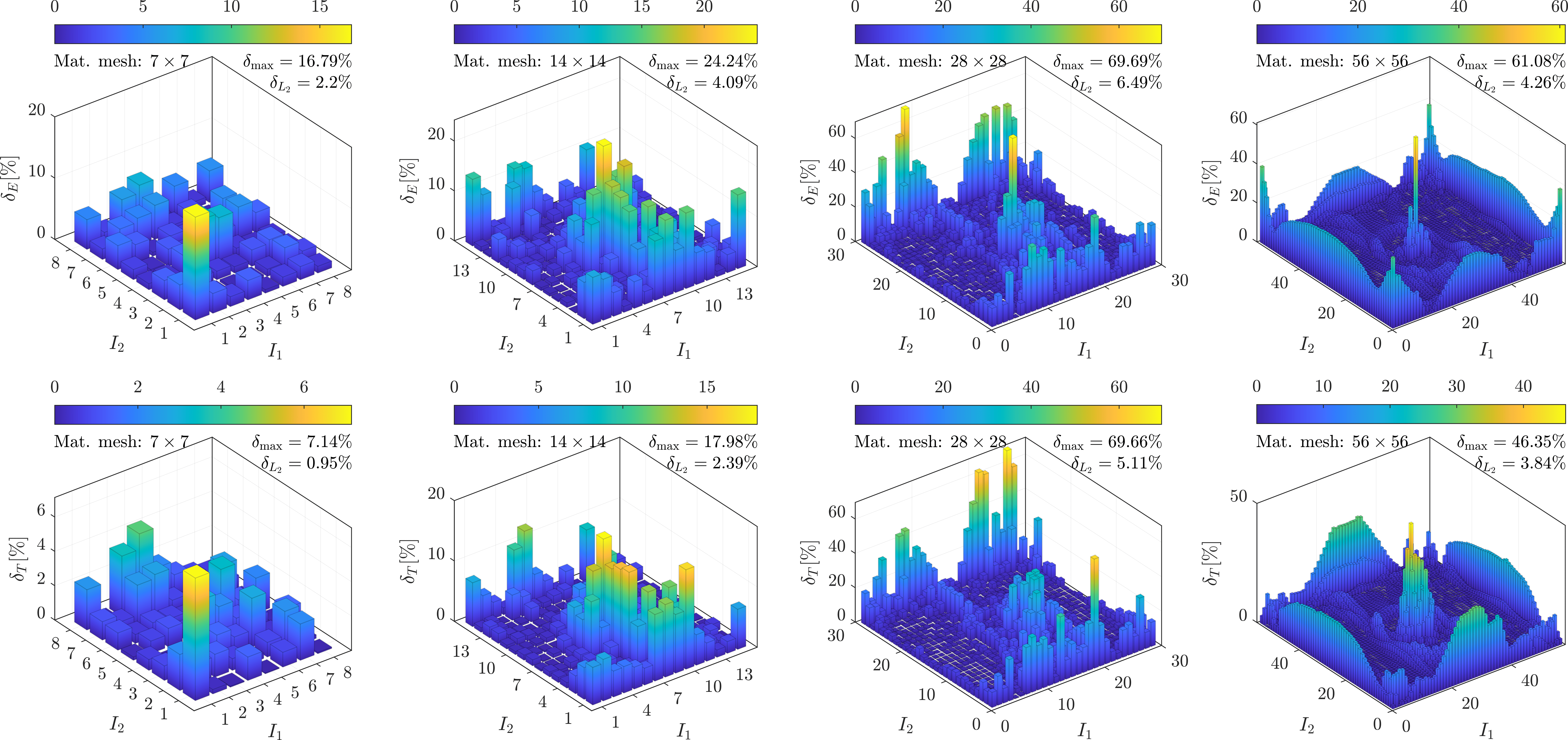}
	\put(-157mm,0mm){\small{a.}}
	\put(-157mm,39.5mm){\small{b.}} 
	\caption{Abdominal wall: Results of CBIA for the probe in position e from Fig.~\ref{fig:abdo1}b. The evolution of the nodal identification error $\delta_I$ for Young's modulus (a.) and thickness (b.) with a uniform material mesh refinement. The cases shown here extend Case 2.1e from Tab.~\ref{tab:abdo_cont_no_noise1}, but do not appear explicitly there. All cases employ $56\times56$ FE mesh. Note that the vertical axes have different scales across the bar plots.} 
	\label{fig:abdoMEmeshConv}
\end{figure}
The convergence of the inverse solution for Case 2.1e w.r.t.~the FE mesh refinement is shown in Fig.~\ref{fig:abdoFEmeshConv}. The maximum nodal identification errors tend to occur at the edges of the material mesh or in the contact area. The results for the last two FE meshes do not differ much. As the FE mesh approaches the one used for data generation ($168\times168$ FE), the last two cases illustrate the combined effects of material approximation and $\epsilon_n$. In turn, Fig.~\ref{fig:abdoMEmeshConv} shows how the inverse solution of Case 2.1e evolves with a refinement of the material mesh. As expected, the identification error increases with the addition of more material nodes, as the model can better fit the inaccurate experimental data (affected only by the systematic error), resulting in larger errors at the edges and in contact areas. 

The observations above raise a question: what measures can be taken to improve CBIA for noise-free data? First, CBIA requires a finer FE mesh than PBIA to achieve comparable accuracy. Since contact induces local deformation, higher FE errors are expected, especially when CBIA employs a uniform FE mesh. Therefore, an adaptive, nonuniform FE mesh would improve CBIA results with minimal computational overhead. Analogously, CBIA is more susceptible to material approximation errors, as the results for $168\times168$ FE mesh in Fig.~\ref{fig:abdoFEmeshConv} show. Thus, a finer ME mesh is also justified. Note that using the same $\epsilon_n$ for both inversion and data generation does not improve the results significantly in this case.

\begin{table}[htbp]
	\raggedright
	\vspace{4mm}	
	\begin{tabular}{@{}lllllllll@{}}
		\toprule
		Case    & FE                & mat.                    & exp.                               & load               	& noise 		& $q(\bX)$        & $\Delta\errM$ 		   & $\Delta\errLe$        \\
		& $n_\mrel$           		& $\bar n_\mrel$          & $n_\mrex/n_\mrll$                  & $n_\mrll$              & {[}$L${]}  	&                 & {[}\%{]}               & {[}\%{]}              \\ \midrule
		2.2all  & $56\times56$      & $7\times7$              & $167^2$                            & $30$    				& 0.02     		& $E$             & $4.62\,\pm\,3.67$      & $0.17\,\pm\,0.20$     \\
		&                   		&                         &                                    &                        &       		& $T$             & $1.28\,\pm\,1.93$      & $0.077\,\pm\,0.074$   \\ \midrule
		2.3all  & $56\times56$      & $7\times7$              & $167^2$                            & $30$    				& 0.04     		& $E$             & $11.19\,\pm\,6.78$     & $0.59\,\pm\,0.35$     \\
		&                  	 		&                         &                                    &                        &       		& $T$             & $4.53\,\pm\,2.65$      & $0.29\,\pm\,0.14$     \\ \midrule
		2.4all  & $56\times56$      & $7\times7$              & $167^2$                            & $30$    				& 0.08     		& $E$             & $25.75\,\pm\,12.87$    & $1.57\,\pm\,0.36$     \\
		&                   		&                         &                                    &                        &       		& $T$             & $11.84\,\pm\,4.63$     & $0.79\,\pm\,0.19$     \\ \bottomrule
	\end{tabular}
	\caption{Abdominal wall: Results of identifying $E(\xi^1,\xi^2)$ and $T(\xi^1,\xi^2)$ from all contact load cases shown in Fig.~\ref{fig:abdo1}c and simultaneously provided to $f(\mq)$. The last two columns show the increases in the identification errors, $\Delta\errM$ and $\Delta\errLe$, w.r.t.~Case 2.1all due to noise in the synthetic experimental data. For all cases, the optimal solution is found in 13--29 iterations (average $15.7$).} 
	\label{tab:abdo_cont_noise}
\end{table}
\begin{figure}[htbp] 
	\centering
	\vspace{4mm}
	\includegraphics[width=142mm]{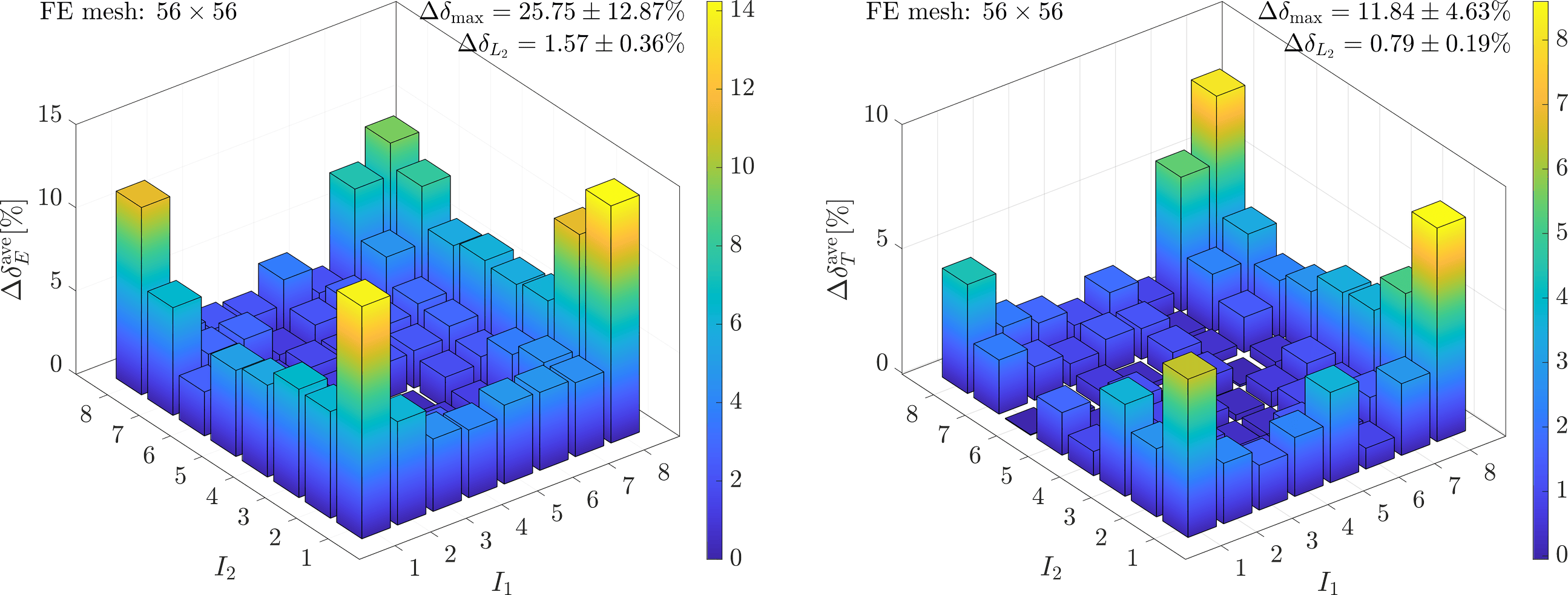}
	\put(-145mm,0mm){\small{a.}}
	\put(-68mm,0mm){\small{b.}} 
	\caption{Abdominal wall: Distributions of the average increase in the nodal identification error, $\Delta\delta_I^\mathrm{ave}$, w.r.t.~Case 2.1all, for Young's modulus (a.) and thickness (b.) in Case 2.4all from Tab.~\ref{tab:abdo_cont_noise}. The corresponding standard deviations of the increases, $\delta_I^{std}$, show a similar distribution, with pronounced errors at the edges and corners. Note that the vertical axes have different scales across the bar plots.} 
	\label{fig:abdo_noise}
\end{figure}
To assess the effect of random noise on the identification, Case 2.1all is repeated 99 times for each of the three noise levels considered for PBIA in Tab.~\ref{tab:abdo_pres}. The results are summarized in Tab.~\ref{tab:abdo_cont_noise}. Owing to an over sevenfold increase in the amount of experimental data, CBIA exhibits remarkably smaller increases in $\errLe$. A reduction in $\Delta\errM$ is also observed for CBIA; however, the maximum identification errors remain high. Further, for CBIA, the shell thickness $T$ is less affected by noise than the Young's modulus $E$. A similar, but much weaker, effect can be observed for the results of PBIA in Tab.~\ref{tab:abdo_pres}. Fig.~\ref{fig:abdo_noise} presents the averaged increases in the nodal identification error w.r.t.~Case 2.1all for Case 2.4all. The largest errors tend to occur at the edges and corners, where the sensitivity of the FE solution to $\mq$ is particularly low. Conversely, the smallest errors occur, on average, in the contact areas and the central zone. Similar observations can be made for Cases 2.1a--i with added noise. These results are not shown here.

\subsection{Block probing}\label{s:iron}

\begin{figure}[htb]
	\begin{center} \unitlength1cm
		\begin{picture}(0,5.5)
			\put(-7.5,0.25){\includegraphics[height=50mm]{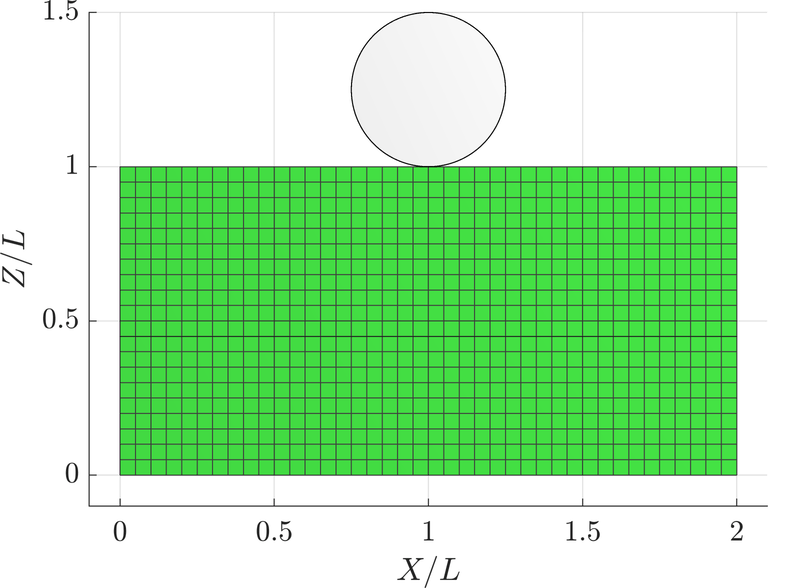}}
			\put(1,0){\includegraphics[height=55mm]{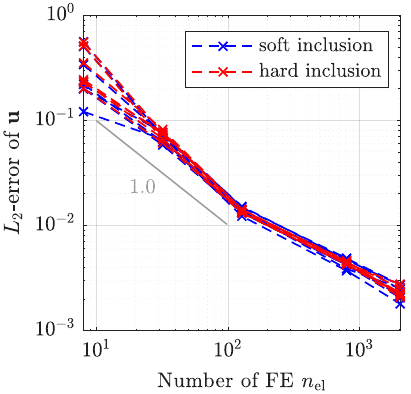}}			
			\put(-7.5,0.25){\small{a.}}
			\put(1,0){\small{b.}}
		\end{picture}
		\caption{Block probing: a.~The undeformed configuration with boundary conditions and an exemplary position of the contact probe; b.~FE convergence of the discrete $L_2$ error for the load cases shown in Figs.~\ref{fig:iron2}~and~\ref{fig:iron3} w.r.t.~the FE solution for 8192 elements. The seven red dashed lines correspond to seven indentation cases.}
		\label{fig:iron1}	
	\end{center}
\end{figure}

\begin{figure}[htb] 
	\centering
	\includegraphics[height=55mm]{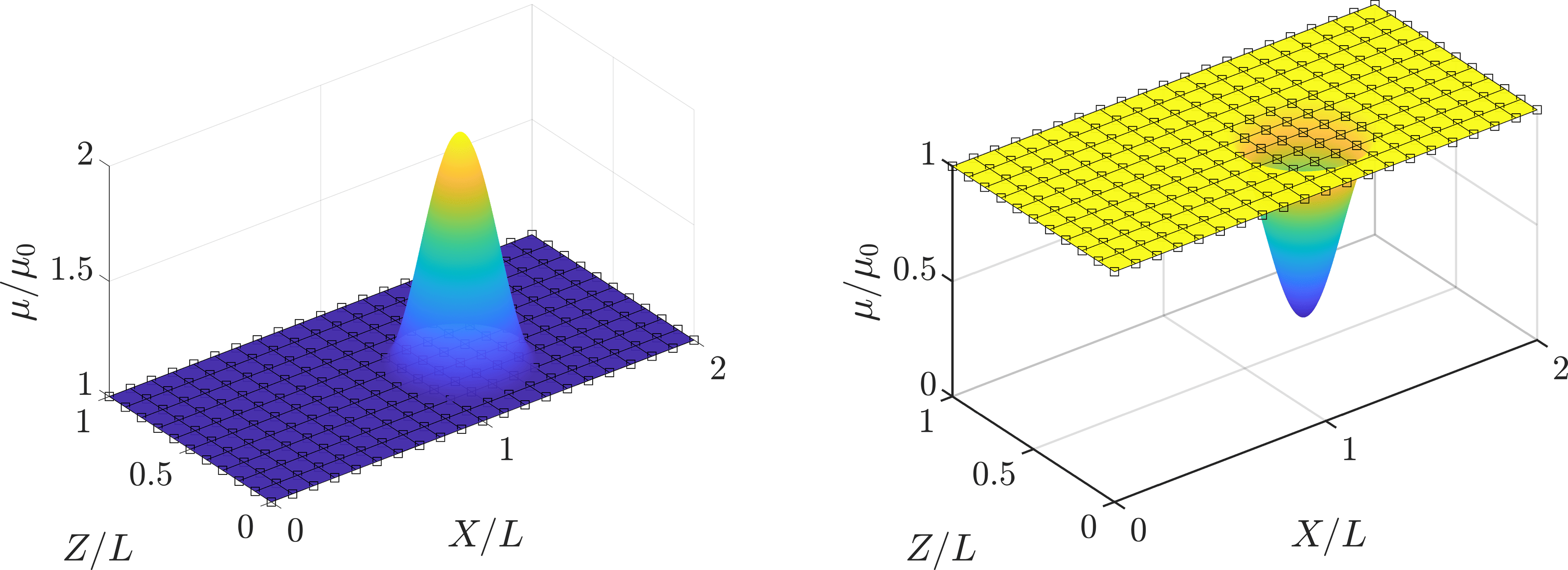}
	\put(-155mm,0mm){\small{a.}}
	\put(-75mm,0mm){\small{b.}} 
	\caption{Block probing: Reference distributions of $\mu$ in the case of hard (a.) and soft (b.) inclusions overlayed with a $20\times10$ uniform bilinear material mesh. Distributions of $\Lambda$ follows analogically.} 
	\label{fig:iron_material} 
\end{figure}

The last example considers a 3D block in plane strain as shown in Fig.~\ref{fig:iron1}a. The block has width $L_x = 2L$ and height $L_z = L$, and is made of compressible Neo-Hookean solid according to Eq.~\eqref{e:sigNH}, with $\Lambda$ and $\mu$ as material parameters.
Consequently, the example couples stretching and shearing deformations. The reference $\mu(X,Y)$ is given by
\eqb{lll}
\mu_\mrrf(X,Y) = \left\{ \begin{array}{lcl}
	\mu_0 & \mbox{for} & R \geq R_0 \vspace{0.3em}\,,\\
	\mu_0 + \ds\frac{\Delta\mu_1}{2}\left(1+\cos\left(\pi\frac{R}{R_0}\right)\right) & \mbox{for} & R \le R_0 \vspace{0.3em}\,,\\
\end{array}\right.
\label{e:ironMat}\eqe
where $R^2 = (X-1.2L)^2 + (Z-0.4L)^2$, and $R_0 = 0.3L$. The reference $\Lambda(X,Y)$ follows analogously. Two values of $\Delta\mu_1$ and $\Delta\Lambda_1$ are considered: $\Delta\mu_1 = \mu_0$ and $\Delta\mu_1 = -0.8\mu_0$ (respectively $\Delta\Lambda_1 = \Lambda_0$ and $\Delta\Lambda_1 = -0.8\Lambda_0$), which correspond to hard and soft inclusions in the material. Both are illustrated in Fig.~\ref{fig:iron_material} together with the selected material mesh. Quantities $L$, $\mu_0$, and $\Lambda_0$ are used for normalization and do not need to be specified.

Displacements of the block are fixed on the bottom ($Z = 0$), and periodic boundary conditions are imposed on the sides, so that $u_i(X = 2L) = u_i(X = 0)$. 
\begin{figure}[htbp] 
	\centering
	\includegraphics[height=32mm]{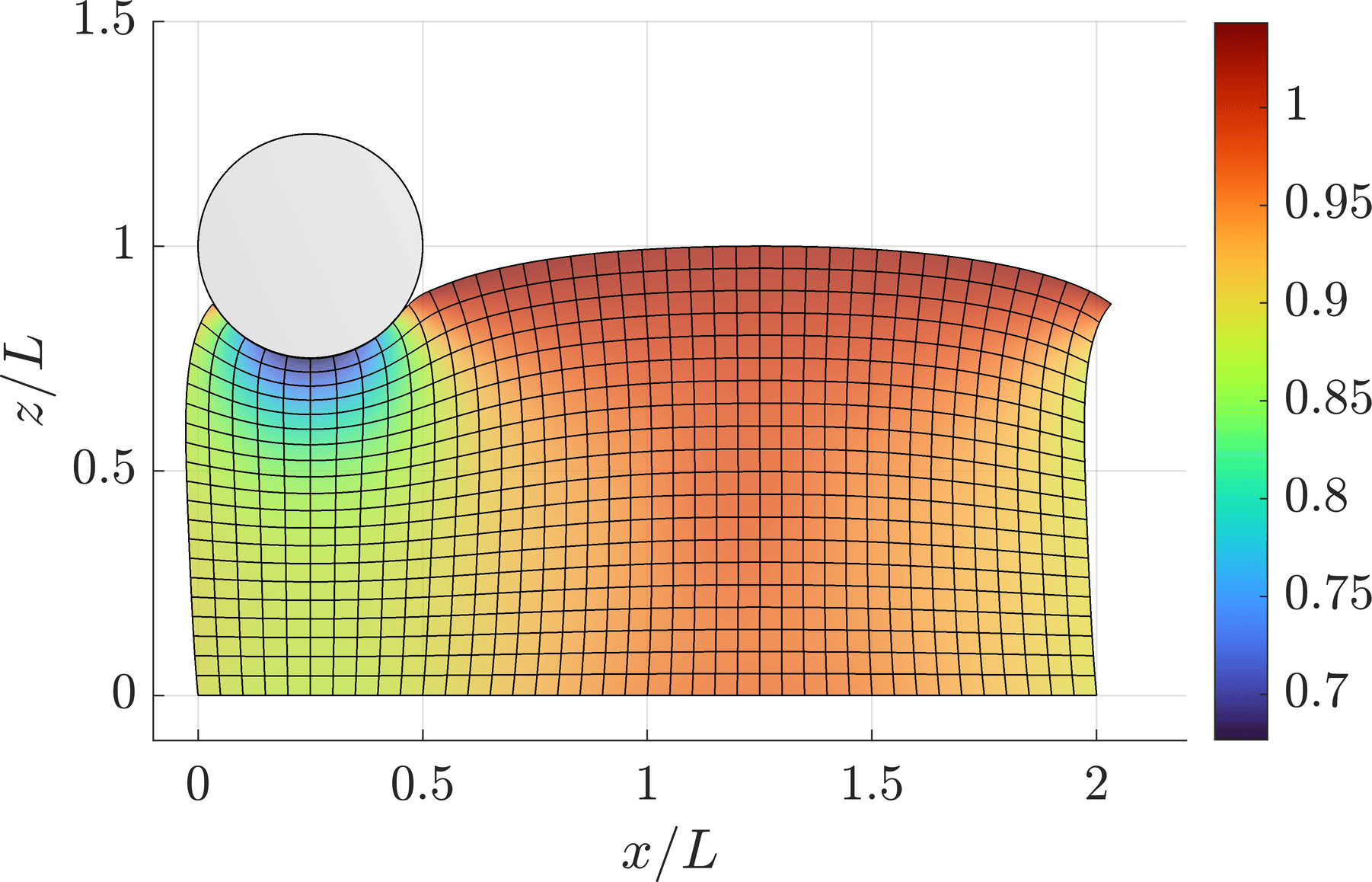}
	\put(-49mm,0mm){\small{a.}}
	\hfil 
	\includegraphics[height=32mm]{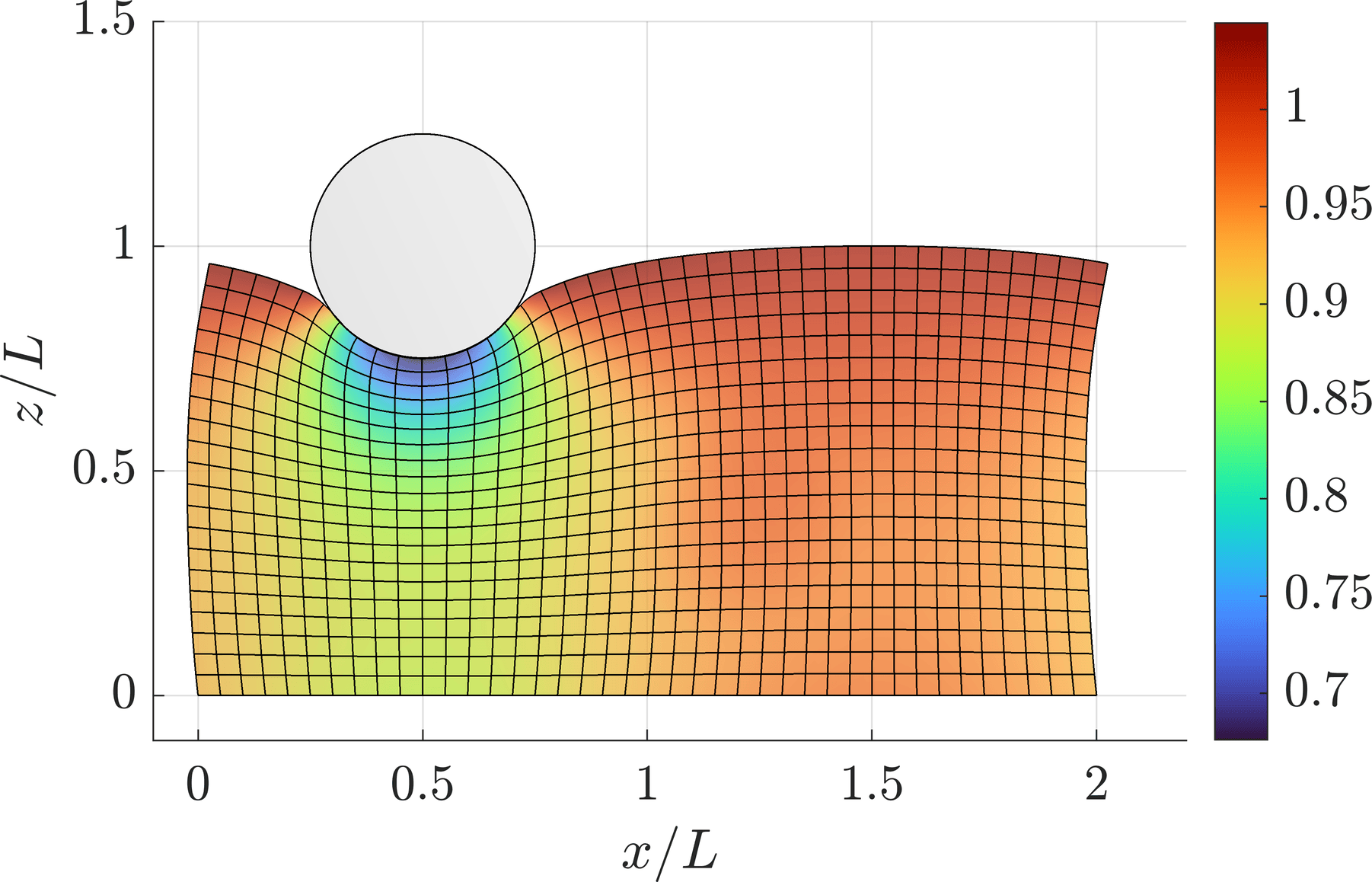}
	\put(-49mm,0mm){\small{b.}}
	\hfil 
	\includegraphics[height=32mm]{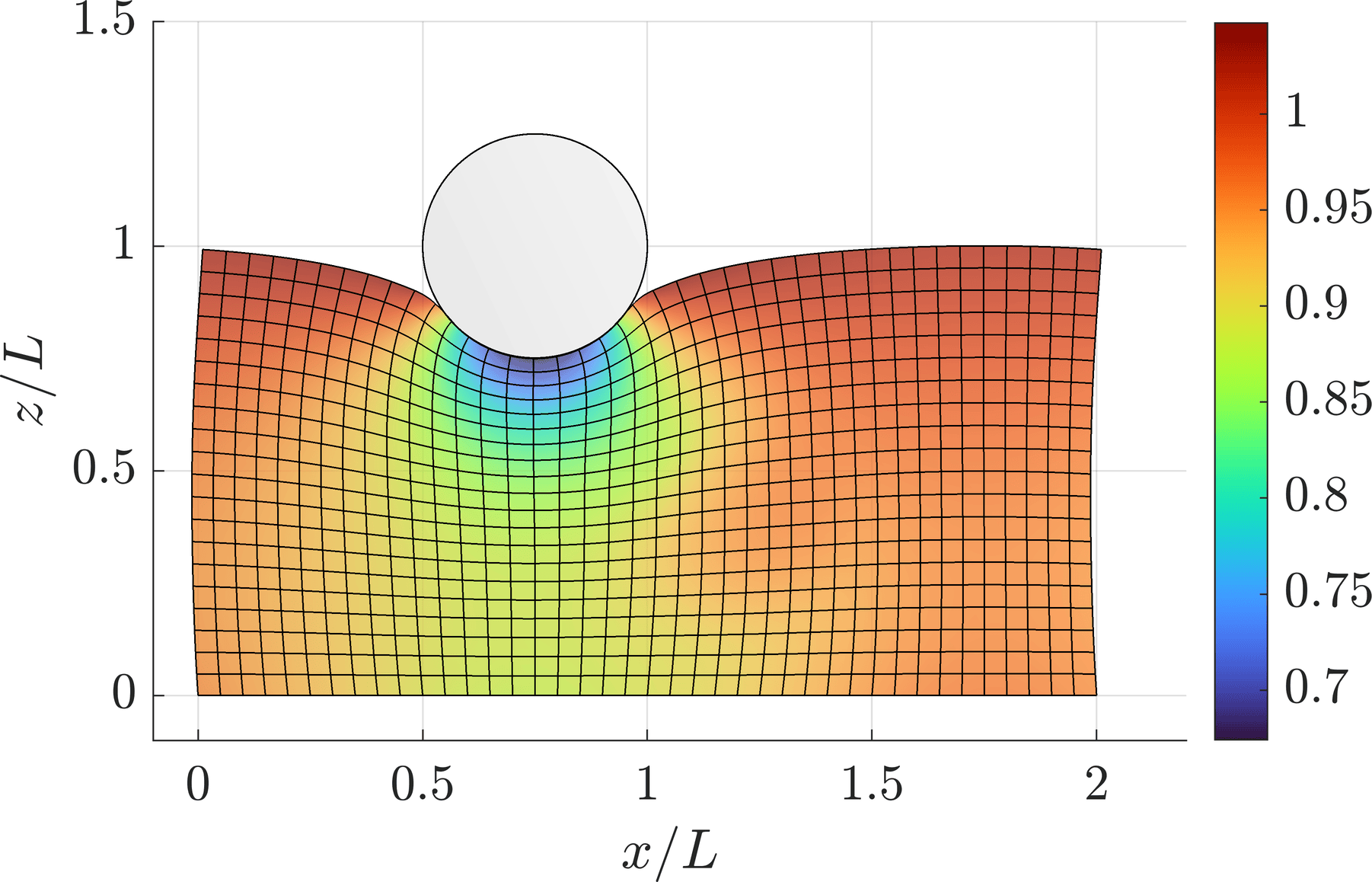} 
	\put(-49mm,0mm){\small{c.}}
	\vspace{3mm}
	\includegraphics[height=32mm]{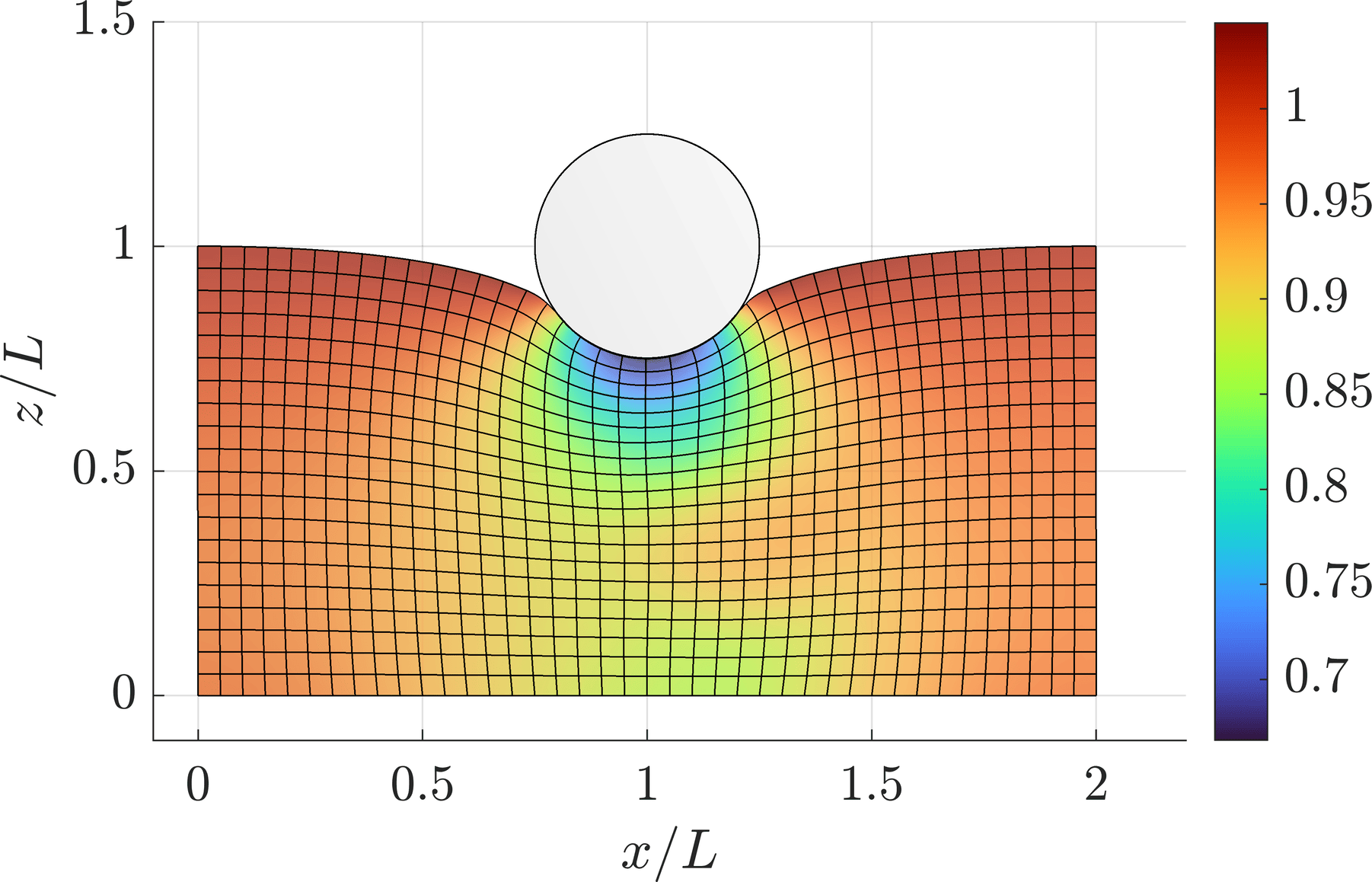}
	\put(-49mm,0mm){\small{d.}}
	\hfil 
	\includegraphics[height=32mm]{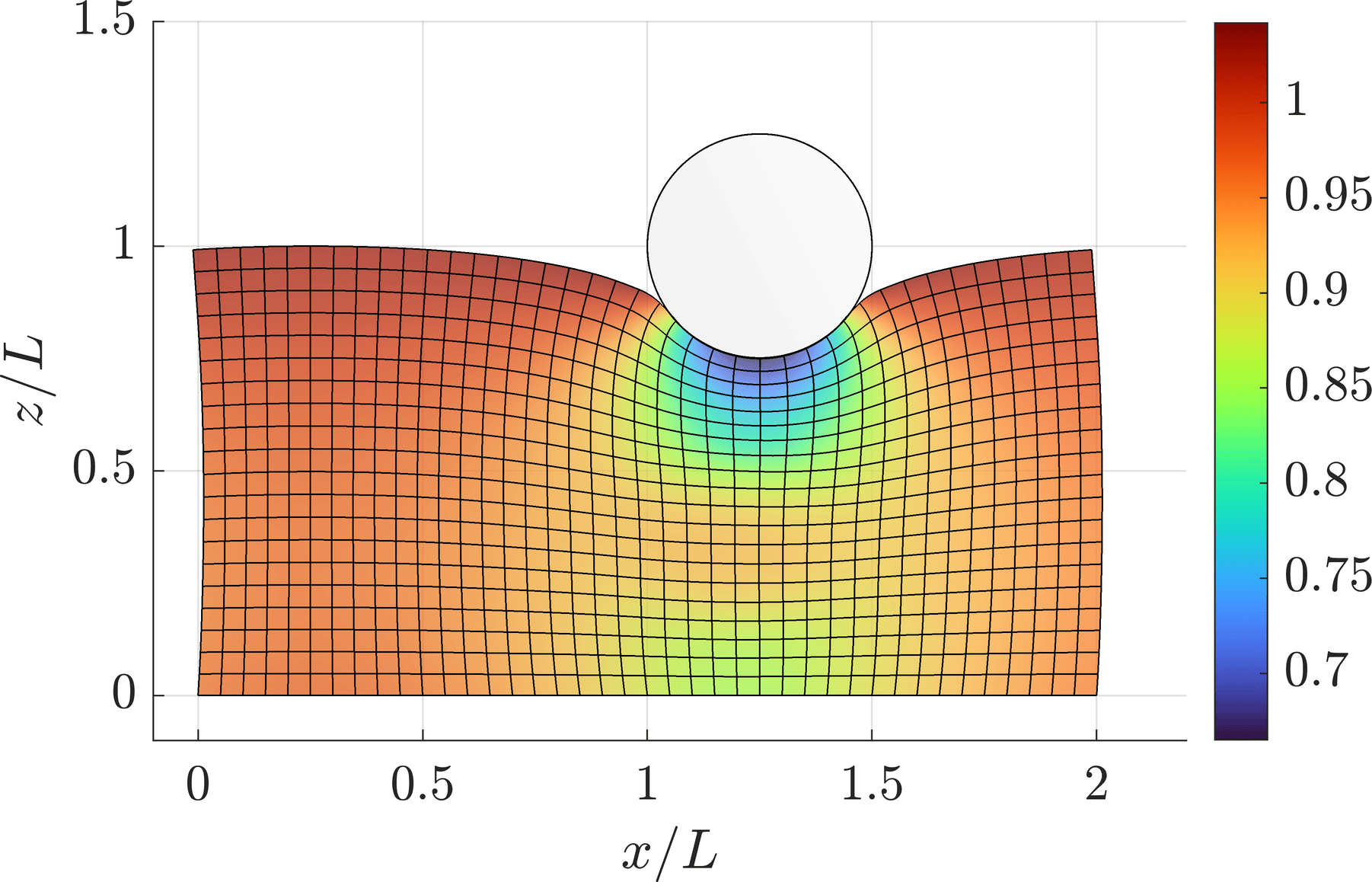}
	\put(-49mm,0mm){\small{e.}}
	\hfil 
	\includegraphics[height=32mm]{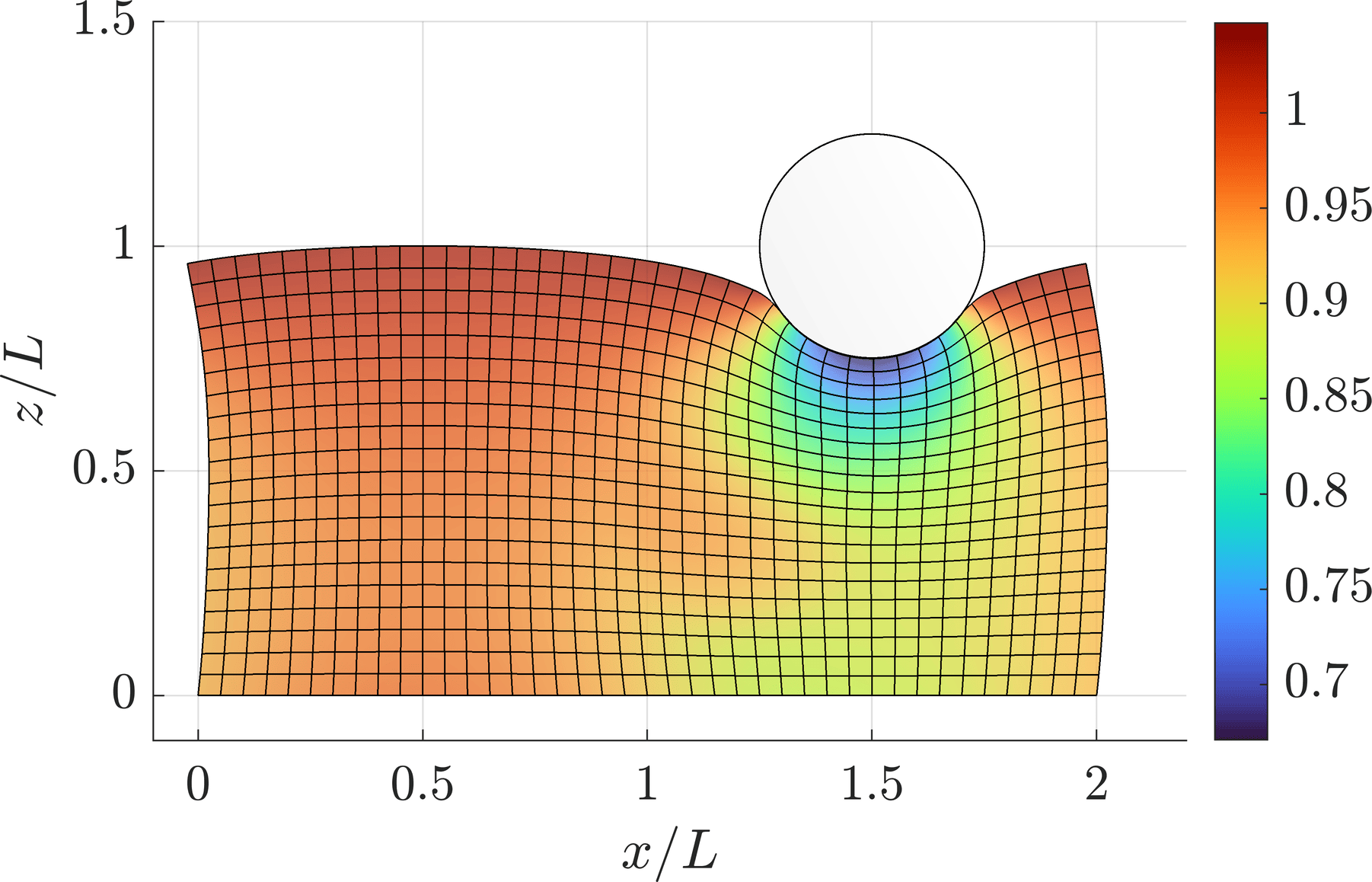} 
	\put(-49mm,0mm){\small{f.}}
	\vspace{3mm}
	\includegraphics[height=32mm]{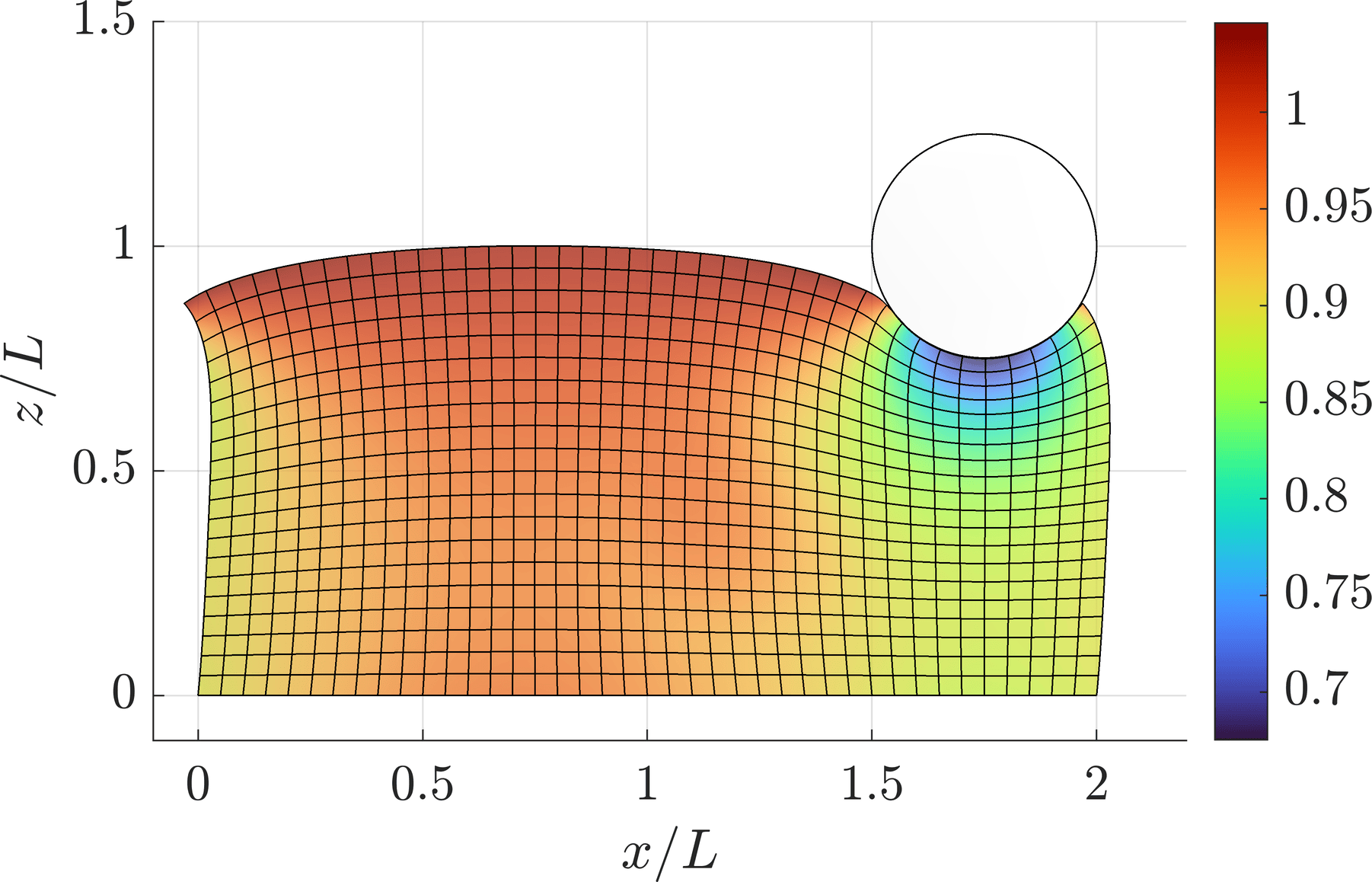}
	\put(-49mm,0mm){\small{g.}}
	\caption{Block probing: Deformed geometry of the block with a hard inclusion for seven load cases corresponding to the probe positions: a.~$X=0.25L$, b.~$X=0.50L$, c.~$X=0.75L$, d.~$X=1.00L$, e.~$X=1.25L$, f.~$X=1.50L$, and g.~$X=1.75L$. Colors represent the local volume change $\tJ$.}
	\label{fig:iron2}
\end{figure}
\begin{figure}[htbp] 
	\centering
	\includegraphics[height=32mm]{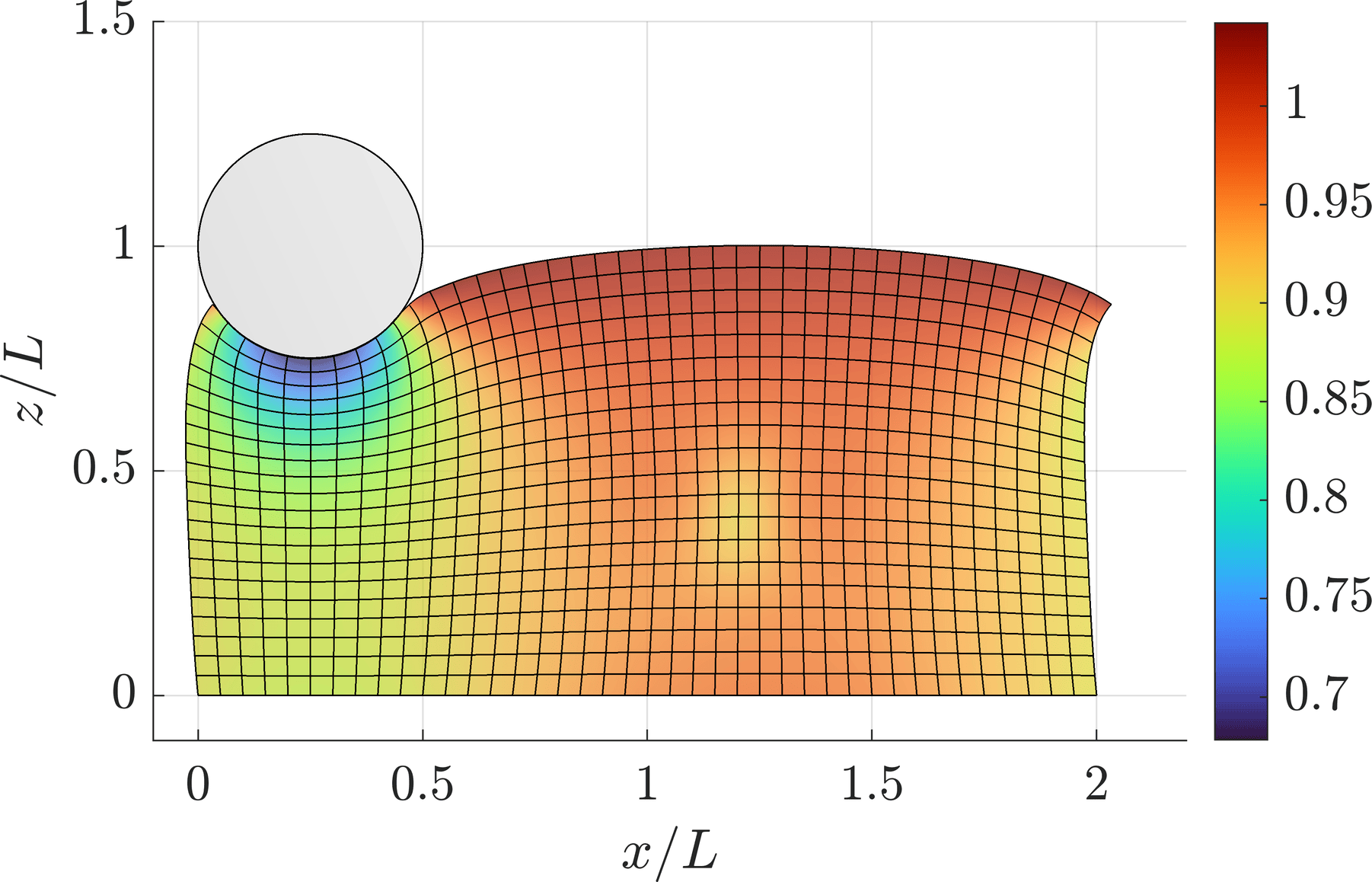}
	\put(-49mm,0mm){\small{a.}}
	\hfil 
	\includegraphics[height=32mm]{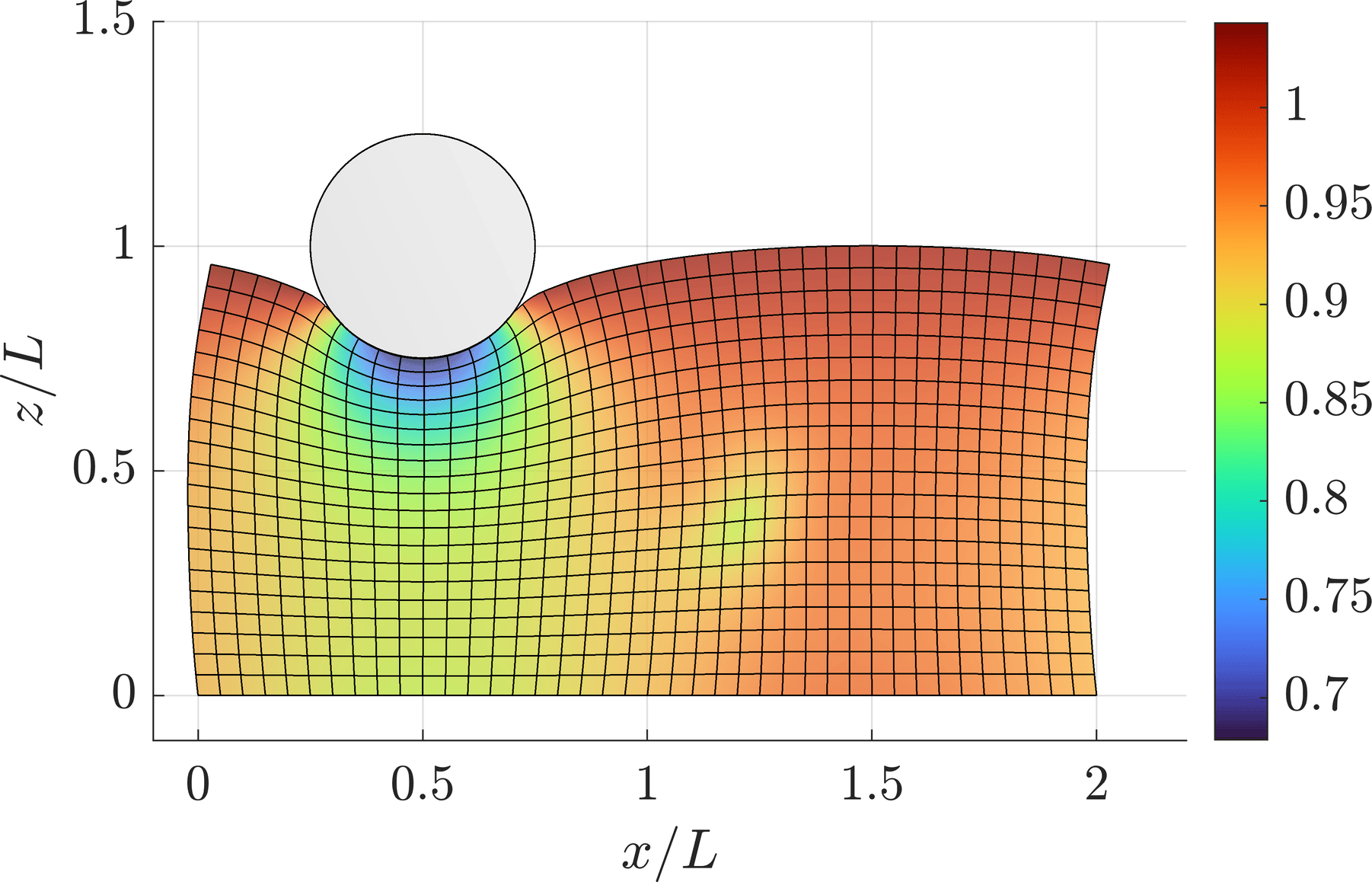}
	\put(-49mm,0mm){\small{b.}}
	\hfil 
	\includegraphics[height=32mm]{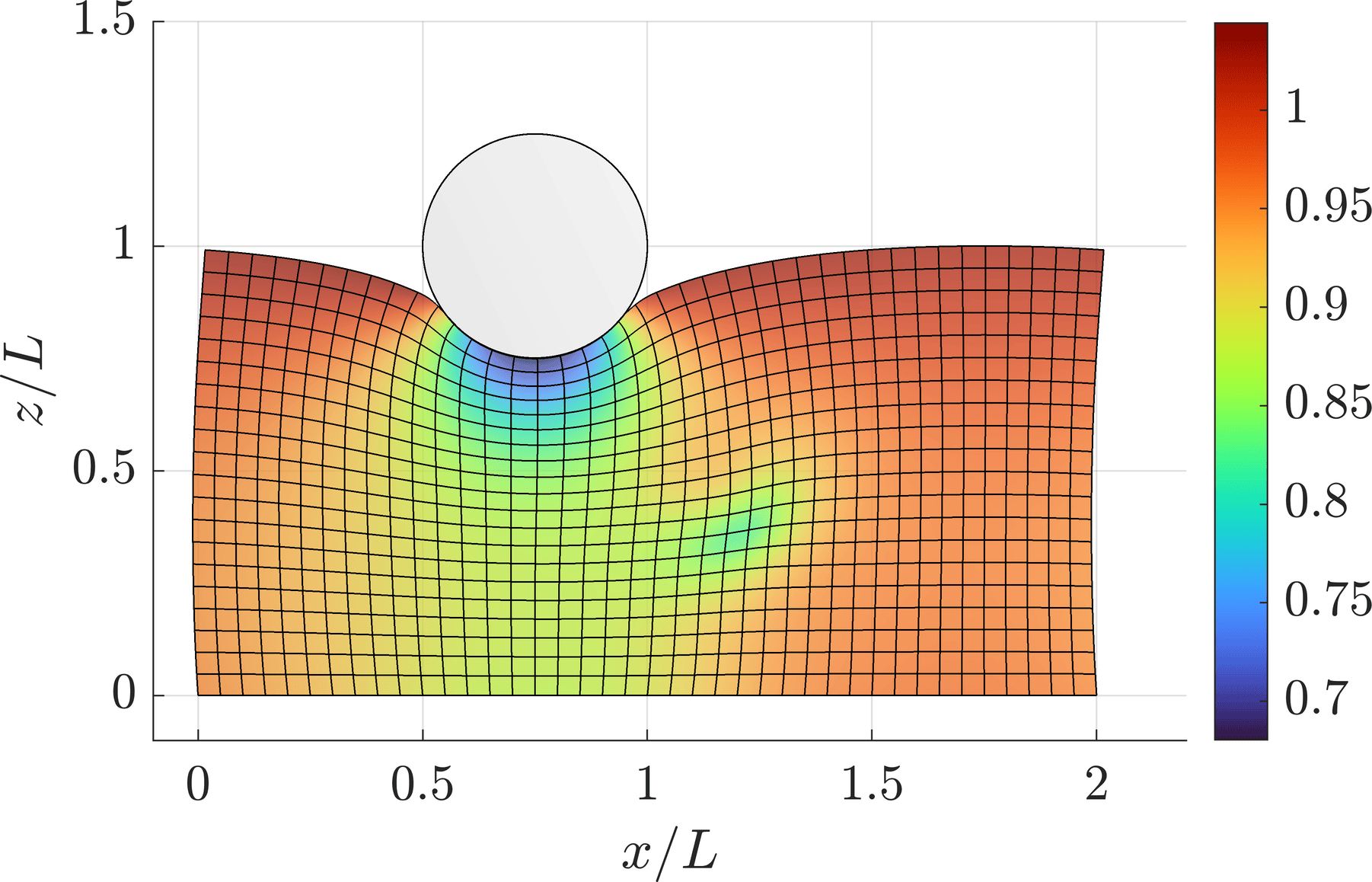} 
	\put(-49mm,0mm){\small{c.}}
	\vspace{3mm}
	\includegraphics[height=32mm]{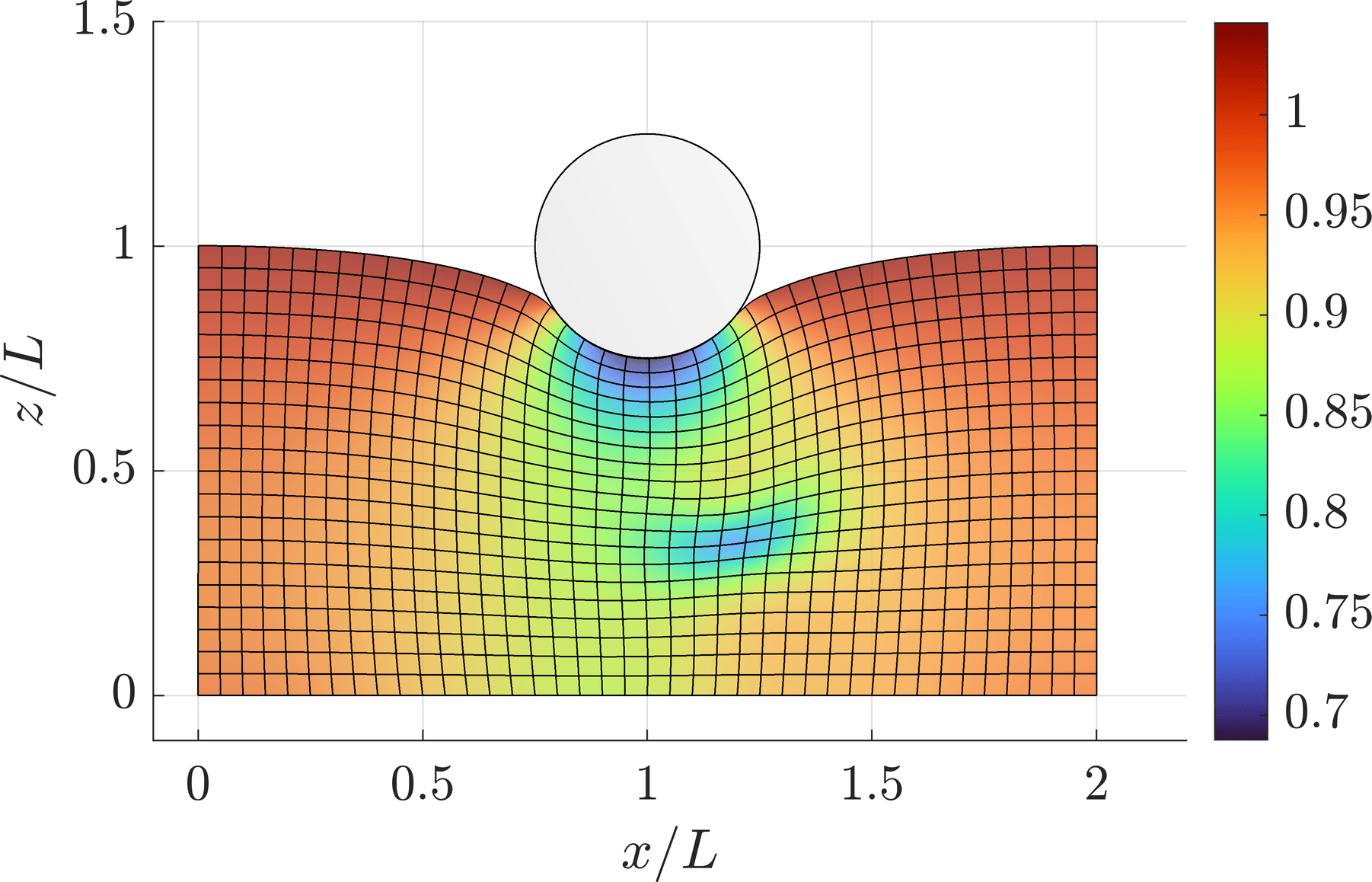}
	\put(-49mm,0mm){\small{d.}}
	\hfil 
	\includegraphics[height=32mm]{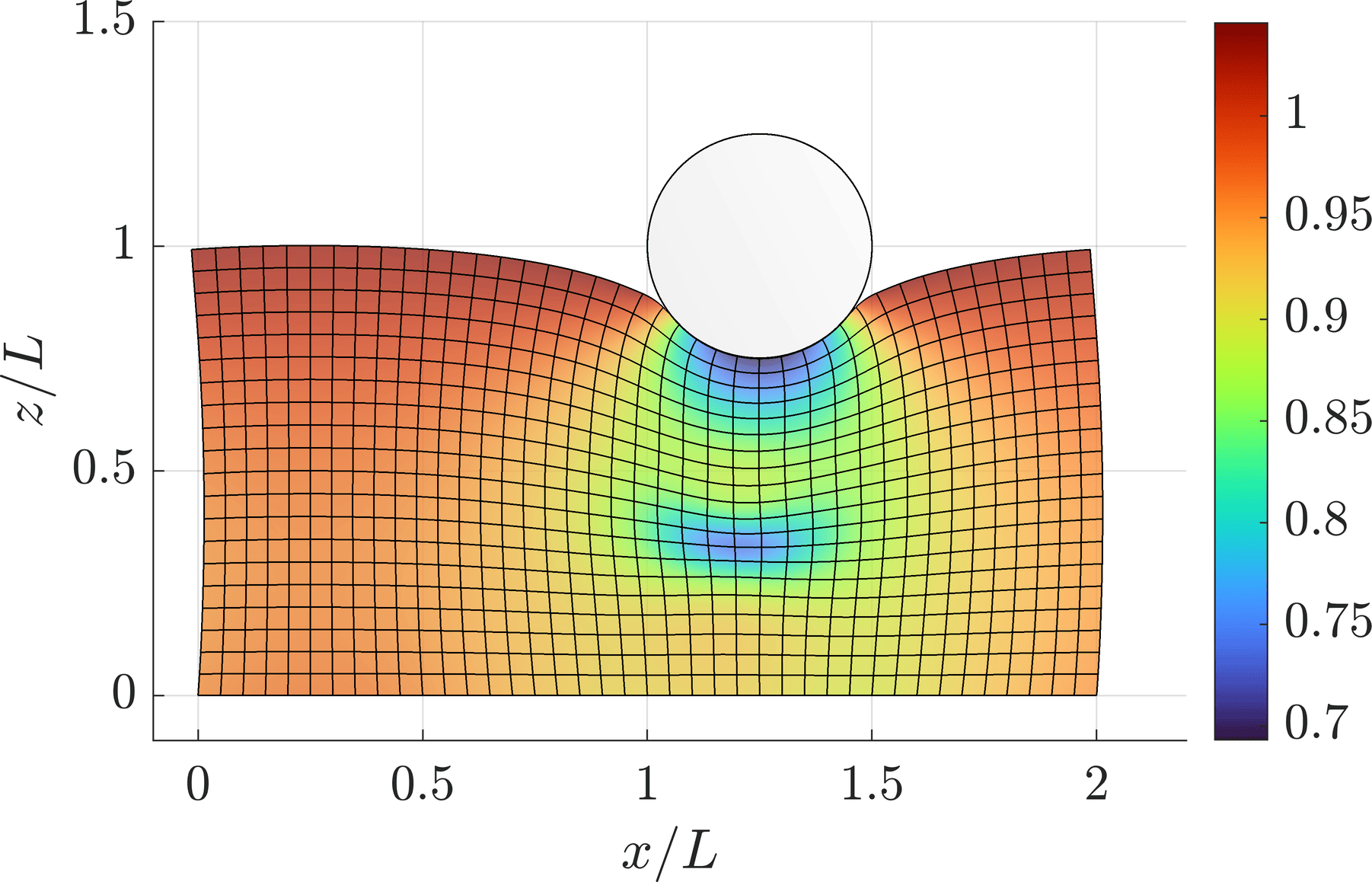}
	\put(-49mm,0mm){\small{e.}}
	\hfil 
	\includegraphics[height=32mm]{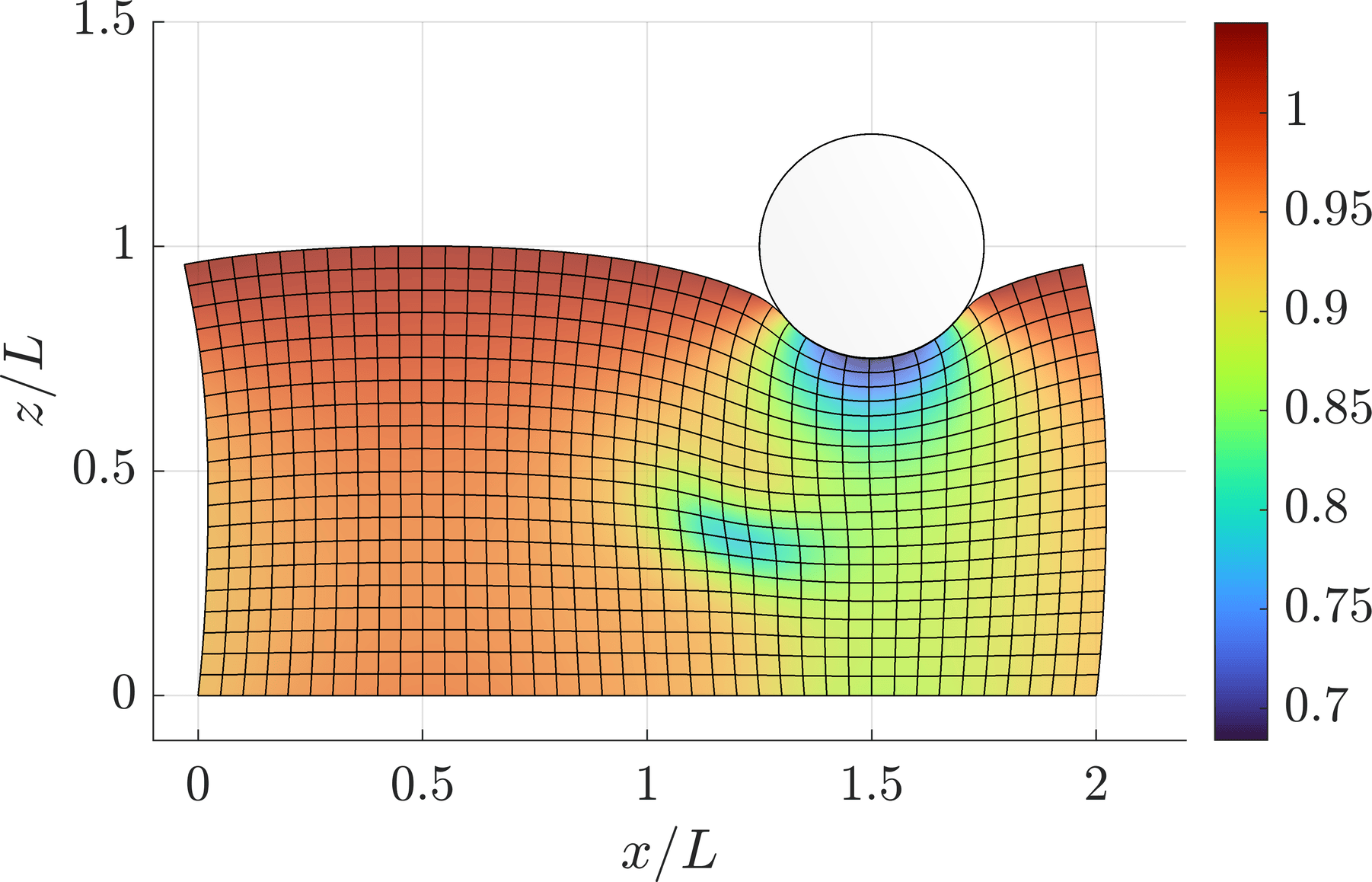} 
	\put(-49mm,0mm){\small{f.}}
	\vspace{3mm}
	\includegraphics[height=32mm]{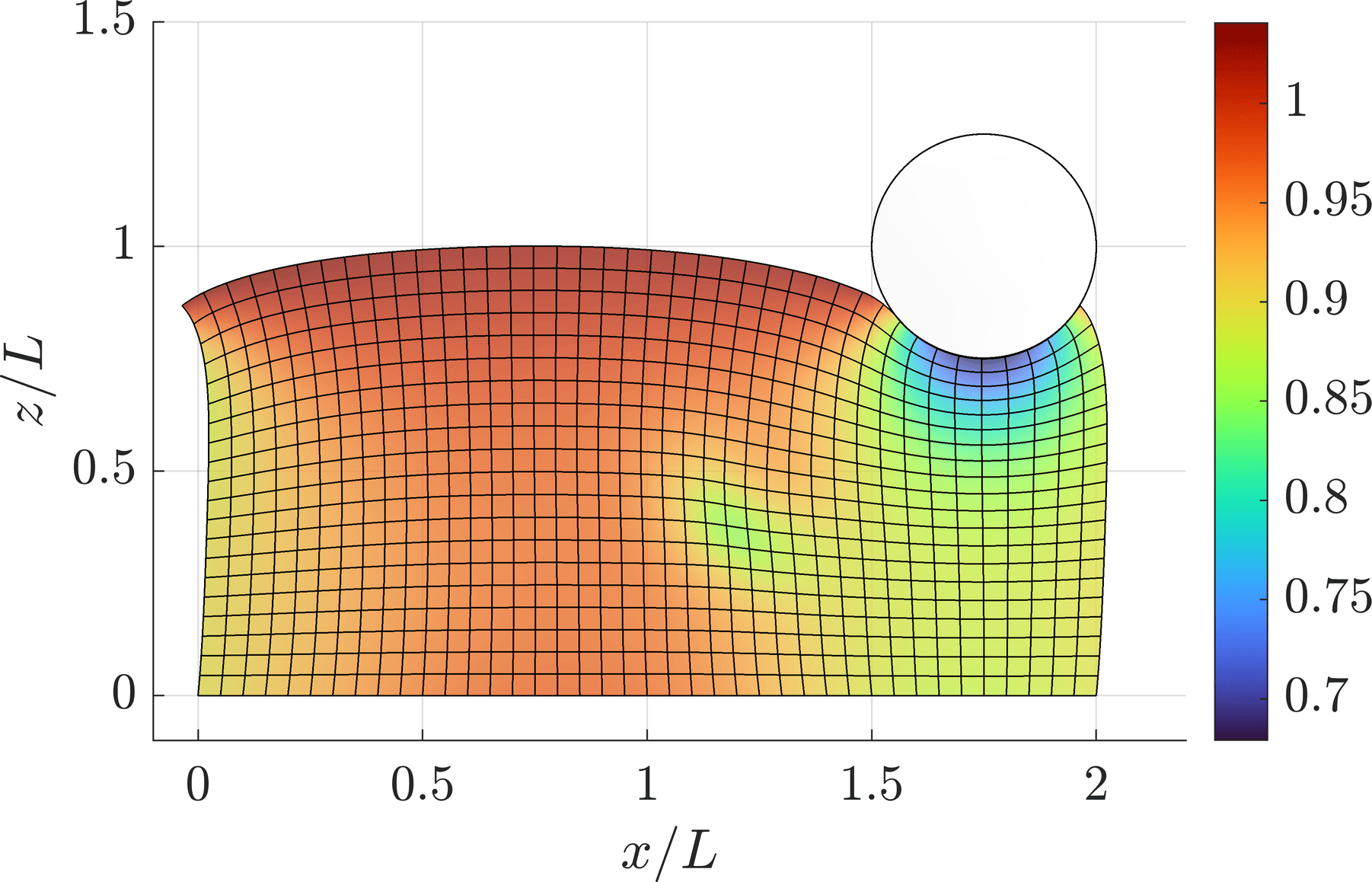}
	\put(-49mm,0mm){\small{g.}}
	\caption{Block probing: Deformed geometry of the block with a soft inclusion for seven load cases corresponding to the probe positions: a.~$X=0.25L$, b.~$X=0.50L$, c.~$X=0.75L$, d.~$X=1.00L$, e.~$X=1.25L$, f.~$X=1.50L$, and g.~$X=1.75L$. Colors represent the local volume change $\tJ$.}
	\label{fig:iron3}
\end{figure}
The periodic boundary conditions are used to reduce insensitivity of the FE model to material parameters on the sides for dense material meshes. A rigid probe of radius $0.25L$ is pushed into the top of the block at seven different positions up to an indentation of $0.25L$. The displacements are recorded every $0.05L$ of indentation, giving five load levels per load case. Figs.~\ref{fig:iron2} and \ref{fig:iron3} show the deformation maps for all considered load cases at an indentation depth of $0.25L$, for hard and soft inclusions, respectively.

The synthetic experimental data is generated using a $128\times64$ FE mesh. Based on the separate convergence study in Fig.~\ref{fig:iron1}b, the inverse analysis employs a $40\times20$ FE mesh. The contact penalty parameter depends on the number of finite elements as $\epsilon_n = 12.5F/L\,\times n_\mrel$. All identification cases simultaneously utilize experimental data from the seven probe locations, giving $7\times5 = 35$ load levels. Since the considered FE problem is purely Dirichlet, the resultant contact forces are included in $f(\mq)$ to ensure a determinable inverse solution. After preliminary studies, the weights in \eqref{e:objfnc} are selected as $w_\mru = 1$ and $w_\mrc = 0.1$. The parameter's box constraints are $[0.01\Lambda_0, 10\Lambda_0]$ and $[0.01\mu_0, 10\mu_0]$. The use of regularization is described separately for the two subsequent cases. The homogeneous preconditioning strategy from Sec.~\ref{s:overview} is always used, with initial homogenous guess picked randomly between the parameter's box constraints.

\subsubsection{Full-field experimental data}\label{s:ironF}

As a first step, the performance of the proposed inverse framework is evaluated with full-field data. All cases below utilize the same uniform experimental grid with $201$ and $101$ points in $X$-~and~$Z$-~directions, respectively. Unlike in Sec.~\ref{s:abdo}, the contact zone is included in the experimental data. No additional regularization is used.

\begin{table}[htb]
	\raggedright
	\begin{tabular}{@{}lllllll@{}}
		\toprule
		Case    & noise 			& $q(\bX)$        & \multicolumn{2}{l}{$\errM/\Delta\errM${ [}\%{]}} 		   & \multicolumn{2}{l}{$\errLe/\Delta\errLe${ [}\%{]}}			\\
		\cmidrule(l){4-5}\cmidrule(l){6-7}	
		& {[}$L${]}  				&                 & hard               	  & soft              	 & hard              	& soft			     	\\ \midrule
		3.1  	& 0     			& $\Lambda$       & $66.54$    	 		  & $57.31$    		     & $2.97$    	 		& $3.36$			    \\
		&       					& $\mu$           & $14.02$      		  & $31.08$    		     & $1.42$      			& $1.81$				\\ \midrule
		3.2  	& 0.0005     		& $\Lambda$       & $+\,8.48\pm22.89$     & $+\,0.62\pm11.30$    & $+\,1.12\pm0.36$     & $+\,0.90\pm0.24$		\\
		&				 			& $\mu$           & $+\,2.76\pm5.56$      & $+\,1.50\pm2.75$     & $+\,0.29\pm0.33$     & $+\,0.22\pm0.065$		\\ \midrule
		3.3  	& 0.001     		& $\Lambda$       & $+\,30.66\pm33.64$    & $+\,13.87\pm17.75$   & $+\,3.24\pm0.42$     & $+\,2.93\pm0.37$		\\
		&       					& $\mu$           & $+\,11.48\pm8.38$     & $+\,5.33\pm6.00$     & $+\,0.86\pm0.13$     & $+\,0.77\pm0.16$		\\ \midrule
		3.4     & 0.002     		& $\Lambda$       & $+\,61.88\pm44.72$    & $+\,50.43\pm26.66$   & $+\,7.59\pm0.74$     & $+\,7.20\pm0.62$		\\
		&       					& $\mu$           & $+\,32.92\pm15.17$    & $+\,18.79\pm11.63$   & $+\,2.26\pm0.24$     & $+\,2.15\pm0.28$		\\ \bottomrule
	\end{tabular}
	\caption{Block probing: results of identifying $\Lambda(\xi^1,\xi^2)$ and $\mu(\xi^1,\xi^2)$ for the hard and soft inclusions. All cases employ $40\times20$ FE mesh, $20\times10$ material mesh, $201\times101$ experimental grid, and $7$ load cases of $5$ load steps. For Case 3.1, the total identification errors $\errM$ and $\errLe$ are reported, while the subsequent cases with noise show only the corresponding identification error increases $\Delta\errM$ and $\Delta\errLe$ w.r.t.~Case 3.1. All cases with noise are repeated at least $96$ times for statistical analysis.} 
	\label{tab:iron_full}
\end{table}
Tab.~\ref{tab:iron_full} summarizes four identification cases with noise level ranging from $0$ to $0.002L$. Each case presents the identification errors for the hard and soft inclusions. The results indicate that both inclusion types are reconstructed with comparable accuracy across all cases. Fig.~\ref{fig:iron_full_no_noise} shows the identified nodal values for Case 3.1 in the absence of noise. Overall, the reconstruction is highly accurate, except at the contact region located in the upper part of the model ($Z = 1L$). As a uniform FE mesh is employed, the FE-based discrepancy between the forward and inverse solvers is concentrated primarily near the contact surface. Refining the FE mesh of the inverse solver in the contact zones would eliminate this inaccuracy.

Fig.~\ref{fig:iron_full_noise} presents the identification results for two representative samples from Case 3.4. Despite the high noise level, the hard and soft inclusions in $\mu(X,Y)$ are accurately captured. In contrast, for $\Lambda(X,Y)$, the inverse solution oscillates, and the inclusions become hardly visible, especially the hard inclusion. Regardless of the case and the noise level, the largest identification errors occur on the boundary of the material mesh.
\begin{figure}[htbp] 
	\centering
	\includegraphics[width=155mm]{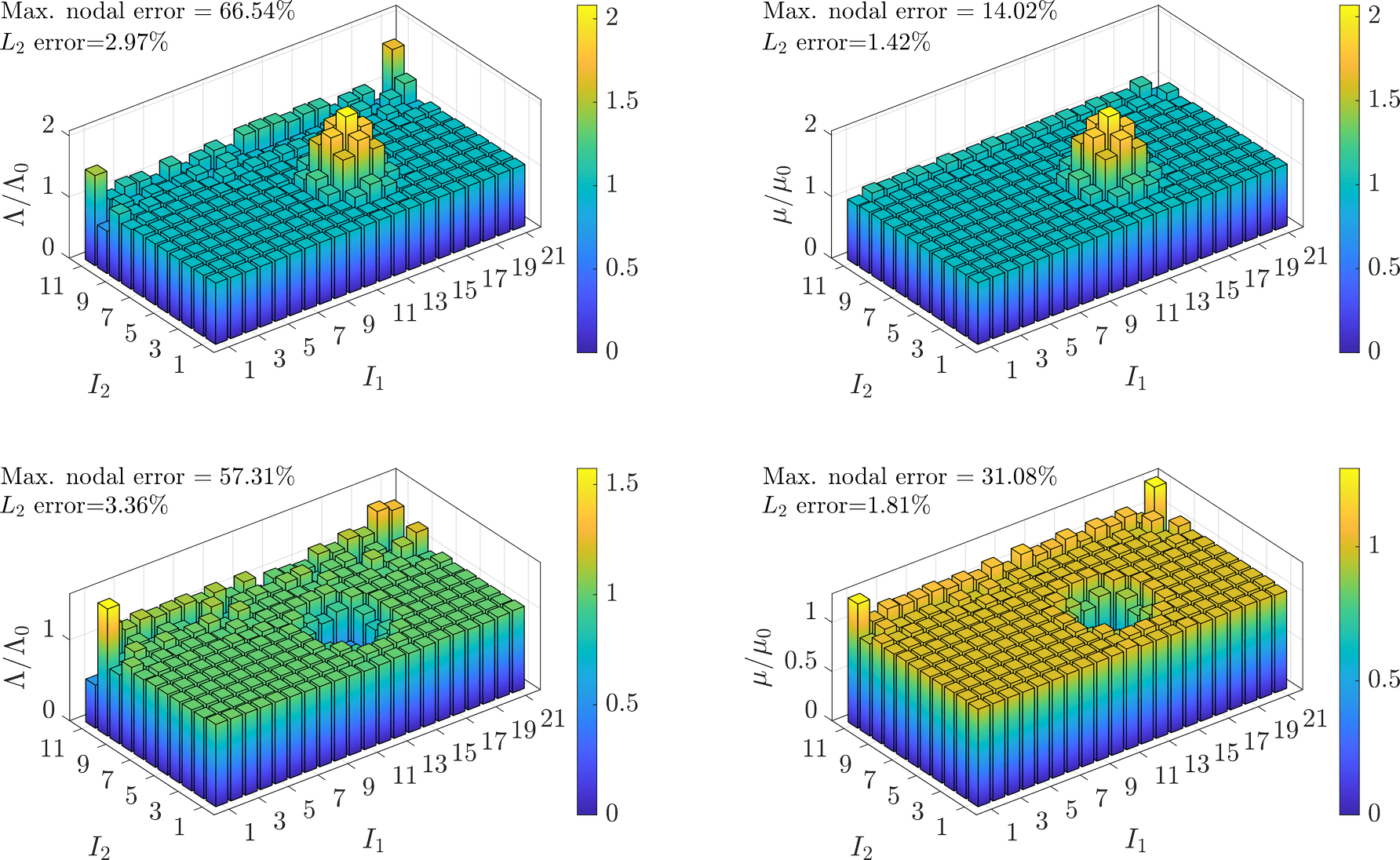}
	\put(-155mm,50mm){\small{a.}} 
	\put(-155mm,0mm){\small{c.}}
	\put(-75mm,50mm){\small{b.}}
	\put(-75mm,0mm){\small{d.}}
	\caption{Block probing, Case 3.1: distributions of identified nodal values of $\Lambda$ and $\mu$ for hard (a.~and~b.) and soft (c.~and~d.) inclusions. The top surface, where contact occurs,~($Z = 1L$) corresponds to $I_2 = 11$.}
	\label{fig:iron_full_no_noise} 
\end{figure}
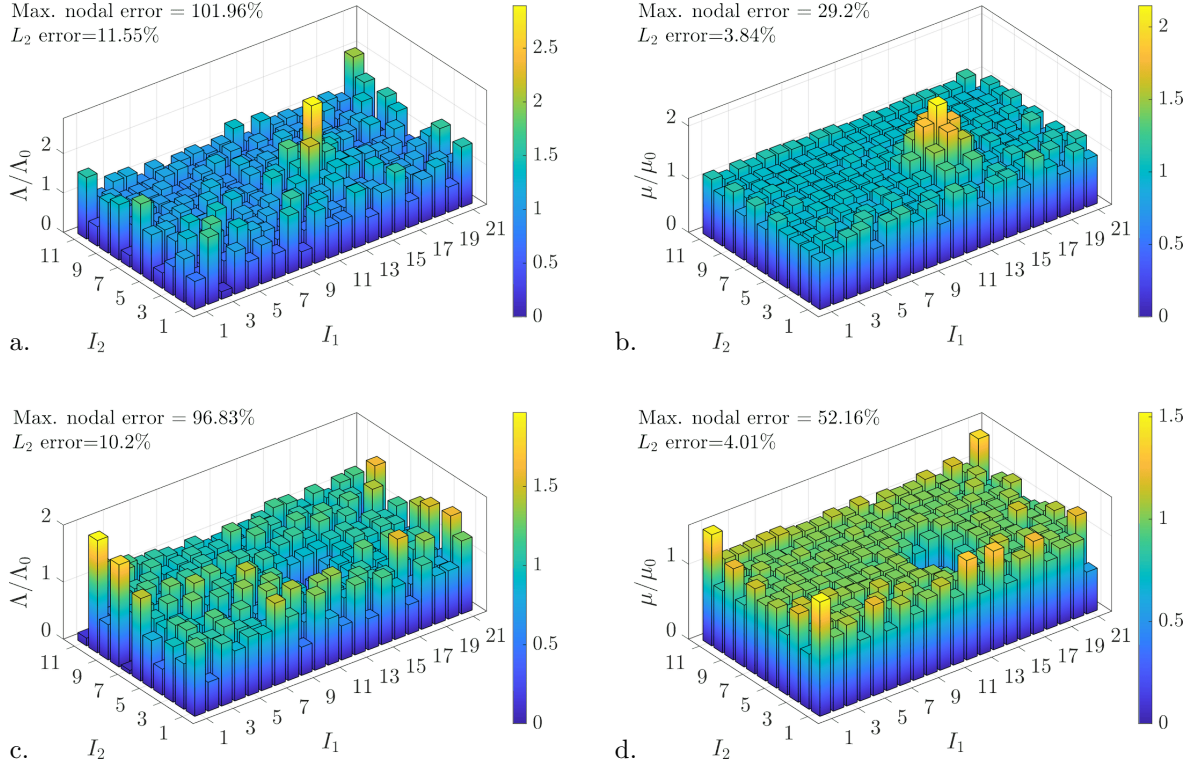
\begin{figure}[htbp] 
	\centering
	\includegraphics[width=155mm]{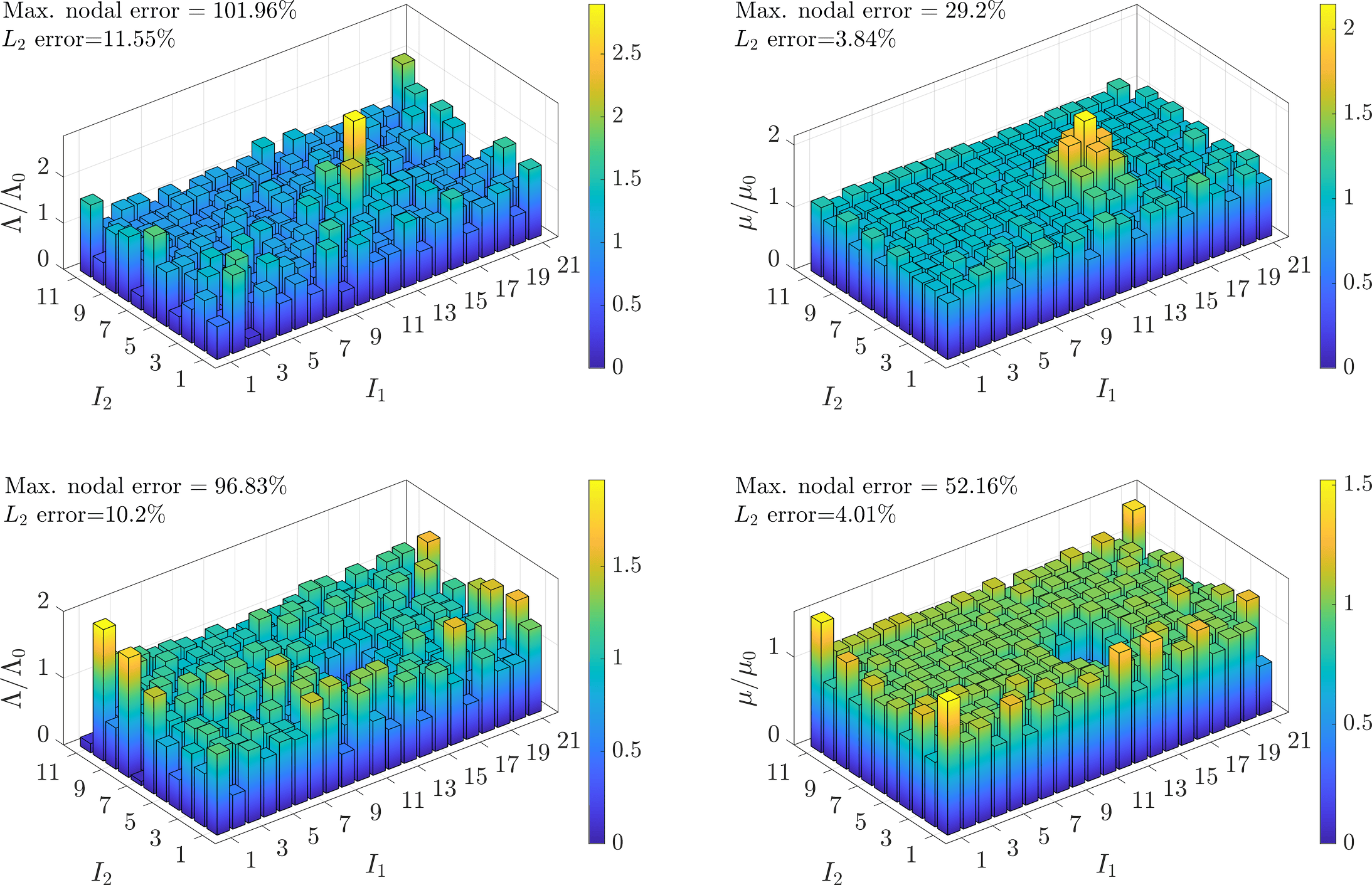}	
	\put(-155mm,0mm){\small{c.}} 
	\put(-75mm,0mm){\small{d.}}
	\put(-155mm,54mm){\small{a.}} 
	\put(-75mm,54mm){\small{b.}}
	\caption{Block probing: a.~and~b.~exemplary sample from Case 3.4 with hard inclusion: distributions of identified nodal values of $\Lambda$ (a.) and $\mu$ (b.). The top surface, where contact occurs,~($Z = 1L$) corresponds to $I_2 = 11$.} 
	\label{fig:iron_full_noise}  
\end{figure}

\begin{figure}[htbp] 
	\centering
	\includegraphics[width=155mm]{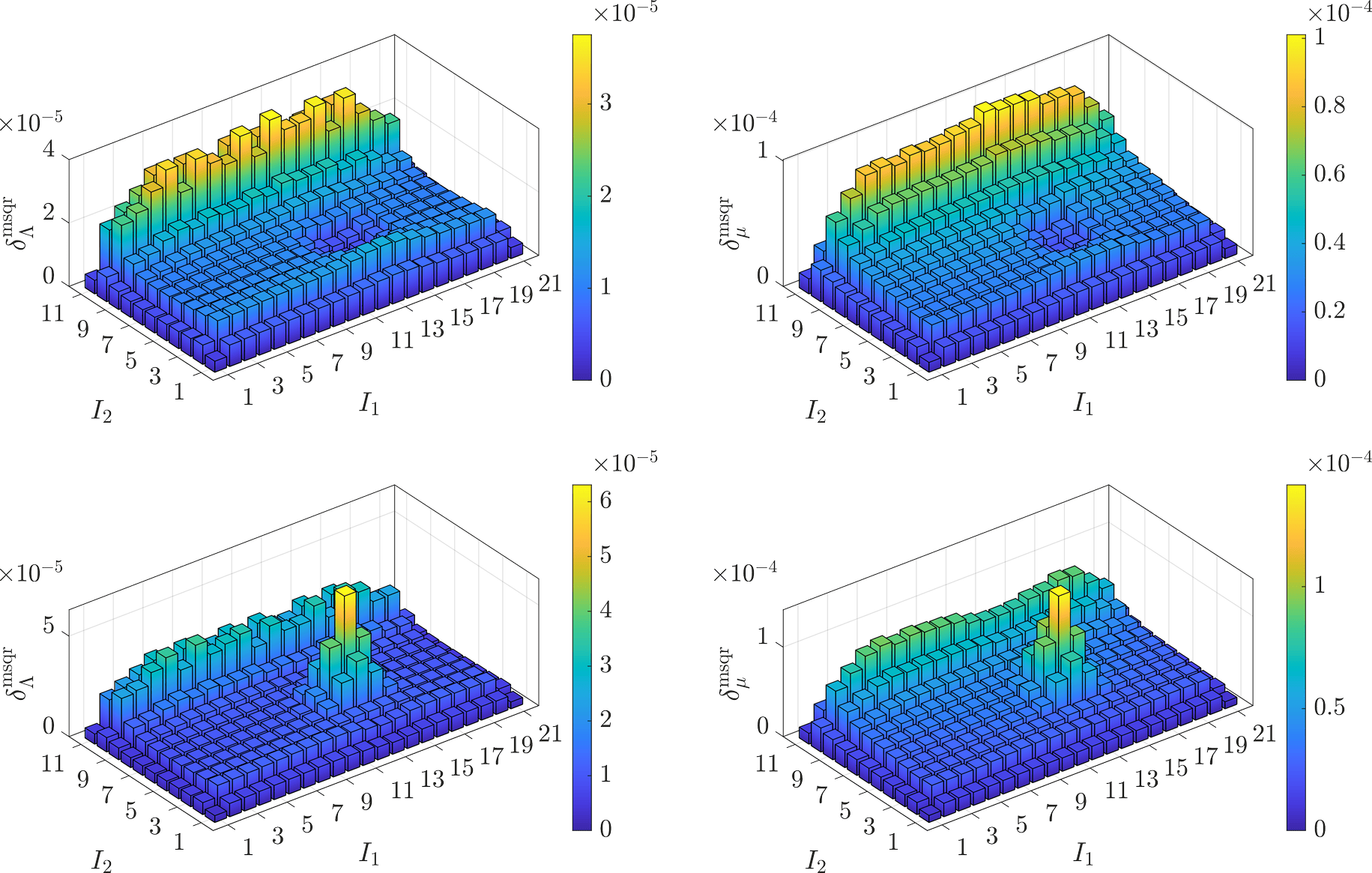}
	\put(-155mm,50mm){\small{a.}} 
	\put(-155mm,0mm){\small{c.}}
	\put(-75mm,50mm){\small{b.}}
	\put(-75mm,0mm){\small{d.}}
	\caption{Block probing: distributions of the sensitivity measures $\delta^\mathrm{msqr}$ evaluated at the optimal solution $\mq_\mrop$ for $\Lambda$ (left) and $\mu$ (right) in Case 3.1 with hard (a.~and~b.) and soft (c.~and~d.) inclusions. The top surface, where contact occurs,~($Z = 1L$) corresponds to $I_2 = 11$.} 
	\label{fig:iron_full_sens} 
\end{figure}
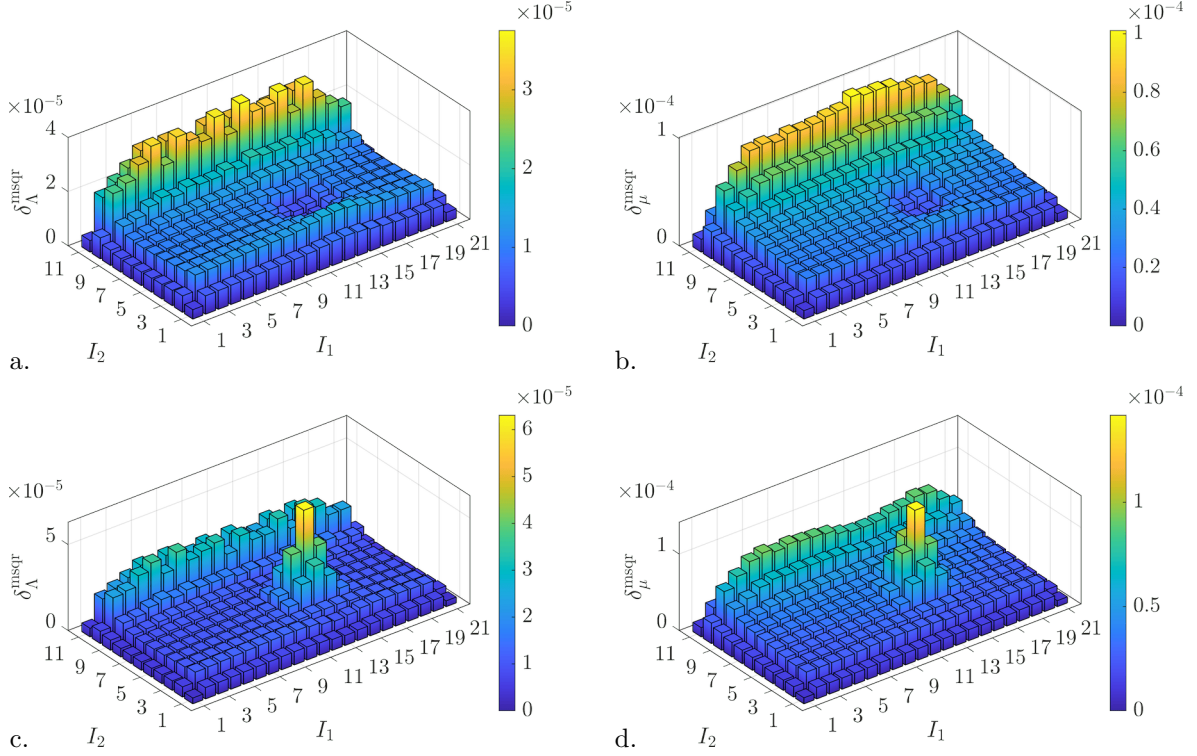

The results indicate that $\Lambda$ is significantly more sensitive to noise. One possible explanation is that displacement noise induces local volume fluctuations, to which $\Lambda$, as a parameter describing the compressibility of the material, is particularly sensitive. To address this issue further, Fig.~\ref{fig:iron_full_sens} presents the sensitivity measure~\citep{Brun2001}, defined for each material node as 
\eqb{l}
	\delta^\mathrm{msqr}_I := \ds\sqrt{\frac{1}{3n_\mrex}\sum_{i=1}^{3n_\mrex} J_{iI}^2} \,,
\label{e:dmsqr}\eqe
where $J_{iI}$ are the components of the Jacobian (see Appendix~\ref{s:deriv})\footnote{Since both $f(\mq)$ and $\mq$ are normalized, the components of $\mJ$ are dimensionless and $\delta^\mathrm{msqr}$ requires no further normalization.}. This measure quantifies the effect of a small perturbation in the material parameter $q_I$ on the model response~\citep{Zhang2022_02}. Hence, parameters with larger values of $\delta^\mathrm{msqr}$ are expected to be the more identifiable. As shown in Fig.~\ref{fig:iron_full_sens}, the values of $\delta^\mathrm{msqr}$ associated with $\Lambda$ are typically three to six times smaller than those corresponding to $\mu$.

All cases in Tab.~\ref{tab:iron_full} exhibit large maximum identification errors, most of which are concentrated near the model boundaries. As discussed above, the identification errors at the top edge originate from the FE difference between the forward and inverse solvers. The remaining boundaries are characterized by remarkably lower values of $\delta^\mathrm{msqr}$, making the corresponding $q_I$ more susceptible to noise, see Fig.~\ref{fig:iron_full_sens}. 

High $\delta^\mathrm{msqr}$ is not sufficient for a parameter to be identifiable, as it does not account for possible interactions between parameters. For a subset $K$ of the set of $n_\mrvr$ material parameters, the collinearity index is defined as
\eqb{l}
	\gamma_K := \ds\frac{1}{\sqrt{\lambda_\mathrm{min}}} \,,
\label{}\eqe
where $\lambda_\mathrm{min}$ is the smallest eigenvalue of $\bar{\mJ}_K^\T\bar{\mJ}_K^{\phantom{\T}}$, and $\bar{\mJ}_K$ is obtained by normalizing the columns of $\mJ$ (see Appendix~\ref{s:deriv}) corresponding to the subset $K$, i.e., $\bar{\mJ}_I = \mJ_I/\norm{\mJ_I}$.

The collinearity index quantifies the extent to which the effect of a change of a parameter $q_I\in K$ on the model response can be compensated by appropriate changes in the remaining parameters in $K$. In particular, such compensation may be up to a level of $1/\gamma_K\times100\%$. Therefore, high $\gamma_K$ indicates that the parameter subset $K$ is poorly identifiable, even if all parameters in $K$ have high $\delta^\mathrm{msqr}$~\citep{Brun2001,Gbor2017}. In general, $\gamma_K$ can be evaluated at different positions in the design space to assess parameter identifiability, especially for nonlinear problems.

Here, $\gamma_K$ is calculated for the whole parameter set for simplicity. At the optimal solution $\mq_\mrop$ of Case 3.1, $\gamma_K = 14.56$ and $\gamma_K = 15.30$ for the hard and soft inclusions, respectively. These values will be used in the following section.

\subsubsection{Surface experimental data}\label{s:ironS}

In the last example, the experimental data are restricted to the surface at $Z = 1L$. Preliminary studies indicated that $7\times5 = 35$ load steps are not sufficient to ensure reliable reconstruction as before. Therefore, the overall number of load cases is increased substantially. Now, the model in Fig.~\ref{fig:iron1}a is indented in $31$ positions, starting from $X = 0.25L$ and ending at $X = 1.75L$, with an increment of $0.05L$. Moreover, the indentation depth is increased from $0.25L$ to $0.40L$, and the displacements are recorded every $0.05L$ of indentation as before. The new setting therefore yields $31\times8 = 248$ load steps in CBIA. For brevity, the new deformation maps are not shown.

Since the experimental data is limited to the surface, poor identifiability of the material parameters is expected. For only surface data, the noise-free Case 3.1 in Tab.~\ref{tab:iron_full} fails to converge, leading to $\gamma_K$ of the order of $10^{16}$. To obtain a stable solution, additional regularization is introduced. The regularization term in Eq.~\eqref{e:objfnc} is selected as
\eqb{l}
	\sR(\alpha,\mq) = \alpha^2\norm{\mq-\mq_{0,\text{\tiny HP}}}^2 \,,
\label{e:regHP}\eqe
where $\mq_{0,\text{\tiny HP}}$ is the initial guess from the homogeneous preconditioning strategy. Consequently, deviations from the homogeneous inverse solution are penalized. 

\begin{figure}[p] 
	\centering
	\includegraphics[height=55mm]{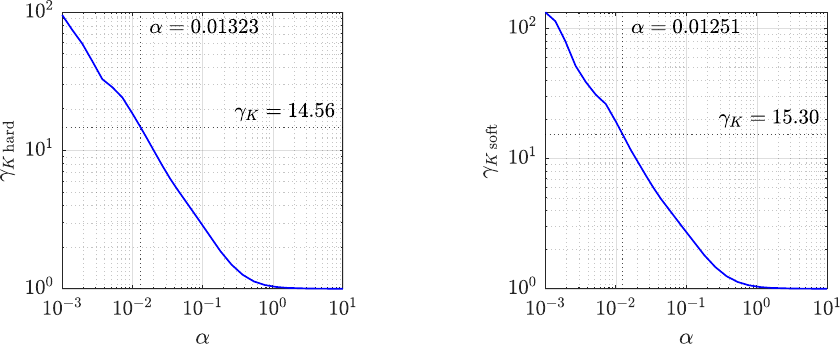}
	\put(-138mm,0mm){\small{a.}} 
	\put(-60mm,0mm){\small{b.}}
	\caption{Block probing using only surface data: Collinearity index at the optimal solution $\mq_\mrop$, as a function of the regularization parameter $\alpha$ for hard~(a.) and soft~(b.) inclusions in CBIA based solely on surface data.} 
	\label{fig:iron_surf_gamma} 
\end{figure}
\begin{figure}[p] 
	\centering
	\includegraphics[width=155mm]{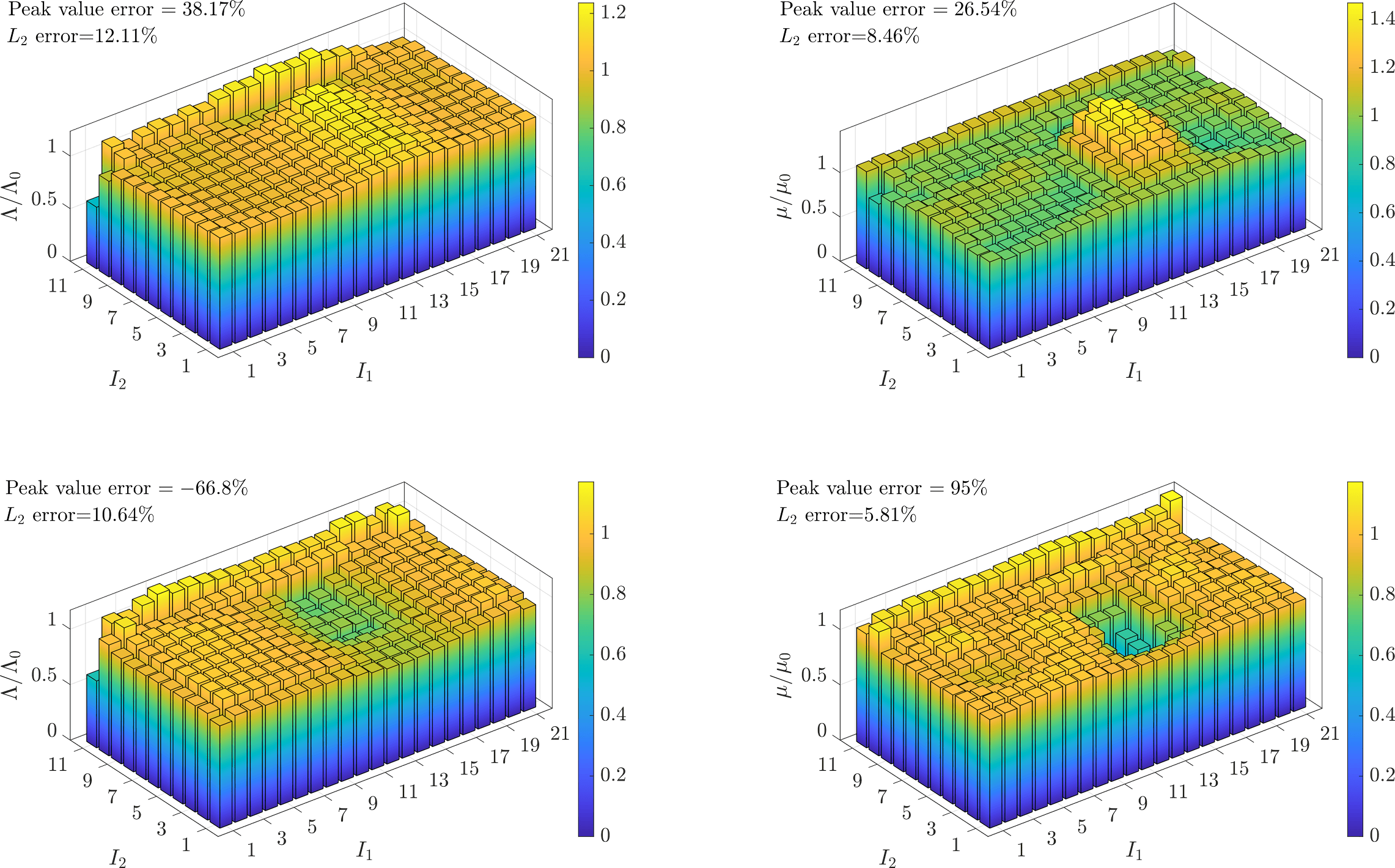}
	\put(-155mm,50mm){\small{a.}} 
	\put(-155mm,0mm){\small{c.}}
	\put(-75mm,50mm){\small{b.}}
	\put(-75mm,0mm){\small{d.}}
	\caption{Block probing using only surface data: Distributions of nodal values of $\Lambda(\xi^1,\xi^2)$ and $\mu(\xi^1,\xi^2)$ identified using CBIA based solely on noise-free surface data for hard~(a.~and b.) and soft~(c.~and d.) inclusions. Both CBIA cases are regularized with $\alpha$ values corresponding to Fig.~\ref{fig:iron_surf_gamma}. Peak error values denote the relative identification error of the peak value of the material parameter within the inclusion; negative values indicate overestimation, and positive values underestimation. The top surface, where contact occurs,~($Z = 1L$) corresponds to $I_2 = 11$.} 
	\label{fig:iron_surf_no_noise} 
\end{figure}
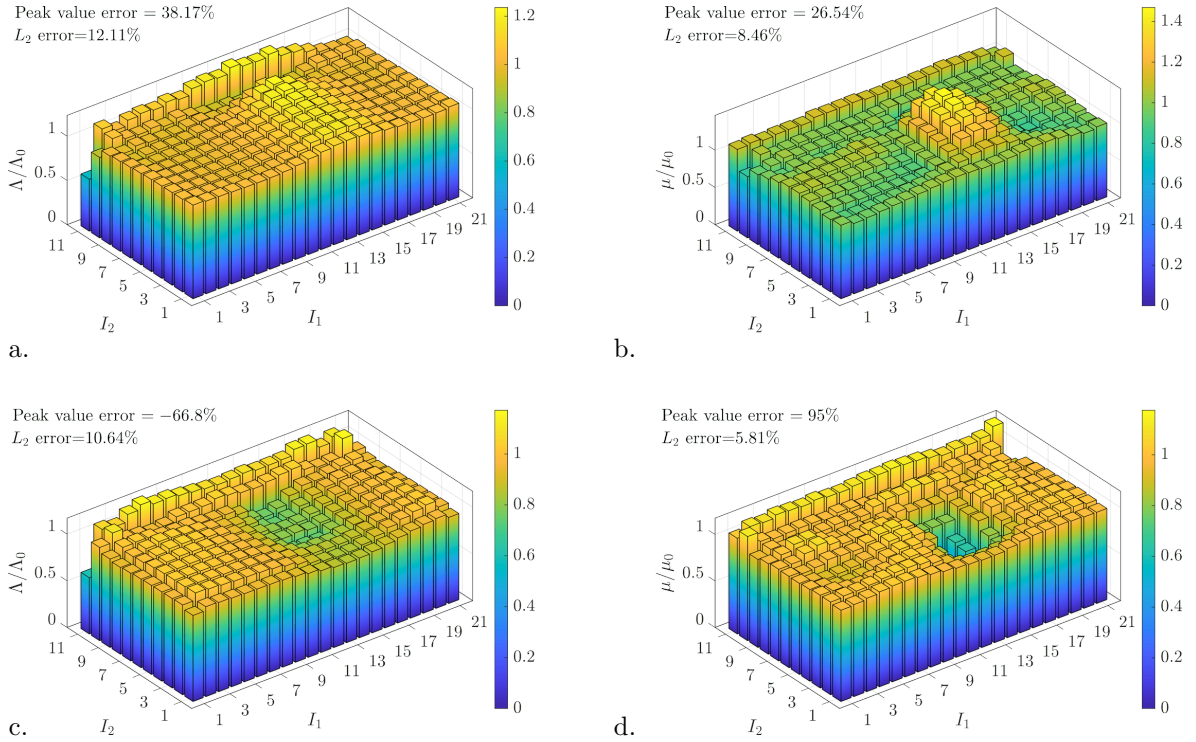
The parameter $\alpha$ in Eq.~\eqref{e:regHP} is selected to match the same level of $\gamma_K$ as the corresponding full-field cases from Tab.~\ref{tab:iron_full} without noise, see Fig.~\ref{fig:iron_surf_gamma}. Here, this criterion is preferred since a popular L-curve method~\citep{Hansen1993} promotes overregularized solutions. 

Fig.~\ref{fig:iron_surf_no_noise} presents the CBIA solutions obtained from noise-free surface data. As expected, the reconstruction accuracy is much lower than that achieved with full-field data (see Fig.~\ref{fig:iron_full_no_noise}), and all cases exhibit high identification errors. Although both hard and soft inclusions are captured in $\mu(\xi^1,\xi^2)$, the inclusions in $\Lambda(\xi^1,\xi^2)$ are oversmoothed and deteriorate in shape. For different values of $\alpha$, CBIA either oversmooths the inclusions or permits spurious oscillations to dominate $\mq_\mrop$. The soft inclusion is reconstructed more accurately than the hard one for both $\Lambda$ and $\mu$, as it has a stronger effect on the model response. For all identified material fields, $\Lambda$ and $\mu$ are overestimated in the surface layer ($I_2=11$). Interestingly, despite the strong smoothing effect of Tikhonov regularization, CBIA predicts a lower value of $\mu$ at the center of the soft inclusion ($0.01\mu_0$) than the reference value ($0.2\mu_0$). As shown, despite the substantial increase in the number of load cases, using surface data alone remains challenging for qualitative reconstruction with CBIA.

\begin{figure}[htbp] 
	\centering
	\includegraphics[height=40mm]{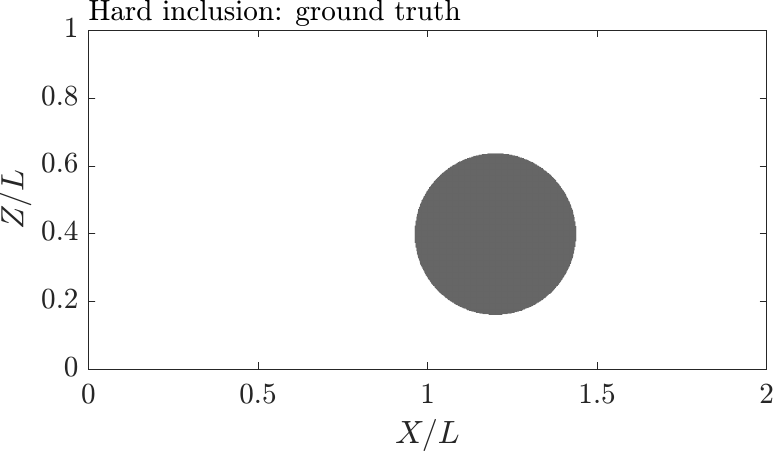}
	\put(-70mm,0mm){\small{a.}}
	\hfil 
	\includegraphics[height=40mm]{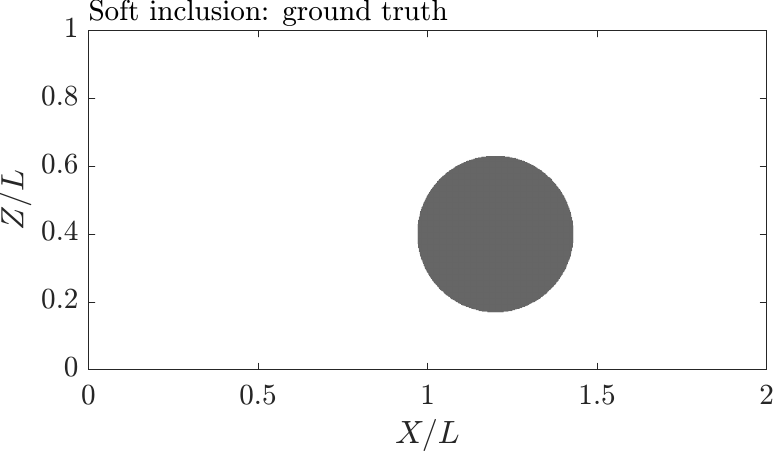}
	\put(-70mm,0mm){\small{b.}}
	\vspace{3mm}
	\includegraphics[height=40mm]{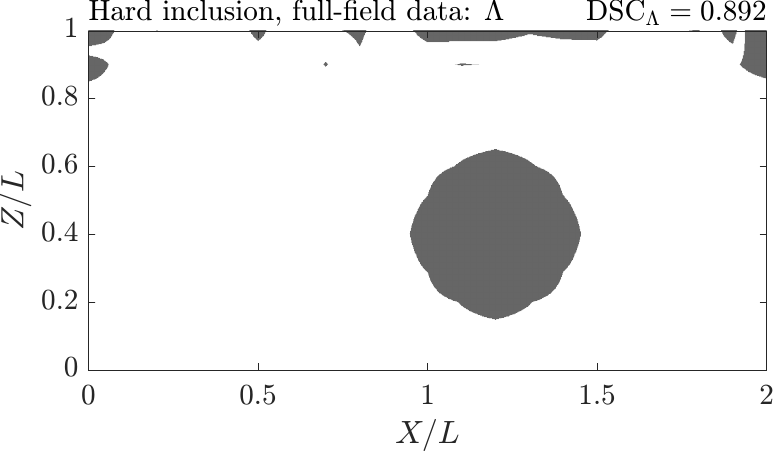}
	\put(-70mm,0mm){\small{c.}}
	\hfil 
	\includegraphics[height=40mm]{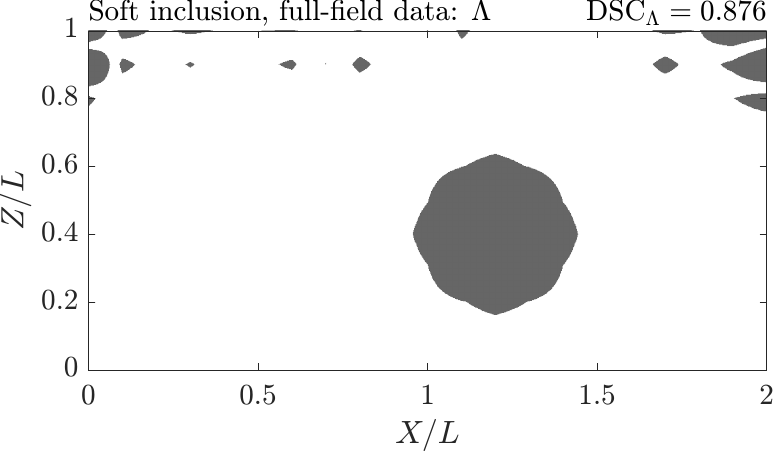} 
	\put(-70mm,0mm){\small{d.}}
	\vspace{3mm}
	\includegraphics[height=40mm]{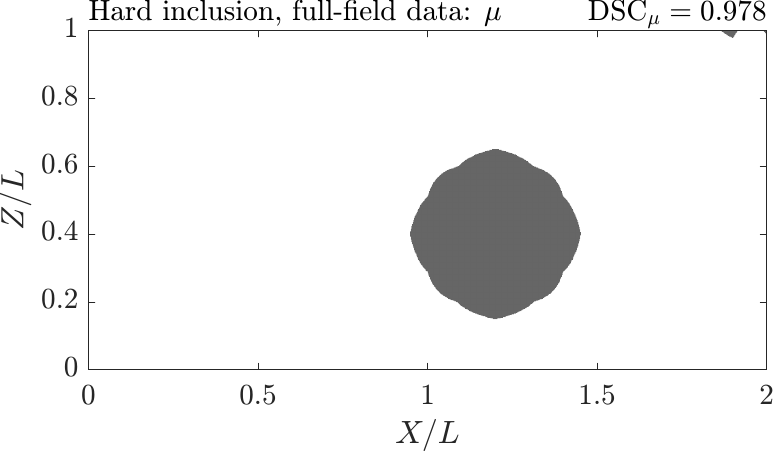} 
	\put(-70mm,0mm){\small{e.}}
	\hfil
	\includegraphics[height=40mm]{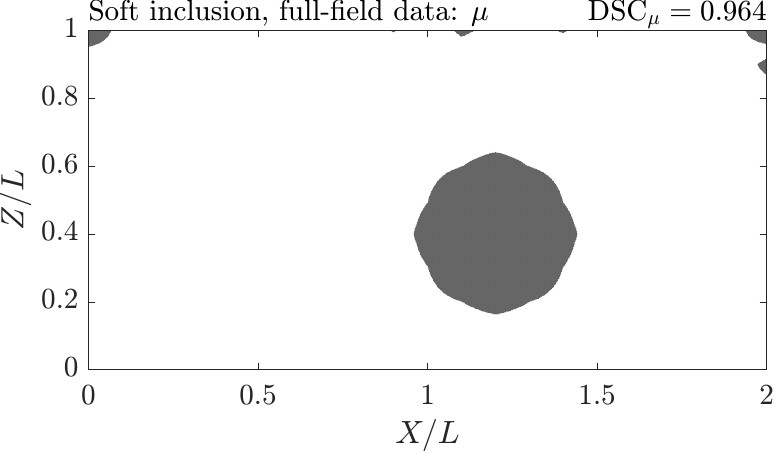} 
	\put(-70mm,0mm){\small{f.}}
	\vspace{3mm}
	\includegraphics[height=40mm]{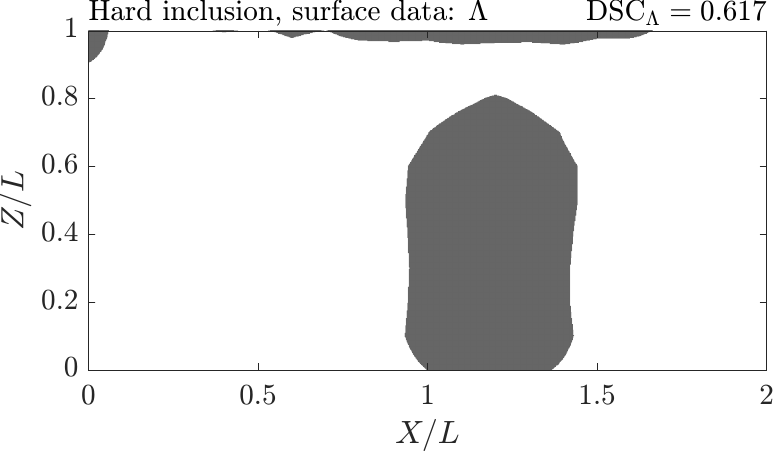} 
	\put(-70mm,0mm){\small{g.}}
	\hfil 
	\includegraphics[height=40mm]{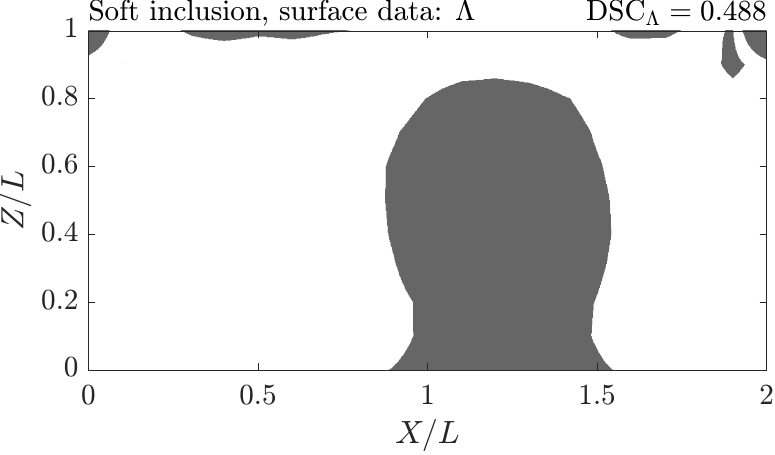} 
	\put(-70mm,0mm){\small{h.}}
	\vspace{3mm}
	\includegraphics[height=40mm]{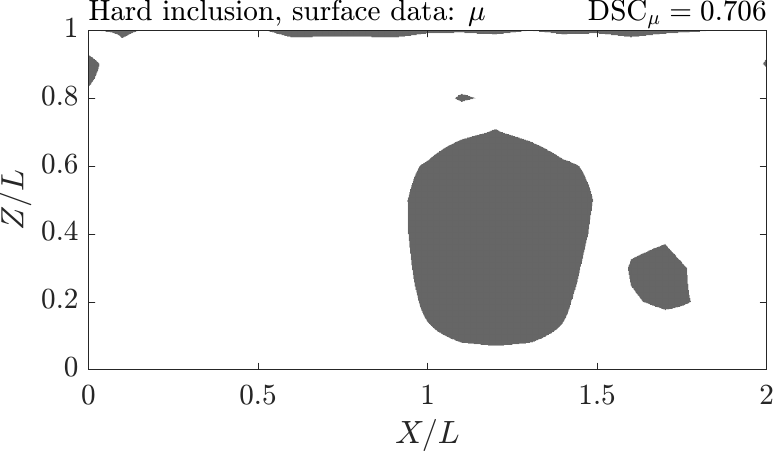} 
	\put(-70mm,0mm){\small{i.}}
	\hfil
	\includegraphics[height=40mm]{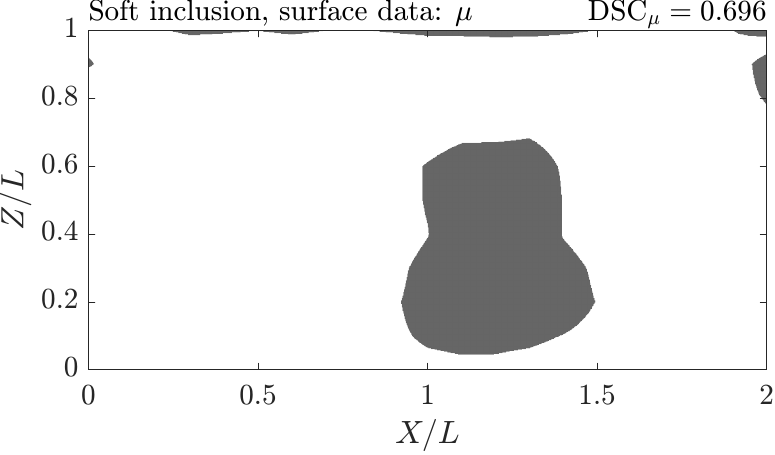}
	\put(-70mm,0mm){\small{j.}}
	\caption{Block probing using only surface data: Segmented images of the identified material fields for the reference distribution in Fig.~\ref{fig:iron_material}~(a.~and~b.) using noise-free full-field data in Fig.~\ref{fig:iron_full_no_noise}~(c--f), and noise-free surface data in Fig.~\ref{fig:iron_surf_no_noise}~(g--j). The selected threshold in Eq.~\eqref{e:thresholding} is $t = 0.1$.}
	\label{fig:iron_img}
\end{figure}
On the other hand, a quantitative assessment of the identified material field may be more relevant than its absolute values. It remains to be investigated whether CBIA can reliably identify the presence of an inclusion using solely surface data. To this end, the reconstructed material fields are segmented into inclusion and surrounding regions, as shown in Fig.~\ref{fig:iron_img}. Subsequently, the similarity between the reference and the identified fields is quantified using the Dice similarity coefficient
\eqb{l}
	\mathrm{DSC} := \ds\frac{2\mathrm{Area}(S_\mrrf\cap S_\mrop)}{\mathrm{Area}(S_\mrrf) +  \mathrm{Area}(S_\mrop)} \,,
\label{}\eqe
where $S_\mrrf$ and $S_\mrop$ denote the reference and identified inclusions, respectively. DSC takes values between 0~(no alignment) and 1~(perfect alignment). The regions of inclusions are obtained via simple thresholding
\eqb{l}
	S_\bullet := \left\{\bx\in\sB:\lvert q_\mrop(\bx) - q_\mrrf(\bx) \rvert > t\right\} \,,
\label{e:thresholding}\eqe
where $t$ is the segmentation threshold. Note that in Eq.~\eqref{e:thresholding}, positive and negative inclusions are detected equivalently due to the absolute value. 

Fig.~\ref{fig:iron_img} compares segmentations of the reference material fields with noise-free CBIA results for both full-field and surface data. As seen, CBIA successfully identifies the inclusion for full-field data, achieving almost perfect overlap for $\mu$ and approximately $90\%$ agreement for $\Lambda$. However, oscillations at the contact surface are still visible, particularly in $\Lambda$. The remaining mismatch between CBIA and the reference is attributed to bilinear interpolation of the material mesh. 

For surface data (Figs.~\ref{fig:iron_img} g--j), the inclusions are smeared in the $Z$-~direction towards the base of the block, leading to reduced DSC, especially for $\Lambda$. In all surface-data-based cases, artifacts at the contact surface are observed, likely due to FE discretization errors concentrated there. Note that the segmentation results are sensitive to the choice of the threshold $t$ in Eq.~\eqref{e:thresholding}. For larger $t$, the soft inclusions yield higher DSC, which follows from Fig.~\ref{fig:iron_surf_no_noise}.

\begin{figure}[htb] 
	\centering
	\includegraphics[height=40mm]{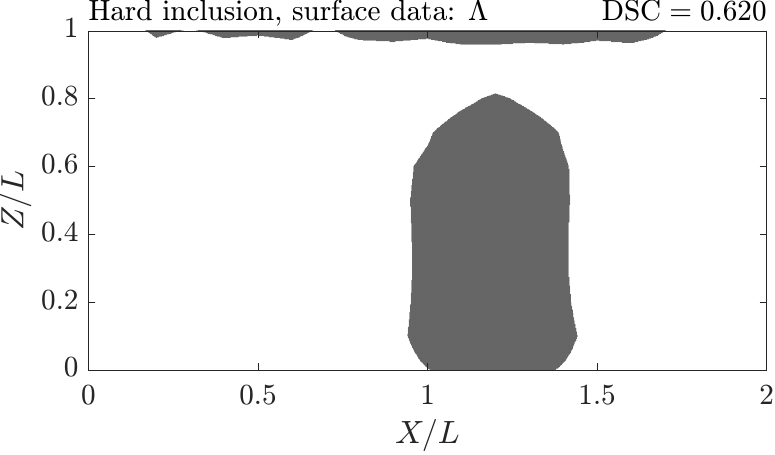}
	\put(-70mm,0mm){\small{a.}}
	\hfil 
	\includegraphics[height=40mm]{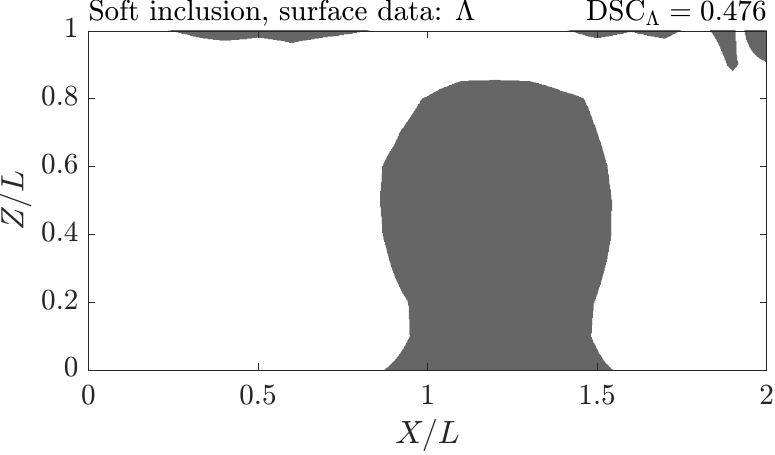}
	\put(-70mm,0mm){\small{b.}}
	\vspace{3mm}
	\includegraphics[height=40mm]{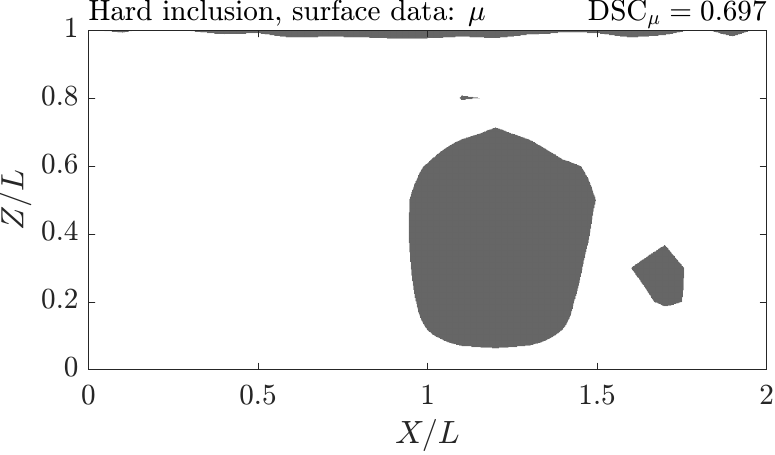}
	\put(-70mm,0mm){\small{c.}}
	\hfil 
	\includegraphics[height=40mm]{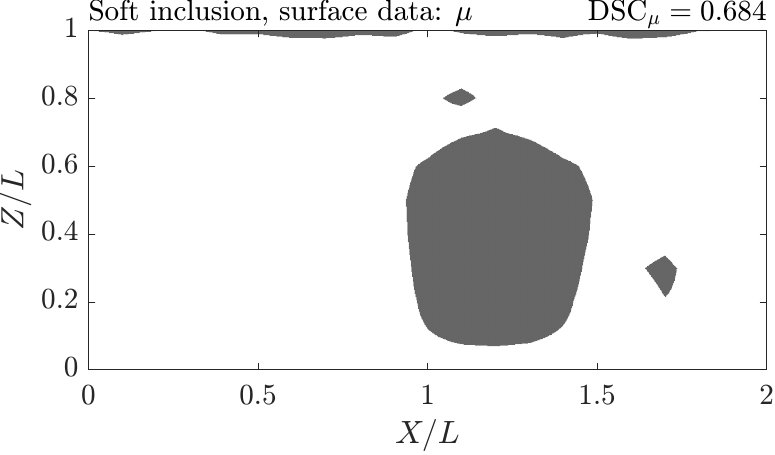} 
	\put(-70mm,0mm){\small{d.}}
	\caption{Block probing using only surface data: Segmented images of the material fields identified with CBIA based solely on surface data corrupted by noise with $\gamma_{Ii} = 0.001L$. The selected threshold in Eq.~\eqref{e:thresholding} is $t = 0.1$.}
	\label{fig:iron_img_noise}
\end{figure}
To investigate the effect of noise on CBIA using solely surface data, the surface displacements are perturbed with noise. Subsequently, the computations are repeated 10 times for statistical analysis. The regularization parameter $\alpha$ is kept identical to that determined in Fig.~\ref{fig:iron_surf_gamma}. Owing to regularization, the influence of noise on the results is marginal, yielding $\mathrm{DSC}_\Lambda = 62.76\pm0.48\%$ and $\mathrm{DSC}_\mu = 71.08\pm0.88\%$ for the hard inclusion, and $\mathrm{DSC}_\Lambda = 48.21\pm0.45\%$ and $\mathrm{DSC}_\mu = 69.84\pm0.58\%$ for the soft inclusion. Fig.~\ref{fig:iron_img_noise} shows the segmentations of the $\Lambda$ and $\mu$ fields corresponding to the worst-case scenarios in terms of DSC. Compared with Figs.~\ref{fig:iron_img}g--j, the inclusions do not differ significantly from those reconstructed with noise-free data. Noticeable changes are observed only in Fig.~\ref{fig:iron_img_noise}d.

\section{Conclusion}\label{s:concl}

This paper proposes a Finite Element Model Updating~(FEMU) framework for identifying spatially heterogeneous material properties in nonlinear solids subjected to mechanical contact under quasi-static conditions. The approach employs isogeometric analysis~(IGA) as a robust discretization method, particularly advantageous for contact computations, and covers both solids and thin shells. Material fields are reconstructed using experimentally measured resultant contact forces and full-field displacements measured at least on the free surface. The theoretical framework is based on a versatile hyperelastic material formulation that can represent a variety of materials. Material heterogeneity is described using low-order Lagrange interpolation independent of the IGA mesh, enabling the representation of non-smooth material fields and flexible adjustment of the inverse problem size to avoid under-~and overfitting. The FEMU least-squares objective is minimized using a gradient-based trust-region reflective algorithm. The computational efficiency of FEMU is improved by using analytical derivatives of the objective and incrementing material parameters instead of load levels between consecutive optimization iterations. The inverse analysis is initiated by assuming homogeneous material properties to obtain a reliable starting point for identifying spatially varying material fields. Several numerical examples based on synthetically generated experimental data contaminated with both systematic and random errors investigate various aspects of the proposed framework and demonstrate the potential of mechanical contact for reconstructing heterogeneous material properties. The key findings of this study are:
\begin{itemize}[noitemsep,topsep=0pt]
	\item Due to its ability to generate more data, contact-based inverse analysis~(CBIA) outperforms its pressure-based counterpart for noisy quasi-experimental data, as demonstrated by the abdominal wall example in Sec.~\ref{s:abdo}. However, CBIA requires a finer FE mesh than pressure-based analysis due to the localized deformations induced by contact.
	\item Accordingly, high identification errors are observed in the contact regions for the CBIA examples in Secs.~\ref{s:abdo} and~\ref{s:iron}. Since these errors originate from the concentration of FE discretization errors in the contact regions, the use of an adaptive nonuniform IGA mesh is a natural direction for future work.  
	\item The results for the Neo-Hookean block presented in Sec.~\ref{s:iron} show that parameters related to material compressibility, such as $\Lambda$, are particularly sensitive to noise in the experimental data. This observation appears to be characteristic of indentation tests and is consistent with the findings of~\citet{Hartmann2018}. 
	\item Full-field surface displacements and resultant contact forces are sufficient to identify the fields of Neo-Hookean model parameters in the bulk of the solid. However, the resulting inverse problem is severely ill-conditioned, deteriorating identifiability of material parameters and requiring regularization. Although the presence of an inclusion can be detected, the reconstructed peak values of the material parameters are subject to large errors, as shown in Sec.~\ref{s:ironS}.
	\item Analytical derivatives of the objective and the material iteration strategy provide a substantial speed-up for FEMU, enabling the solution of inverse problems with $n_\mrvr = 462$ material unknowns in Sec.~\ref{s:iron} and up to $n_\mrvr = 6498$ material unknowns in Sec.~\ref{s:abdo}, see Fig.~\ref{fig:abdoMEmeshConv}. However, for $n_\mrvr = 462$, the optimization algorithm becomes the computational bottleneck, even when Jacobian-vector products (see Appendix~\ref{s:jacvec}) are used.
\end{itemize}

The presented method applies to a broad range of identification problems involving mechanical contact and full-field measurements, including in-vivo biomechanics, in-situ structural engineering, and laboratory material testing. As in-silico simulations for medical diagnosis and treatment are expected to become increasingly common, there is a growing demand for robust and efficient inverse methods for identifying inhomogeneous material properties. 

The proposed framework can be extended in several ways. Extending the framework to frictional contact is a natural next step. However, as the material iteration strategy is limited to conservative systems, this is nontrivial. More reliable and efficient contact computations require adaptive nonuniform IGA discretizations. Nevertheless, the most relevant direction is the development of a robust material mesh selection strategy that mitigates both under-~and overfitting. For large inverse problems, more efficient optimization algorithms, such as L-BFGS, may provide further improvements. Realistic modeling of soft tissues requires extending the formulation beyond isotropic hyperelasticity to more complex material behaviors, including viscoelasticity and prestrain. Moreover, since the stress-free geometry is generally unavailable for soft bodies, future studies should investigate the simultaneous reconstruction of spatially varying material properties, stress-free geometry, and prestrain-inducing loading.

\bigskip

{\Large{\bf Acknowledgments}}

The authors thank Prof.~Izabela Lubowiecka for her comments and gratefully acknowledge financial support from the Research School of Ruhr University Bochum through their PhD Exchange Program and from the Doctoral School at Gda\'{n}sk University of Technology.

\appendix

\section{Derivatives of the objective function}\label{s:deriv}

This appendix presents two alternative approaches for computing the derivatives of $f(\mq)$ w.r.t. the design variable vector $\mq$, required by the TRR optimization algorithm. As usual in NLSQ problems, the derivatives are expressed in terms of the \textit{Jacobian} of the residual. The standard approach computes the Jacobian explicitly and is well-suited for relatively small inverse problems. The approach based on Jacobian-vector products avoids explicit formation of the Jacobian, and is hence suitable for larger inverse problems. It is important to note that all derivatives shown below are derived assuming a fixed contact active set $\mathcal{A}$, i.e.,
\eqb{l}
	\ds\pa{(\bullet)}{\mq} = \pa{(\bullet)}{\mq}\bigg|_{\mathcal{A} = \text{const.}} \,.
\label{e:A1_0}\eqe

\subsection{Standard formulation}\label{s:derivS}

Given the residuals
\eqb{l}
	\mbU_{\mrr\,i} := \mbU_{\mrex\,i} - \mbU_{\mrfe\,i}(\mq) \,, \qquad
	\mq_\mrrg := \alpha\mL(\mq - \mcc) \,, 
\label{e:A1_1}\eqe
with
\eqb{l}
 	\mbU_{\mrex\,i} :=	\begin{bmatrix}
 							\begin{array}{l}
 							 	\sqrt{w_\mrU}\, \mU_{\mrex\,i} / \norm{\mU_{\mrex\,i}} \\[1mm]
 								\sqrt{w_\mrF}\, \mF_{\mrex\,i} / \norm{\mF_{\mrex\,i}}
 							\end{array}
 						\end{bmatrix} \,,\qquad
 	\mbU_{\mrfe\,i}(\mq) :=	\begin{bmatrix}
 								\begin{array}{l}
 									\sqrt{w_\mrU}\, \mU_{\mrfe\,i} / \norm{\mU_{\mrex\,i}} \\[1mm]
									\sqrt{w_\mrF}\, \mF_{\mrfe\,i} / \norm{\mF_{\mrex\,i}}
 								\end{array}
 							\end{bmatrix} \,,
\label{e:A1_2}\eqe
the objective function in \eqref{e:objfnc} can be written in compact form as
\eqb{l}
	f(\mq) = \ds\sum_{i=1}^{n_\mrlc} \mbU_{\mrr\,i}^\T \mbU_{\mrr\,i} + \mq_\mrrg^\T\,\mq_\mrrg^{\phantom{\T}} \,.
\label{e:A1_3}\eqe
The derivatives of $f(\mq)$ are decomposed into residual and regularization parts. First, the contribution of a single load case to the residual part is considered. Accordingly, the subscript $i$ is dropped here for clarity. The \textit{gradient} of $f(\mq)$ is given by
\eqb{l}
	\mg(\mq) = \ds\left(\pa{f(\mq)}{\mq}\right)^{\!\!\T} = 2\,\mJ(\mq)^\T \mbU_\mrr(\mq) \,,
\label{e:A1_4}\eqe
where
\eqb{l}
	\mJ(\mq) = \ds\pa{\mbU_\mrr(\mq)}{\mq} = 	\begin{bmatrix}
													\mJ_\mrU \\
													\mJ_\mrF
												\end{bmatrix} \,,
\label{e:A1_5}\eqe
is the \textit{Jacobian} of the residual that consists of blocks
\eqb{l}
	\mJ_\mrU = -\ds\frac{\sqrt{w_\mrU}}{\norm{\mU_\mrex}} \pa{\mU_\mrfe}{\mq} \,,
\label{e:A1_6}\eqe
and
\eqb{l}
	\mJ_\mrF = -\ds\frac{\sqrt{w_\mrF}}{\norm{\mF_\mrex}} \pa{\mF_\mrfe}{\mq} \,.
\label{e:A1_7}\eqe
The \textit{Hessian} is given by
\eqb{l}
	\mH(\mq) = \ds\paq{f(\mq)}{\mq} =  2\,\mJ^\T\,\mJ + 2\ds\sum_{i=1}^{n_\mrex} [\brU_\mrr]_i \paq{[\brU_\mrr]_i}{\mq} \approx 2\,\mJ^\T\,\mJ \,,
\label{e:A1_8}\eqe
where the second term is neglected; thus, the Hessian is computed based only on the Jacobian. This simplification is common among algorithms solving NLSQ problems (see, e.g.,~\citet{hansen2013least}).

{\small \textbf{Remark 5}: A common strategy is to scale the design variables to improve the conditioning of the optimization problem, i.e., $\mq := \mD\,\mbq$ with $\mD := \diag\left[q_1/q_0\,, q_2/q_0\,, \dots\,, q_{n_\mrvr}/q_0\right]$. Consequently, $f$ becomes a function of the normalized design vector $\mbq$, and its derivatives are
\eqb{l}
	\ds\pa{(\bullet)}{\mbq} = \pa{(\bullet)}{\mq}\pa{\mq}{\mbq} = \pa{(\bullet)}{\mq}\,\mD\,,
\label{e:A1_9}\eqe
e.q., $\mJ(\mbq) = \mJ(\mq)\,\mD$.}\par

To compute \eqref{e:A1_6}, one needs to expand
\eqb{l}
	\ds\pa{\mU_\mrfe}{\mq} = \pa{\mU_\mrfe}{\muu}\pa{\muu}{\mq} \,,
\label{e:A1_10}\eqe
where $\muu$ is given in Eq.~\eqref{e:u}, and $\partial\mU_\mrfe/\partial\muu$ can be found in~\citet{Borzeszkowski2022}. 

{\small \textbf{Remark 6}: It is worth noting that due to the linear structure of Eq.~\eqref{e:uFE}, the matrix $\partial\mU_\mrfe/\partial\muu$ can be used to conveniently project kinematic quantities from the FE mesh onto the experimental grid, for example
\eqb{l}
	\mU_\mrfe = \ds\pa{\mU_\mrfe}{\muu}\,\muu\,,\qquad \mx_\mrex = \ds\pa{\mU_\mrfe}{\muu}\,\mx\,.
\label{e:A1_10b}\eqe}\par

The derivative $\partial\muu/\partial\mq$ follows from differentiating the equilibrium equation~\eqref{e:weakFE}
\eqb{l}
	\ds\pad{\mf(\muu(\mq),\mq)}{\mq} = \ds\pad{}{\mq}\left(\mf_\mrint(\muu(\mq),\mq) + \mf_\mrc(\muu(\mq)) - \mf_\mrex(\muu(\mq))\right) = \bf0 \,,
\label{e:A1_11}\eqe
which leads to a set of linear systems
\eqb{l}
	\ds\mK\pa{\muu}{\mq} = -\pa{\mf_\mrint}{\mq} = -\mS \,,
\label{e:A1_12}\eqe
where $\mK$ is the tangent stiffness matrix, given by
\eqb{l}
	\ds\mK =  \pa{\mf_\mrint}{\muu} + \pa{\mf_\mrc}{\muu} - \pa{\mf_\mrext}{\muu}\,,
\label{}\eqe
and $\mS = \partial\mf_\mrint / \partial\mq$ is the global sensitivity matrix, see \citet{Borzeszkowski2022} for further details. The sensitivities must be derived for each identified constitutive model, which is detailed in Appendix.~\ref{s:sensi}. The influence of the global volume constraint $g_V = V_0 - V = 0$, employed in Sec.~\ref{s:abdo}, is taken into account implicitly in $\mK$.

As the resultant contact force vector $\mF_\mrfe = \mP_\mrc\mf_\mrc$ depends on $\mq$ only through $\muu(\mq)$, expanding \eqref{e:A1_7} yields
\eqb{l}
	\ds\pa{\mF_\mrfe(\mq)}{\mq} = \mP_\mrc\,\pa{\mf_\mrc}{\muu}\pa{\muu}{\mq} = \mP_\mrc\,\mK_\mrc\pa{\muu}{\mq} = \mK_{\mrP\mrc}\pa{\muu}{\mq} \,,
\label{e:A1_13}\eqe
where $\mP_\mrc := [\bone,\bone,\dots,\bone]$ is a matrix consisting of $n_\mrno$ blocks that sums nodal contact forces over all Euclidean directions in the $d$-dimensional space, $\mK_\mrc$ is the contact contribution to the tangent stiffness matrix, and $\mK_{\mrP\mrc} := \mP_\mrc\,\mK_\mrc$. 

Finally, the contributions of the regularization are derived. Differentiating the regularization part in Eq.~\eqref{e:A1_3} gives
\eqb{l}
\	\mg_\mrrg(\mq) = 2\,\mJ^\T_\mrrg\mq_\mrrg \,,
\label{e:A1_14}\eqe
where
\eqb{l}
\	\mJ_\mrrg = \alpha\,\mL \,.
\label{e:A1_15}\eqe
The Hessian becomes
\eqb{l}
\	\mH_\mrrg(\mq) = 2\,\mJ^\T_\mrrg\mJ_\mrrg = \alpha^2\,\mL^\T\mL\,.
\label{e:A1_16}\eqe
In practice, all residuals in \eqref{e:A1_1} are concatenated into a single vector, and Eq.~\eqref{e:A1_3} reduces to a scalar product. The corresponding Jacobians are concatenated accordingly.

{\small \textbf{Remark 7}: If projections are applied on the FE or material dofs within the FE solver (e.g., due to periodic boundary conditions or reduction of a 2D material mesh to 1D), they can be directly incorporated into the tangent stiffness and global sensitivity matrices as follows
\eqb{l}
	\mK_\mrrd = \mP_\mrfe^{\phantom{\T}}\,\mK\,\mP_\mrfe^\T \,,\qquad \mS_\mrrd =  \mP_\mrfe^{\phantom{\T}}\,\mS\,\mP_\mrm^\T \,,
\label{}\eqe
where $\mP_\mrfe$ and $\mP_\mrm$ are FE and material projection matrices, respectively.}\par

\subsection{Jacobian-vector products}\label{s:jacvec}

Instead of computing and storing the Jacobian directly, the TRR algorithm can operate on Jacobian-vector products, i.e., $\mJ\,\mhq$, $\mJ^\T\mhU$, and $\mJ^\T(\mJ\,\mhq)$. All formulas below are given for a single load case contribution. Vectors $\mhq$ and $\mhU$ have length of $n_\mrvr$ and $n_\mrex$, respectively. Combining Eqs.~\eqref{e:A1_6}, \eqref{e:A1_10}, and \eqref{e:A1_12} for the displacements part, and Eqs.~\eqref{e:A1_7}, \eqref{e:A1_12}, and \eqref{e:A1_13} for the contact part, the $\mJ\,\mhq$ product becomes
\eqb{lll}
\begin{cases}
	\mJ_\mrU\,\mhq = \ds\frac{\sqrt{w_\mrU}}{\norm{\mU_\mrex}}\pa{\mU_\mrfe}{\muu}\,\blam_\mrq \,,\\[5mm]
	\mJ_\mrF\,\mhq = \ds\frac{\sqrt{w_\mrF}}{\norm{\mF_\mrex}}\mK_{\mrP\mrc}\,\blam_\mrq \,,\\[4mm]
	\mK\,\blam_\mrq = \mS\,\mhq \,,
\end{cases} 
\label{}\eqe
where $\blam_\mrq$ is a vector with $d\,n_\mrno$ components. Correspondingly, using Eqs.~\eqref{e:A1_6}, \eqref{e:A1_10}, and \eqref{e:A1_12}, the displacements part of $\mJ^\T\mhU$ is given by
\eqb{lll}
\begin{cases}
	\mJ^\T_\mrU\,\mhU = \ds\frac{\sqrt{w_\mrU}}{\norm{\mU_\mrex}}\mS^\T\,\blam_{\mrU\mrU} \,,\\[5mm]
	\mK^\T\,\blam_{\mrU\mrU} = \ds\left(\pa{\mU_\mrfe}{\muu}\right)^{\!\!\T}\,\mhU \,,
\end{cases} 
\label{}\eqe
while using Eqs.~\eqref{e:A1_7}, \eqref{e:A1_12}, and \eqref{e:A1_13}, the contact part of $\mJ^\T\mhU$ takes the form
\eqb{lll}
\begin{cases}
	\mJ^\T_\mrF\,\mhU = \ds\frac{\sqrt{w_\mrF}}{\norm{\mF_\mrex}}\mS^\T\,\blam_{\mrU\mrF} \,,\\[5mm]
	\mK^\T\,\blam_{\mrU\mrF} = \mK_{\mrP\mrc}^\T\,\mhU \,,
\end{cases} 
\label{}\eqe
where $\blam_{\mrU\mrU}$ and $\blam_{\mrU\mrF}$ are vectors of length $d\,n_\mrno$. Note that in the presence of contact $\mK$ may not be symmetric, especially for friction. The regularization parts are given by 
\eqb{l}
	\mJ_\mrrg\,\mhq = \alpha\,\mL\mhq \,,\qquad \mJ^\T_\mrrg\,\mhU = \alpha\,\mL^\T\mhU \,.
\label{}\eqe
The mixed product $\mJ^\T(\mJ\mhq)$ is computed as a composition of the two above. 

{\small \textbf{Remark 8}: In a similar manner, the contribution of a single load case to the gradient of $f(\mq)$ can be efficiently computed via the adjoint formulation
\eqb{lll}
	\begin{cases}
		\mg^\T(\mq) = \blam^\T\mS \,,\\
		\mK^\T\blam = -2\ds\left(\pa{\mbU_\mrr}{\muu}\right)^{\!\!\T}\mbU_\mrr \,,
	\end{cases} 
\label{}\eqe
with
\eqb{l}
	\ds\pa{\mbU_\mrr}{\muu} = 
	\begin{bmatrix}
		\begin{array}{l}
			-\ds\frac{\sqrt{w_\mrU}}{\norm{\mbU_\mrex}}\pa{\mbU_\mrfe}{\muu} \\[4mm]
			-\ds\frac{\sqrt{w_\mrU}}{\norm{\mbU_\mrex}}\mK_{\mrP\mrc} 
		\end{array}
	\end{bmatrix} \,,
\label{}\eqe
see, e.g.,~\citet{GIVOLI2021}. These formulas are applicable to optimization algorithms that rely solely on the gradient, such as L-BFGS.}\par 

\section{Analytical sensitivities}\label{s:sensi}

Computation of the objective function derivatives in any form requires differentiating the internal FE forces vectors in Eqs.~\eqref{e:fint3D},~\eqref{e:FEvectortau}, and~\eqref{e:FEvectorM} w.r.t.~the material unknowns vector $\mq$. This derivative is referred to as \textit{sensitivity matrix}, and its elemental contribution is given by
\eqb{l}
	\mS^{e\bre}_\mrint := \ds\pa{\mf^e_\mrint}{\mq_{\bre}} \,.
\label{e:sens}\eqe
As this formula incorporates both FE and material meshes, a mapping between their elements is required, see Sec.~\ref{s:matFE}. In the case of the compressible Neo-Hookean solid model, by applying Eq.~\eqref{e:qdiscr} to $\Lambda$ and $\mu$, the change of the internal force vector due to a change in material unknowns becomes
\eqb{l}
	\Delta\mf^e_\mrint = \ds\pa{\mf^e_\mrint}{\mLam^{\bre}}\Delta\mLam^{\bre} + 
	\pa{\mf^e_\mrint}{\mmu^{\bre}}\Delta\mmu^{\bre} = 
	\mS_\Lambda^{e\bre}\Delta\mLam^{\bre} + \mS_\mu^{e\bre}\Delta\mmu^{\bre} \,,
\eqe
where 
\eqb{l}
	\mS_\Lambda^{e\bre} := \ds\intoe {\mB^e}^\T\,\mI_\Lambda\,\bar\mN^{\bre} \,\dif v 
\label{e:sensNHlam}\eqe
and
\eqb{l}
	\mS_\mu^{e\bre} := \ds\intoe {\mB^e}^\T\,\mI_\mu\,\bar\mN^{\bre} \,\dif v 
\label{e:sensNHmu}\eqe
are the Neo-Hookean sensitivities for $\Lambda$ and $\mu$, respectively. The two vectors in \eqref{e:sensNHlam}~and~\eqref{e:sensNHmu} are given by
\eqb{lll}
	\mI_\Lambda	\dis	\ds\frac{\ln \tJ}{\tJ}\begin{bmatrix}1 & 1 & 1 & 0 & 0 & 0\end{bmatrix}^\T \,, \\[4mm]
	\mI_\mu		\dis	\ds\frac{1}{\tJ}\begin{bmatrix}\tB^\ast_{11} & \tB^\ast_{22} & \tB^\ast_{33} & \tB^\ast_{12} & \tB^\ast_{13} & \tB^\ast_{23}\end{bmatrix}^\T \,,
\eqe
where $\tbB^\ast = \tbF\tbF^\T - \bI$. 

Following \citet{Borzeszkowski2022}, the analytical sensitivities for the Koiter shell model follow from
\eqb{l}
\Delta\mf^e_\mrint = \ds\pa{\mf^e_\mrint}{\mE^{\bre}}\Delta\mE^{\bre} + 
\pa{\mf^e_\mrint}{\mT^{\bre}}\Delta\mT^{\bre} = 
\mS_{E}^{e\bre}\Delta\mE^{\bre} + \mS_{T}^{e\bre}\Delta\mT^{\bre} \,,
\eqe
with 
\eqb{l}
\mS_{E}^{e\bre} := \ds\intooe \left(
\frac{T}{1+\nu}{\mN^e}^{\T}_{\!\!\!,\alpha}\,\Iabgd_\nu\,\eps_{\gamma\delta}\,\ba_\beta + 
\frac{T^3}{12(1+\nu)}{\mN^e}^{\T}_{\!\!\!;\alpha\beta}\,\Iabgd_\nu\,\kappa_{\gamma\delta}\,\bn
\right)\bar\mN^{\bre} \,\dif A \,,
\label{e:sensKoitE}\eqe
and
\eqb{l}
\mS_{T}^{e\bre} := \ds\intooe \left(
\frac{E}{1+\nu}{\mN^e}^{\T}_{\!\!\!,\alpha}\,\Iabgd_\nu\,\eps_{\gamma\delta}\,\ba_\beta + 
\frac{E\,T^2}{4(1+\nu)}{\mN^e}^{\T}_{\!\!\!;\alpha\beta}\,\Iabgd_\nu\,\kappa_{\gamma\delta}\,\bn
\right)\bar\mN^{\bre} \,\dif A \,,
\label{e:sensKoitT}\eqe
where
\eqb{l}
\Iabgd_\nu = \ds\frac{\nu}{1-\nu}\Iabgd_\mathrm{dil} + \Iabgd_\mathrm{dev} \,,
\label{e:Iabgd}\eqe
and 
\eqb{l}
\eps_{\alpha\beta} = \ds\frac{1}{2}(\auab - \Auab) \,, \qquad \kappa_{\alpha\beta} = \buab - \Buab \,.
\label{}\eqe
Analytical sensitivities for the Canham bending model can be found in \citet{Borzeszkowski2022}. A convenient implementation of analytical sensitivities can be achieved following the approach shown in \citet{Duong2017}.

All $\mS^{e\bre}_\bullet$ are of size $dn_e \times \brn_e$ and require numerical integration over the element $\Omega^e$, followed by the local assembly with respect to $\brd$ unknown material parameters, and the global assembly for all $e = 1,\dots,n_\mrel$ and $\bre = 1,\dots,\brn_\mrel$. This leads to the global sensitivity matrix $\mS$ with dimensions $dn_\mrno \times \brd\bar n_\mrno$. Since all FEs are mapped to the corresponding ME, the elemental sensitivities can be computed within a standard element subroutine; thus, reducing the computational effort.

{\small \textbf{Remark 9}: In fact, the sensitivities should be calculated in a separate element loop run after the NR method converges. However, using $\mS$ and $\mK$ from the penultimate NR iteration to evaluate the objective function derivatives does not affect their accuracy in the author's experience.}\par

\bibliographystyle{apalike} 
\bibliography{references}

\end{document}